\documentclass[11pt]{article}

\usepackage{times, mathptmx}

\usepackage[letterpaper,left=2.5cm,top=0.5cm,right=2.5cm,bottom=3cm]{geometry}
\usepackage{amsmath, amssymb}
\usepackage{amscd,amsmath,latexsym}
\usepackage{mathrsfs}
\usepackage{bm,bbm}
\usepackage{dsfont}
\usepackage{lmodern}
\usepackage{amsfonts, multirow, epsfig, floatrow}
\usepackage{graphicx, pdflscape, verbatim, enumerate, colortbl, setspace}
\usepackage{setspace, color,bm}
\usepackage[normalem]{ulem}
\usepackage[authoryear, round]{natbib}
\usepackage{cite}
\usepackage{multirow}
\usepackage{algorithm}
\usepackage{algorithmic}
\usepackage{booktabs,array}
\PassOptionsToPackage{hyphens}{url}\usepackage[colorlinks=true, citecolor=blue]{hyperref}
\usepackage{float}
\graphicspath{{./art/}}
\usepackage{caption}
\usepackage{subcaption}
\usepackage{adjustbox}
\usepackage{tikz}
\usepackage{tikz}
\usepackage{makecell}
\usetikzlibrary{patterns}

\newtheorem{theorem-new}{Theorem} 
\newtheorem{defi}{Definition}
\newtheorem{theorem}{Theorem} 

\newtheorem{corollary}{Corollary}

\newtheorem{lemma}{Lemma} 
\newtheorem{assumption}{Assumption}
\newtheorem{proposition}{Proposition} 

\newtheorem{remark}{Remark}  

\usepackage{array}
\newcolumntype{M}[1]{>{\centering\arraybackslash}m{#1}}
\providecommand{\keywords}[1]
{
  \small	
  \textbf{\textit{Keywords---}} #1
}

\usepackage{enumerate}

\title{L-2 Regularized maximum likelihood for $\beta$-model in large and sparse networks}
\author{Meijia Shao$^{\spadesuit}$, 
        Yu Zhang$^{\flat}$, Qiuping Wang$^\diamondsuit$,
        Yuan Zhang$^{\spadesuit}$\footnote{Correspondence author, email \href{yzhanghf@stat.osu.edu}{yzhanghf@stat.osu.edu}.},
        Jing Luo$^\#$ and
        Ting Yan$^\flat$
        {\footnote{{The first two authors contributed equally.}}}
        \\
        \ 
        \\
        {\small $\spadesuit$} Department of Statistics, The Ohio State University, U.S.A.
        \\
        {\small $\flat$} Department of Statistics, Central China Normal University, China \\
        {\small $\diamondsuit$} School of Mathematics and Statistics, Zhaoqing University, China \\
        {\small $\#$} School of Mathematics and Statistics, South-Central Minzu University, China
}
\date{}

\newcommand{\ep}{\mathbb{E}}

\newcommand{\pr}{\mathbb{P}}

\newcommand{\var}{\mathrm{Var}}

\newcommand{\tr}{\mathrm{Tr}}

\newcommand{\proj}{{\cal P}}
\newcommand{\projperp}{{\cal P}_\perp}

\renewcommand{\hat}{\widehat}
\renewcommand{\tilde}{\widetilde}

\definecolor{yuancolor1}{rgb}{0.56, 0.0, 1.0}
\definecolor{yuancolor2}{rgb}{0.0, 0.42, 0.24}
\definecolor{yuancolor3}{rgb}{0.13,0.55,0.13} 

\definecolor{yuancolor-revision-1}{rgb}{0.06, 0.3, 0.57}

\definecolor{meijia-revision-1}{RGB}{131, 50, 168}

\definecolor{yuancolor-revision-1}{rgb}{0, 0, 0}
\definecolor{meijia-revision-1}{rgb}{0, 0, 0}

\begin{document}
\maketitle

\begin{abstract}
The $\beta$-model is a powerful tool for modeling large and sparse networks driven by degree heterogeneity, where many network models become infeasible due to computational challenge and network sparsity.
However, existing estimation algorithms for $\beta$-model do not scale up.
Also, theoretical understandings remain limited to dense networks.
This paper brings several significant improvements over existing results to address the urgent needs of practice.
We propose a new $\ell_2$-penalized MLE algorithm that can comfortably handle sparse networks of millions of nodes with much-improved memory parsimony.
We establish the first rate-optimal error bounds and high-dimensional asymptotic normality results for $\beta$-models, under much weaker network sparsity assumptions than best existing results.
Application of our method to large COVID-19 network data sets discover meaningful results.
\end{abstract}

\section{Introduction}
\label{section::introduction}

The $\beta$-model \citep{chatterjee2011random} is a heterogeneous exponential random graph model (ERGM) with the degree sequence as the only sufficient statistic.
It is a popular choice for characterizing networks mainly driven by degree heterogeneity.  
Under this model, an undirected and binary network of $n$ nodes, represented by its adjacency matrix $A=(A_{i,j})_{1\leq \{i,j\}\leq n}\in\{0,1\}^{n\times n}$, is generated by
\begin{equation}
    \pr(A_{i,j}=1) = \dfrac{e^{\beta_i^*+\beta_j^*}}{1+e^{\beta_i^*+\beta_j^*}},  
    \quad
    \textrm{where }
    A_{i,j}=A_{j,i},
    \quad
    \textrm{for all }
    1\leq i<j\leq n,
    \label{model::original-beta-model}
\end{equation}
where $\beta^*=(\beta_1^*,\ldots,\beta_n^*)$ denotes the vector of true model parameters.
Assume that $A_{i,j}$' s are mutually independent and there is no self-loop: $A_{i,i}\equiv0$ for all $i$. 
The negative log-likelihood function is
\begin{align}
    \mathcal{L}({\beta})
    = &
    \sum_{1\leq i<j\leq n}\log\big(1+e^{\beta_{i}+\beta_{j}}\big)-\sum_{i=1}^{n}\beta_{i}d_{i}.
    \label{neg-log-likelihood::vanilla}
\end{align}
where $d_1,\ldots,d_n$ are observed node degrees: $d_i=\sum_{j\in[1:n]\backslash \{i\}}A_{i,j}$.

The $\beta$-model is a simple, yet expressive tool for describing degree heterogeneity \citep{babai1980random, fienberg2012brief}.
Compared to other network models, it enjoys two attractive features.
First, as is evident from \eqref{neg-log-likelihood::vanilla}, fitting a $\beta$ model requires only the knowledge of degrees, not the detailed adjacency matrix, allowing it to offer significant privacy protection \citep{fan2020asymptotic}. 
For example, \citet{elmer2020students} collected social network data from a group of Swiss students before and during COVID-19 lockdown. 
To protect privacy, they released only node degrees rather than the full adjacency matrix. 
The $\beta$-model is particularly suitable for analyzing this data set, while most other popular models that require the adjacency matrix cannot be applied.
Second, it can scale up to very large and sparse networks. 
In Section \ref{section::data-examples}, we will analyze a massive COVID-19 knowledge graph \citep{covid_kg} that contains $n\approx 10^7$ non-isolated nodes with just $2.1\times 10^7$ edges, where our method demonstrates high speed and memory parsimony.

Since its introduction, the $\beta$-model has attracted considerable research interest.  
The early works \citet{holland1981exponential, chatterjee2011random,park2004statistical,rinaldo2013maximum, hillar2013maximum} studied basic model properties and established existence, consistency and asymptotic normality of the maximum likelihood estimator (MLE).
\citet{chen2013directed, yan2016asymptotics, yan2019statistical, stein2021sparse} extended the model for directed and bipartite networks.
\citet{karwa2016inference} studied differential privacy in $\beta$-model. 
\citet{graham2017econometric, su2018note, gao2020nonparametric,yan2019statistical, stein2020sparse, stein2021sparse} incorporated nodal or edge-wise covariates.
\citet{wahlstrom2017beta} established the Cramer-Rao bound with repeated network observations.
\citet{mukherjee2018detection} studied a different variant of $\beta$-model for sparse networks with a known sparsity parameter.

Despite significant research advances, several important challenges in $\beta$-model analysis remain unaddressed. These include:
handling very sparse networks, 
developing scalable estimation algorithms, 
improving theoretical understanding of the estimator's finite-sample and asymptotic behaviors, 
and  determining fundamental limits of estimation accuracy. 
This paper aims to address these challenges.

\subsection{ Regularization schemes for the $\beta$-model}
\label{subsec::intro::regularized-beta-models}

Recently, regularization has been introduced as a promising technique to address many of the aforementioned challenges. 
The pioneering works \citet{chen2019analysis} and \citet{stein2020sparse} introduced $\ell_0$- and $\ell_1$-regularized $\beta$-models, respectively.
 To streamline the presentation, we first describe the modeling assumptions in \citet{chen2019analysis} and \citet{stein2020sparse}, and then review their methods.
\citet{chen2019analysis} proposed the \emph{sparse $\beta$-model (S$\beta$M)}. 
Suppose there exists a set ${\cal S}\subseteq[1:n]$, called the \emph{active set}, such that
\begin{align}
    \beta_i^* =
    \begin{cases}
        -(\gamma/2) \log n + o(\log n),
        &\textrm{ if }i\notin {\cal S},
        \\
        (\alpha_i-\gamma/2) \log n + o(\log n),
        &\textrm{ if }i\in {\cal S},
    \end{cases}
    \label{setting::chen-kato-leng}
\end{align}
where \citet{chen2019analysis} assumes that $\alpha_i= \alpha$ for all $i$ and that both $\alpha$ and $\gamma$ are \emph{known} constants, satisfying  the following conditions to ensure that the network is not too sparse:
\begin{align}
    \gamma\in[0,2),
    \quad
     \alpha_i\equiv~\alpha\in [0,1),
    \quad\textrm{and}\quad
    0\leq \gamma-\alpha<1.
    \label{assumption-chen-kato-leng}
\end{align} 
Another key modeling assumption in \citet{chen2019analysis} is:
\begin{align}
    |{\cal S}|\ll n^{1-\alpha}.
    \label{beta-sparsity-assumption}
\end{align}
This assumption ensures that the vast majority of nodes lie outside the active set and share similar degrees, so that the relatively few active nodes with markedly different degrees stand out and can be accurately estimated.
Such structure assumed in both \citet{chen2019analysis} and \citet{stein2020sparse} can be summarized as a "$\beta$-sparsity" assumption, as follows.
\begin{assumption}[$\beta$-sparsity, \citet{chen2019analysis, stein2020sparse}]
    \label{assumption::beta-sparsity}
    Most true parameter $\beta_i^*$'s share a common value, except for indices in a so-called \emph{``active set''}, denoted by ${\cal S}\subset \{1,\ldots,n\}$.
    The active set ${\cal S}$ is small in size, as specified in \eqref{beta-sparsity-assumption}.
    All nodes $i$ outside ${\cal S}$ have nearly identical true values of $\beta_i^*$, as specified in \eqref{setting::chen-kato-leng}.
\end{assumption}
In summary, S$\beta$M is formally specified by combining equations \eqref{model::original-beta-model}, \eqref{setting::chen-kato-leng}, \eqref{assumption-chen-kato-leng} and \eqref{beta-sparsity-assumption}.
Equations \eqref{model::original-beta-model} and \eqref{setting::chen-kato-leng} formulate the model, while \eqref{assumption-chen-kato-leng} and \eqref{beta-sparsity-assumption} impose key assumptions.
We describe them together here for convenience in Sections \ref{subsec::theory::upper-bounds} and \ref{subsec::theory::lower-bounds}.
In a follow-up work, \citet{stein2020sparse} relaxed this model by dropping the requirement that $\alpha_i\equiv \alpha$.

Next, we review the regularization methods.
\citet{chen2019analysis} proposed a provably consistent parameter estimation method assuming a \emph{knownn} active set ${\cal S}$.
They then proposed a method to consistently estimate ${\cal S}$.
Conceptually, their approach can be understood as an $\ell_0$-regularized MLE:
\begin{align}
    \arg\min_{\beta,\beta_0}
    {\cal L}_{\lambda;\beta_0}(\beta)
    :=&~
    {\cal L}(\beta) + \lambda\big\|\beta - \beta_0\cdot\mathbbm{1}\big\|_0,
    \label{method::L0-MLE}
\end{align}
where $\mathbbm{1}$ is an all-one vector and $\lambda$ is a tuning parameter,
despite \eqref{method::L0-MLE} was not explicitly stated in \citet{chen2019analysis}.
Later, \citet{stein2020sparse} proposed an $\ell_1$-regularized MLE:
\begin{align}
    \arg\min_{\beta,\beta_0}
    {\cal L}_{\lambda;\beta_0}(\beta)
    :=&~
    {\cal L}(\beta) + \lambda\big\|\beta - \beta_0\cdot\mathbbm{1}\big\|_1.
    \label{method::L1-MLE}
\end{align}
 This formulation serves as a softer version of \eqref{method::L0-MLE}.
Both \eqref{method::L0-MLE} and \eqref{method::L1-MLE} pursue $\beta$-sparsity as in Assumption \ref{assumption::beta-sparsity}, but in some data examples, this assumption might not hold.
This motivates us to propose an $\ell_2$-regularized MLE:
\begin{align}
    \arg\min_\beta
    {\cal L}_\lambda(\beta)
    :=&~
    {\cal L}(\beta) + \frac\lambda2
    \big\|\beta-\bar\beta\cdot\mathbbm{1}\big\|_2^2,
    \label{our-method}
\end{align}
where $\bar\beta = n^{-1}\sum_{i=1}^n \beta_i$ replaces the "$\beta_0$'' in \eqref{method::L1-MLE}, since  under the $\ell_2$ penalty, the optimal $\beta_0$ equals $\beta_0=\bar\beta$.
 We will show, both theoretically and numerically, that our method offers more flexibility and better parameter estimation accuracy across many settings, in the presence or absence of $\beta$-sparsity.

\subsection{Our contributions}
\label{subsec::intro::our-contributions}

This paper substantially advances the methodology and theoretical foundations in the $\beta$-model literature.
Table \ref{table::comparison} provides a detailed comparison with the current state-of-the-art: \citet{chen2019analysis} and \citet{stein2020sparse}.
Here, we do not repeat Table \ref{table::comparison}, but instead highlight big-picture insights.

Historically, regularization in $\beta$-models was introduced to address network sparsity.
 Importantly, imposing the $\beta$-sparsity structure (Assumption \ref{assumption::beta-sparsity}) reduces the model's parameter complexity from $O(n)$ to $O(|{\cal S}|)$, thus effectively handling much sparser networks than earlier works.
However, their theoretical results left the impression that $\beta$-sparsity is the key to this success.

Surprisingly, our findings are very different.
We showed that even \emph{slightly} penalized MLE  is sufficient to address much sparser network regimes than the network sparsity regimes described in \citet{chen2019analysis} and \citet{stein2020sparse}, \emph{without} the restrictive $\beta$-sparsity assumption.
Moreover, imposing a nonzero amount of regularization has its merit -- we proved that \emph{any} nonzero amount of regularization suffices to guarantee the existence of the MLE.
This result substantively improves upon the complicated MLE existence conditions in \citet{rinaldo2013maximum, chen2019analysis}, which cannot be feasibly verified in practice.
 Therefore, we propose $\ell_2$-regularized MLE.
We chose $\ell_2$ over other $\ell_p$ penalty terms for its analytical simplicity.
Our  analysis techniques are novel and yield much sharper results  than the previous state-of-the-art \citet{chen2019analysis, stein2020sparse}.
 For instance, our error bound does not critically depend on $|{\cal S}|$ and can handle the non-$\beta$-sparse case.
Notably, we also established lower bounds for the first time and proved that our error bounds are rate-optimal.

In summary, the key takeaway is \emph{not} that ``$\ell_2$ is better than $\ell_0$ or $\ell_1$'', but that \emph{substantive regularization is unnecessary, while a slight $\ell_2$ penalty ensures MLE existence and lets us thoroughly analyze the estimator's behaviors -- something $\ell_0$ and $\ell_1$ regularization cannot do as elegantly}.

For practitioners, we offer a  comprehensive and significantly enhanced toolbox.
First, we propose a novel fast algorithm that can easily handle networks with millions of nodes by leveraging network sparsity and the monotonicity property of the MLE for the $\beta$-model.
Second, we provide a data-driven tuning method for selecting the appropriate amount of regularization.
Third, when the $\beta$-sparsity assumption holds, we offer a post-estimation thresholding algorithm for recovering the active set ${\cal S}$.
Our method is more flexible than the methods proposed by \citet{chen2019analysis, stein2020sparse}, as we allow nodes in ${\cal S}$  to have degrees both above and below the median degree.

We apply our method to real-world data sets and obtained interesting and meaningful results.

\begin{table}[h]
\renewcommand{\arraystretch}{1.2}
   \aboverulesep=0ex 
   \belowrulesep=0ex 
 \adjustbox{max width=1.1\linewidth,width=1.1\linewidth,center=\linewidth}{
    \centering
    \normalsize
    \begin{tabular}{c | M{4cm} | M{4cm} | M{4cm} | M{3cm}}
        \toprule  
& \citet{chen2019analysis}
& \citet{stein2020sparse}
& Our paper
& Discussed in
\\\bottomrule
Regularization type
& $l_0$   
& $l_1$ 
& $l_2$ 
& Section \ref{subsec::intro::regularized-beta-models}
\\\hline
Model assumption &
\eqref{setting::chen-kato-leng}, with $\alpha_i\equiv\alpha$, $|{\cal S}|\ll n^{1-\max_i\alpha_i}$
&
\eqref{setting::chen-kato-leng}, with $|{\cal S}|\ll n^{1-\max_i\alpha_i}$
&
\eqref{model::original-beta-model} (contains \eqref{setting::chen-kato-leng} as a special case)
&
Section \ref{subsec::intro::regularized-beta-models}
\\\hline
Parameter sign constraints    
& \emph{Known} $\gamma$ and $\alpha$; $\alpha >0$
& $\alpha_i > 0$ 
& None 
& Section \ref{subsec::intro::regularized-beta-models}
\\\hline
MLE existence &
Not always guaranteed &
Yes &
Yes
&
Section \ref{section::our-method}
\\\hline
Computation cost  &
$O(n^3)$ per iteration&
$O(n^3)$ per iteration&
$O(n\cdot \bar{d})$ + ($O(m^2)$ per iteration)
\footnote{: $\bar{d}$ is the average degree; $m$ is the number of different unique degrees.  See Section \ref{section::our-method}.}
&
Section \ref{section::our-method}
\\ \hline 
Network sparsity requirement \footnote{: All methods require denser networks as degrees become unbalanced.  The table present the best cases.} &
$\gg n^{-1/2}$ & 
$\gg n^{-1/6}$ & 
$\gg n^{-1}$
&
Section \ref{subsec::theory::upper-bounds}
\\\hline
Finite-sample error bound &
No  & 
Yes & 
Yes 
&
Section \ref{subsec::theory::upper-bounds}
\\\hline
Recover ${\cal S}$ (under \eqref{setting::chen-kato-leng}) &
Yes & Yes & Yes
&
Section \ref{subsec::theory::upper-bounds}
\\\hline
Lower bound result  &
No  &
No  &
Yes 
&
Section \ref{subsec::theory::lower-bounds}
\\\hline
Local rate-optimality &
No  &
No  &
Yes 
&
Section \ref{subsec::theory::lower-bounds}
\\\hline
Asymptotic normality  &
Fixed-dimensional&Fixed-dimensional&
High-dimensional 
&
Section \ref{subsec::high-dim-asymp-norm}
\\ \hline
Data-driven tuning of $\lambda$ &
Yes & Yes & Yes
&
Section \ref{section::tuning-parameter}
\\\hline
Empirical scalability &
$n\lesssim 10^{3}$  &
$n\lesssim 10^{3}$  &
At least $n\approx 10^{7}$  
&
Sections \ref{section::numerical-studies}
and \ref{section::data-examples}
\\\bottomrule
    \end{tabular}
    }
   \caption{Comparison table between \citet{chen2019analysis}, \citet{stein2020sparse} and our paper.}
   \label{table::comparison}
\end{table}

\subsection{Notation}
\label{section::introduction::subsec::notation}
We  adopt the standard asymptotic notation $O(\cdot)$, $o(\cdot)$, $\lesssim$ and $\asymp$ from calculus.
Let $\mathbbm{1}=(1,\ldots,1)^T\in\mathbb{R}^{n\times 1}$ and define the projection matrices
$
    \proj = n^{-1}\mathbbm{1}\mathbbm{1}^T
$
and
$
    \projperp = I - n^{-1}\mathbbm{1}\mathbbm{1}^T.
$
For any vector $u\in\mathbb{R}^n$ and matrix $U\in\mathbb{R}^{n\times n}$, define matrices $V(u)$ and $V(U)$ as follows: for $i\neq j$, set $\{V(u)\}_{i,j} = e^{u_i+u_j}/(1+e^{u_i+u_j})^2$ and $\{V(U)\}_{i,j} = e^{U_{i,j}}/(1+e^{U_{i,j}})$;
while in both $V=V(u)$ and $V=V(U)$, the diagonal elements are given by $V_{i,i} = \sum_{j:j\neq i} V_{i,j}$.
For a matrix $J\in\mathbb{R}^{n\times n}$, define $\|J\|_\infty = \sup_{x\neq 0}\|Jx\|_\infty/\|x\|_\infty = \max_{1\leq i\leq n}\sum_{j=1}^n|J_{i,j}|$.
Write $\|\cdot\|_F$ for the Frobenius norm and $\|\cdot\|_{\rm op}$ for the spectral norm.
Finally, we import two concepts of "sparsity" from \citet{chen2019analysis}: \emph{network sparsity},  defined by $\rho_n := \sum_{1\leq i<j\leq n}\ep[A_{i,j}]/\binom{n}2$, and \emph{$\beta$-sparsity},  which is formally described in Assumption \ref{assumption::beta-sparsity}.

\section{L-2 regularization and fast parameter estimation}
\label{section::our-method}

 We now formulate our proposed $\ell_2$-regularized MLE, ${\cal L}_\lambda(\beta)$, as in \eqref{our-method}. 
Denote the gradient of ${\cal L}_\lambda(\beta)$ by $F(\beta) = (F_1(\beta), \ldots, F_n(\beta))$.
 Here,
\begin{align}
    F_i(\beta) = & \frac{\partial {\cal L}_\lambda(\beta)}{\partial \beta_i} = \sum_{\substack{1 \leq j \leq n \\ j \neq i}} \frac{e^{\beta_i + \beta_j}}{1 + e^{\beta_i + \beta_j}} - d_i + \lambda(\beta_i - \bar\beta), 
    \quad 
    1\leq i \leq n.
    \label{eqn::unknown-mu::gradient}
\end{align}
The estimator $\hat{\beta}_\lambda$ is the solution to the following set of equations:
\begin{equation}
    F_i(\hat{\beta}_\lambda) = 0,
    \quad
    1\leq i\leq n.
    \quad
    \textrm{( Equivalently,  $F(\hat\beta_\lambda)=0$.)}
    \label{eqn::estimation-equation-main}
\end{equation}
The uniqueness of $\hat{\beta}_\lambda$ follows from the convexity of ${\cal L}_\lambda(\beta)$.
The existence of $\hat{\beta}_\lambda$,
however, has historically required complicated conditions \citep{hillar2013maximum, rinaldo2013maximum, chen2019analysis}\footnote{For example, Proposition 2.1 of \citet{hillar2013maximum} and Appendix B of \citet{chen2019analysis} assume that the degree vector $d$ is an interior point of $B$, where $B$ is solved from $d = B\cdot {\rm vec}(A)$ for vec$(A)$ being the vectorization of $A$.}.
In contrast, we  present a much cleaner result in Lemma \ref{lemma::MLE-existence}.
\begin{lemma}
    \label{lemma::MLE-existence}
    Suppose $0\max_i d_i>0$ and $\min_i d_i<n-1$.
    For any $\lambda>0$,
    there exists a unique $\hat\beta_\lambda\in\mathbbm{R}^n$ such that $F(\hat\beta_\lambda)=0$.
\end{lemma}

We now present a scalable algorithm to  obtain $\hat\beta_\lambda$.
Existing works \citep{chen2019analysis, stein2021sparse} typically treat the problem as a logistic regression of sample size $O(n^2)$ and delegate parameter estimation to standard GLM packages such as {\tt glmnet}. 
This requires $O(n^3)$ computation per iteration \citep{AmazonH2O, Hastie2009}, and their algorithms can only scale up to $10^3$ nodes.
A better method is to solve \eqref{eqn::estimation-equation-main} by gradient descent, Newton's method,  or fixed-point iteration (cf. \citet{chatterjee2011random, yan2015asymptotic, yan2016asymptotics}), yet still incur an $O(n^2)$ per-iteration cost.
Here, we propose a novel algorithm that  leverages the special structure of the $\beta$-model and the sparsity of large networks, inspired by the following monotonicity lemma.
\begin{lemma}
    \label{lemma::degree-monotonicity}
    The MLE $\hat\beta_\lambda$, defined as the solution to \eqref{eqn::estimation-equation-main}, satisfies that 
    $
    \hat\beta_{\lambda;i} = \hat\beta_{\lambda;j}
    $
    if and only if $d_i=d_j$, for any $1\leq i<j\leq n$, where $\hat\beta_{\lambda;i}$ denotes the $i$th element of $\hat\beta_\lambda$.
\end{lemma}
Lemma \ref{lemma::degree-monotonicity} is not new.
The phenomenon of monotonicity was first discovered by \citet{hillar2013maximum} (Proposition 2.4) and later reiterated by \citet{chen2019analysis}.
However, \citet{hillar2013maximum} did not exploit this result for computation, and \citet{chen2019analysis} used it only for model selection  rather than parameter estimation.
In contrast, we show that Lemma \ref{lemma::degree-monotonicity} can greatly reduce parameter dimensionality, leading to a significant improvement in both speed and memory efficiency.
Let
$
    d_{(1)}<d_{(2)}<\cdots<d_{(m)}
$
be the sorted unique values of observed degrees, where $m = \big| \textrm{Unique}\big(\{d_1,\ldots,d_n\}\big) \big|$.
For each $k\in\{1,\ldots,m\}$, 
let ${\cal D}_k := \{i_1^{(k)},\ldots,i_{n_k}^{(k)}\}\subseteq\{1,\ldots,n\}$ be the set of degree-$d_{(k)}$ nodes.
Define $n_k:=|{\cal D}_k|$.
For instance, consider a 6-node network with degree sequence $(d_1,\ldots,d_6) = (3,4,3,2,3,3)$.
We have $m=3$ and
$
    \big(
        d_{(1)}, d_{(2)}, d_{(3)}
    \big)
    =
    (2,3,4).
$
 Since $d_{(2)}=3$ and ${\cal D}_2=\{1,3,5,6\}$ (because these four nodes have degree 3),  we have $n_2=|{\cal D}_2|=4$.
By Lemma \ref{lemma::degree-monotonicity}, in the MLE, all $\hat \beta_j$ for $j\in {\cal D}_k$ should be equal, so we set them to a common value, denoted by $\delta_k$.
The objective function ${\cal L}_\lambda(\beta)$ can be rewritten as a function  of $\delta$, denoted by $\tilde{\cal L}_\lambda(\delta)$,  with \emph{degree-indexed} parameters $\delta:=(\delta_1,\ldots,\delta_m)$.
 Specifically, we have
\begin{align}
    {\cal L}_\lambda(\beta)
    =
    \tilde{\cal L}_\lambda(\delta)
    := &~
    \sum_{1\leq k< \ell \leq m} n_k n_\ell \log\Big(
        1+e^{\delta_k+\delta_\ell}
    \Big)
    +
    \sum_{1\leq k\leq m} \dfrac{n_k(n_k-1)}{2}\log\Big(
        1+e^{2\delta_k}
    \Big)
    \notag\\
    & -
    \sum_{k=1}^m n_k d_{(k)} \delta_k + \dfrac\lambda 2 \sum_{k=1}^m n_k (\delta_k - \tilde \delta)^2,
    \label{eqn::reparameterization-1}
\end{align}
where 
$
    \tilde \delta
    :=
    n^{-1} \sum_{k=1}^m n_k\cdot \delta_k
$
is a weighted average of $\delta_k$'s.
On the RHS of \eqref{eqn::reparameterization-1}, the first three terms  together correspond to the negative log-likelihood \eqref{neg-log-likelihood::vanilla}, and the last term corresponds to the $\ell_2$ regularization term in \eqref{our-method}.
The gradient and Hessian of $\tilde{\cal L}_\lambda(\delta)$ can be efficiently computed; see Appendix \ref{section::additional-details-computation} for details.
Existence and uniqueness of the optimum of \eqref{eqn::reparameterization-1}  follow from Lemma \ref{lemma::MLE-existence},  since there is a one-to-one  mapping between $\delta$ and $\beta$.

Our algorithm costs $O(\rho_n\cdot n^2)$ in a one-pass computation of all degrees, plus a per-iteration cost of $O(m^2)$.
This is much cheaper  than the  $O(n^2)$ or $O(n^3)$ per-iteration cost of existing methods.
In large and sparse networks, $m$ is typically  smaller than $n$ by several orders of magnitude.
For example,
the COVID-19 knowledge graph data set \citet{covid_kg} contains $n\approx1.3$ million nodes, but only $m=459$ different degrees, see Table \ref{tab::data-2::low-degree-frequencies}.
 Our algorithm makes applying Newton method to networks of this scale feasible, whereas existing methods would find it computationally prohibitive.
\begin{table}[ht!]
    \centering
    \caption{Frequencies of low degrees in the data set \citet{covid_kg}, $n=1304155$.}
    \begin{tabular}{c|ccccc}\hline
        Degree &  
            1 & 2 & 3 & 4 & 5\\\hline
        Number of nodes with this degree & 
            684003 & 132123 & 48126 & 24586 & 15189\\\hline
        Percentage of nodes, unit: \% &
            52.45 & 10.13 & 3.69 & 0.19 & 0.12\\\hline
    \end{tabular}
    \label{tab::data-2::low-degree-frequencies}
\end{table}

\section{Theory and statistical inference}
\label{section::theory}

\subsection{Consistency and finite-sample error bounds}
\label{subsec::theory::upper-bounds}

Following \citet{chen2019analysis, stein2020sparse}, we assume that the true parameters satisfy 
\begin{align}
    \beta^*\in {\cal I}^* = [a_1^*\log n-M^*, a_2^*\log n+M^*]^n
    \label{eqn::assumption-0}
\end{align}
for some constants $a_1^*\leq a_2^*$ and $M^*>0$.  
To understand why the range of $\beta^*$ may scale as $O(\log n)$,  observe that if $(\beta_1^*,\beta_2^*)=(a_1\log n, a_2\log n)$ for $a_1,a_2$  with $a_1+a_2< 0$, then $\pr(A_{1,2}=1) \asymp n^{a_1+a_2}$.
The difficulty of parameter estimation is closely related to quantities such as $\var(A_{i,j})$ and $\var(d_i)$.
To better align with the notation in the existing $\beta$-model literature, we will set up symbols for the inverses of these quantities rather than the quantities themselves.
\begin{defi}
    \label{def::bn-cn-qn}
    For the true parameter value $\beta^*$, define
    \begin{align}
        b_n := \max_{1\leq i<j\leq n}
        &\dfrac{(1+e^{\beta_i^*+\beta_j^*})^2}{e^{\beta_i^*+\beta_j^*}}
        , \quad
        c_n := \min_{1\leq i<j\leq n} \dfrac{(1+e^{\beta_i^*+\beta_j^*})^2}{e^{\beta_i^*+\beta_j^*}},
        \label{defeqn::b_n,c_n}
        \\
        \quad\textrm{ and }\quad &~
        q_n
        :=
        \Bigg\{
            \max_{1\leq i\leq n}(n-1)^{-1}
            \sum_{\substack{1\leq j\leq n\\j\neq i}}
            \dfrac{e^{\beta_i^*+\beta_j^*}}{(1+e^{\beta_i^*+\beta_j^*})^2}
        \Bigg\}^{-1}
        .
    \end{align}
\end{defi}
By definition, the quantities in Definition \ref{def::bn-cn-qn} can be interpreted as follows.
\begin{align}
    b_n^{-1} =&~ \min_{1\leq i<j\leq n}\var(A_{i,j}),
    \quad
    c_n^{-1} = \max_{1\leq i<j\leq n}\var(A_{i,j}),
    \quad\textrm{and}\quad
    q_n^{-1}(n-1)= \max_{1\leq i\leq n}\var(d_i).
    \label{interpretation-bn-cn-qn}
\end{align}
In a sparse network with no high-degree nodes,  i.e., $a_2^*\leq 0$, 
we have $\var(A_{i,j})\approx \ep[A_{i,j}]$.
Then we can approximately interpret $b_n^{-1}$ and $c_n^{-1}$ as  minimum and maximum edge probabilities, respectively; and $q_n^{-1}(n-1)$ as the maximum expected degree.

Since most $\beta$-model papers focus on $\ell_\infty$ bounds, we first establish an $\ell_\infty$ error bound.
\begin{theorem}
\label{Main-theorem::Linf-bound}
    Assume the $\beta$-model as in \eqref{model::original-beta-model} and \eqref{eqn::assumption-0}.
    Define
    \begin{align}
        \Gamma(n, ~& \beta^*;\lambda)
        := 
        \dfrac
            {(q_n^{-1}n\log n)^{1/2}
            +
            \{
                \|\projperp\beta^*\|_\infty
                +
                \log^{1/2}n / (b_n^{-1}n)
            \}
            \lambda}
            {b_n^{-1}n+\lambda}.
        \label{eqn::main-theorem-2::error-bound}
    \end{align}
    Suppose our choice of $\lambda$ satisfies $\lambda>0$ and
    \begin{align}
        (b_n/ q_n
        )
        \cdot
        \Gamma(n,\beta^*;\lambda)
            ~\leq 1/20.
        \label{assumption::main-theorem-Linf-1}
    \end{align}
    Then our $\ell_2$-regularized MLE $\hat{\beta}_{\lambda}$ satisfies
    \begin{align}
        \pr\Big\{
        \|\hat{\beta}_\lambda - \beta^*\|_\infty
        \leq & ~
        C_1
        \Gamma(n,\beta^*;\lambda)
        \Big\}
        \geq 
        1- n^{-C_2}
        \label{eqn::main-theorem-2::main-result}
    \end{align}
    for some constants $C_1,C_2>0$, where $C_1$ depends on $C_2$  and they both depend on $(a_1^*,a_2^*)$.
\end{theorem}
{\bf Interpreting the main assumption of Theorem \ref{Main-theorem::Linf-bound}.}
Let us consider an illustrative special case.
Suppose the true model is
\begin{align}
    \beta_i^* =
    \begin{cases}
      (\alpha-\gamma/2)\,\log n, & i\in\mathcal S,\\
      -(\gamma/2)\,\log n,       & i\notin\mathcal S.
    \end{cases}
    \label{setting::simplified-chen-kato-leng}
\end{align}
Clearly, \eqref{setting::simplified-chen-kato-leng} is a simplified version of \citet{chen2019analysis}'s model \eqref{setting::chen-kato-leng}.
For simplicity, we set $\lambda=O(1)$.
Consequently, condition \eqref{assumption::main-theorem-Linf-1} becomes 
\begin{align}
    (q_n^{-1} n)^3/(b_n^{-1} n)^4 \cdot\log n \to 0.
    \label{eqn::simplified-condition::theorem-1}
\end{align}
Under the conditions \eqref{assumption-chen-kato-leng} and \eqref{beta-sparsity-assumption} from \citet{chen2019analysis},
Equation \eqref{eqn::simplified-condition::theorem-1} simplifies into
\begin{align}
    \frac
        {
            \big(
                \max_i \ep[d_i]
            \big)^3
            \cdot \log n
        }
        {
            \big(
                \min_i \ep[d_i]
            \big)^4
        }
    \stackrel{n\to\infty}\to 0,
    \label{eqn::simplified-condition::theorem-1-version-degree}
\end{align}
because in this setting, $\max_i \ep[d_i] \asymp q_n^{-1}n \asymp n^{\alpha-\gamma+1}$ and $\min_i \ep[d_i] \asymp b_n^{-1}n \asymp n^{-\gamma+1}$.
Now, using this simplified model specified by \eqref{model::original-beta-model}, \eqref{setting::simplified-chen-kato-leng}, \eqref{assumption-chen-kato-leng} and \eqref{beta-sparsity-assumption}, we compare the assumption of Theorem~\ref{Main-theorem::Linf-bound} to the conditions required by \citet{chen2019analysis} and \citet{stein2020sparse}.
\begin{itemize}
    \item Our Theorem \ref{Main-theorem::Linf-bound} requires \eqref{eqn::simplified-condition::theorem-1}  (or equivalently, \eqref{eqn::simplified-condition::theorem-1-version-degree}), which simplifies to $\gamma+3\alpha<1$.
    \item In \citet{chen2019analysis}, the condition \eqref{assumption-chen-kato-leng} and condition (7) in their Lemma 2 yield that $\alpha+\gamma\leq 1/2$, which further implies that $\alpha\leq 1/6$ and $\gamma+3\alpha\leq 5/6$,  a strictly stronger restriction than ours.
    \item In \citet{stein2020sparse}, Assumption 2 can be translated into $3\gamma+2\alpha\leq 1/2$,  again a stronger condition than ours.
\end{itemize}
Consequently, the sparsest networks that these three methods can handle, measured by average edge probabilities and ignoring $\log n$ factors, are listed as follows.
$$
    \rho_n = \text{mean}(\ep[A_{i,j}])
    \asymp n^{-\gamma}
    \gg
    \begin{cases}
        n^{-1}, & \text{ required by our method,}
        \\
        n^{-1/2}, & \text{ required by \citet{chen2019analysis},}
        \\
        n^{-1/6}, & \text{ required by \citet{stein2020sparse}.}
    \end{cases}
$$
More precisely, our method requires $\rho_n\gg n^{-1}\log n$.
This nearly matches the well-known minimal sparsity conditions required in some other network analysis problems, such as community detection \citep{bickel2009nonparametric,zhao2012consistency} and network method-of-moments \citep{zhang2020edgeworth}.
\smallskip

\noindent {\bf Interpreting the error rate of Theorem \ref{Main-theorem::Linf-bound}.}
We continue to adopt the simplified model, specified by \eqref{model::original-beta-model}, \eqref{setting::simplified-chen-kato-leng}, \eqref{assumption-chen-kato-leng} and \eqref{beta-sparsity-assumption}.
Under this setting, the error rate of our Theorem \ref{Main-theorem::Linf-bound} is simplified as follows.
\begin{align}
    \|\hat\beta_\lambda - \beta^*\|_\infty
    \stackrel{\rm w.h.p.}\lesssim&~
    \frac
        {
            \big(
                \max_i \ep[d_i] \log n
            \big)^{1/2}
            +
            \|\beta^* - \bar\beta^*\|_\infty\cdot \lambda
        }
        {
            \min_i \ep[d_i]
            +
            \lambda
        }
    \stackrel{\lambda=O(1)}\asymp
    \frac
        {
            \big(
                \max_i \ep[d_i] \log n
            \big)^{1/2}
        }
        {
            \min_i \ep[d_i]
        }
    \notag\\
    \Big( \text{\eqref{beta-sparsity-assumption}} &\Rightarrow \text{``$b_n^{-1}\asymp \rho_n$''} \Big)
    \asymp 
    n^{(\gamma+\alpha-1)/2} \log^{1/2}n
    \asymp
    \frac{q_n^{-1/2}}{\rho_n}\cdot\sqrt{\frac{\log n}n}.
    \label{eqn::interpret-theorem-1-error-rate}
\end{align}
We compare this error rate to existing works.
\citet{chen2019analysis} only proved asymptotic results.
\citet{stein2020sparse} established an $\ell_1$ error rate for their estimator (denoted by $\hat\beta^{\rm(SL2020)}$):
$$
    n^{-1}\cdot\|\hat\beta^{\rm(SL2020)} - \beta^*\|_1
    \lesssim 
    n^{(2\gamma-1)/2} \log^{1/2}n
    \asymp
    \frac{1}{\rho_n}\cdot\sqrt{\frac{\log n}n}.
$$
Comparing this to \eqref{eqn::interpret-theorem-1-error-rate}, we see that our method's error bound is superior.\footnote{Recall that \eqref{assumption-chen-kato-leng} assumes $\alpha\leq \gamma$.}
Moreover, as discussed earlier, \citet{stein2020sparse} requires a substantially denser network than ours.

Above, we directly plugged the simplified model's configurations directly into the error rate of Theorem \ref{Main-theorem::Linf-bound}.
However, in fact, our method achieves a better rate in this setting.
\begin{proposition}
    \label{proposition::theorem-1-simplified-model}
    Under the simplified model specified by \eqref{model::original-beta-model}, \eqref{setting::simplified-chen-kato-leng}, \eqref{assumption-chen-kato-leng} and \eqref{beta-sparsity-assumption}, replace condition \eqref{assumption::main-theorem-Linf-1} of Theorem \ref{Main-theorem::Linf-bound} by 
    \begin{align}
        \label{assumption::proposition-1}
        (b_n/q_n)\cdot (b_n^{-1}n)^{-1/2} \leq 1/20.
    \end{align}
    For any fixed $\lambda>0$, with probability at least $1-o(1)$, it holds that
    \begin{align}
        \|\hat\beta_\lambda - \beta^*\|_\infty
        \lesssim&~
        \frac{\log^{1/2}n}{(b_n^{-1}n)^{1/2}}.
    \end{align}
\end{proposition}

\smallskip

\noindent{\bf Recovery of the active set ${\cal S}$.}
While our method does not require $\beta$-sparsity as in Assumption \ref{assumption::beta-sparsity}, for interpretation purposes, one might be interested in recovering this structure when it exists.
The idea is very simple: first estimate the majority common value among $\beta_i^*$'s.
This can be achieved by considering nodes with degrees ranked in the middle $\zeta_0$-quantile for any constant $\zeta_0\in(0,1)$.
Then the active set is estimated by collecting all nodes whose estimated $\beta$ parameter is significantly different than this common value.
Algorithm~\ref{algorithm::thresholding} formally describes this procedure.
\begin{algorithm}[htbp]
    \caption{Active set estimation}
    \label{algorithm::thresholding}
    \begin{algorithmic}
        \REQUIRE $A$; user-specified constant $\zeta_0\in(0,1)$
        \ENSURE Estimated active set $\hat {S}\subseteq [1:n]$
        \STATE\label{algorithm-step::1} 
        1. Run our $\ell_2$-regularized MLE with a $\lambda=O(1)$, produce $\hat \beta_\lambda$.
        \STATE\label{algorithm-step::2} 
        2. Compute all node degrees $d_{[1:n]}$.
        Let ${\cal S}_{\zeta_0}$ be the set of nodes whose degrees are in the middle-$\zeta_0$ proportion, that is,
        $
            {\cal S}_{\zeta_0}
            :=
            \big\{
                i:
                d_i
                \in
                \big(
                    \textrm{Quantile}(d_{[1:n]}; (1-\zeta_0)/2),
                    \textrm{Quantile}(d_{[1:n]}; (1+\zeta_0)/2)
                \big)
            \big\}.
        $\\
        Then set
        $
            \bar A_{\zeta_0}
            :=
            \binom{|{\cal S}_{\zeta_0}|}2^{-1}
            \sum_{1\leq i<j\leq n} (A_{\zeta_0})_{i,j},
        $
        and compute
        $
            \hat b_n
            :=
            \{
                \bar A_{\zeta_0} (1-\bar A_{\zeta_0})
            \}^{-1}.
        $
        \STATE\label{algorithm-step::3} 
        3. 
        Output
        $
            \hat {\cal S}
            :=
            \big\{
                j: 
                \big|
                    \hat \beta_{\lambda;j} - \log \{\bar A_{\zeta_0}/(1 - \bar A_{\zeta_0})\}/2 
                \big| 
                > 
                \{d_{\max}\cdot \log n\}^{1/2} / (\hat b_n^{-1}n)
            \big\},
        $
        where
        $d_{\max} := \max_{1\leq i\leq n}d_i$ is the maximum observed degree; $\hat b_n$ is computed by plugging $\hat\beta_\lambda$ into \eqref{defeqn::b_n,c_n}.        
    \end{algorithmic}
\end{algorithm}

\begin{corollary}[Consistency of our Algorithm \ref{algorithm::thresholding}]
    \label{corollary::sparsistency}
    Under the model specified by \eqref{model::original-beta-model}, \eqref{setting::simplified-chen-kato-leng}, \eqref{assumption-chen-kato-leng} and \eqref{beta-sparsity-assumption}, suppose the assumptions of Theorem \ref{Main-theorem::Linf-bound} hold. 
    With $\lambda=O(1)$, the output $\hat {\cal S}$ of Algorithm \ref{algorithm::thresholding} satisfies $\pr(\hat {\cal S} = {\cal S})\to 1$.
\end{corollary}

Our Algorithm \ref{algorithm::thresholding} is computationally efficient. 
\citet{chen2019analysis} estimates ${\cal S}$ by  refitting the model several times, with different candidate sizes of the active set ${\cal S}$, treating it as a model selection problem.
In contrast, our method fits the model only once. 
Although \citet{stein2020sparse} did not explicitly propose an algorithm for estimating $\mathcal S$, support recovery is a key step in their estimation routine.
Our Algorithm~\ref{algorithm::thresholding} requires no such assumption that \(\beta_i>\beta_j\) for all \(i\in\mathcal S\) and \(j\notin\mathcal S\), unlike both \citet{chen2019analysis} and \citet{stein2020sparse}.

Corollary~\ref{corollary::sparsistency} remains valid when each \(i\in\mathcal S\) has its own \(\alpha_i\) satisfying \eqref{beta-sparsity-assumption} with \(\min_{i\in\mathcal S}\alpha_i>0\).

For the final discussion of Theorem \ref{Main-theorem::Linf-bound} regarding large $\lambda$, we return to the general $\beta$-model.
\smallskip

 \noindent{\bf Large $\lambda$.}
We rarely need to choose a large $\lambda$.
 One such scenario occurs when $q_n^{-1}n \asymp b_n^{-1}n \lesssim \log n$, in that case, \eqref{eqn::simplified-condition::theorem-1} fails.
By setting $\lambda\gg n$, \eqref{eqn::simplified-condition::theorem-1} simplifies into
$
    \|\projperp\beta^*\|_\infty
    +
    \log^{1/2}n / (b_n^{-1}n)
    \to 0.
$
 Here, we need two additional assumptions:
(i) \(\|\projperp\beta^*\|_\infty\to0\)  (i.e., all \(\beta_i^*\) are nearly equal);
and (ii) $b_n^{-1}n \gg \log^{1/2}n$.
If either assumption  fails, no existing $\beta$-model method can handle this setting.

 We now conclude our discussion of Theorem \ref{Main-theorem::Linf-bound} and introduce the next result: the $\ell_2$ error bound.
\begin{theorem}
    \label{Main-theorem::L2-upper-bound}
    
    Under the $\beta$-model formulated by  \eqref{model::original-beta-model} and \eqref{eqn::assumption-0}, we have
    \begin{enumerate}[(i)]
        \item \label{main-theorem::L2-bound::small-lambda}
        When $\lambda = O(1)$, suppose we perform a constrained MLE with $\hat\beta_\lambda \in [a_1\log n-M, a_2\log n+M]^n$ for $(a_1,a_2,M)$ satisfying $a_1\leq a_1^* \leq a_2^* \leq a_2$ and $M\geq M^*$.  Then we have
        \begin{align}
            n^{-1/2}\cdot \|\hat\beta_\lambda-\beta^*\|_2
            \lesssim &~
            \Bigg\{
                \frac{b_n'}{q_n}\cdot \frac{\log n}{(b_n')^{-1}n},
            \Bigg\}^{1/2}
            \label{eqn::error-bound::main-theorem::L2-bound::small-lambda}
        \end{align}
        where $b_n':= \max\big\{(1+e^{2a_1})^2/e^{2a_1}, (1+e^{2a_2})^2/e^{2a_2}\big\}$.
        \item \label{main-theorem::L2-bound::large-lambda}
        When $\lambda \geq C_0 q_n^{-1}n$ for a large enough constant $C_0>0$, we have
        \begin{align}
            n^{-1/2}\cdot \|\hat\beta_\lambda-\beta^*\|_2
            \lesssim &~
            (b_n/c_n)^2 \cdot n^{-1/2}\big\|\projperp\beta^*\big\|_2
            +
            (b_n/q_n)\cdot n^{-1}\log^{1/2}n
            .
        \end{align}
    \end{enumerate}

\end{theorem}

\begin{remark}
    \label{remark::L2-bound-guess-a1-a2}
    Part (i) of Theorem \ref{Main-theorem::L2-upper-bound} requires guessing $(a_1^*, a_2^*)$ with $(a_1,a_2)$, which is similar to \citet{chen2019analysis} and \citet{stein2020sparse} in that they also need to guess the parameter space.
    In general, this guess is not easy.
     An incorrect guess may result in inconsistent estimation.
    Fortunately, under the simplified model \eqref{model::original-beta-model}, \eqref{setting::simplified-chen-kato-leng}, \eqref{assumption-chen-kato-leng} and \eqref{beta-sparsity-assumption},
    we can accurately estimate $a_1^*$ and $a_2^*$.
    First, run Step \ref{algorithm-step::2} in Algorithm \ref{algorithm::thresholding} and obtain ${\cal S}_{\delta_0}$ and $\bar{A}_{\delta_0}$.
    Then we can estimate $a_1^*$ by $a_1 := {\rm logit}(\bar A_{\delta_0})/(2\log n)$\footnote{Recall that logit$(u):=\log \{u/(1-u)\}$.}, and it is not difficult to show that $a_1 = a_1^* + o_p(\log^{-1}n)$.
    Next, let $d_{\max, S_{\delta_0}} := \max_{j:j\neq {\cal S}_{\delta_0}} \sum_{i\in{\cal S}_{\delta_0}} A_{i,j}$, which can be interpreted as the maximum degree in ${\cal S}_{\delta_0}$.
    Finally, we can estimate $a_2^*$ by $a_2 := {\rm logit}(d_{\max,{\cal S}_{\delta_0}}/|{\cal S}_{\delta_0}|)/\log n - a_1$, which satisfies $a_2 = a_2^*+o_p(\log^{-1}n)$.
\end{remark} 
{\bf Interpreting the result of Theorem \ref{Main-theorem::L2-upper-bound}.}
Similar to the discussion of Theorem \ref{Main-theorem::Linf-bound}, we adopt the simplified model.
In view of Remark \ref{remark::L2-bound-guess-a1-a2}, we may assume that we correctly guessed $(a_1^*,a_2^*)$, i.e., $(a_1,a_2)=(a_1^*,a_2^*)=(-\gamma/2,\alpha-\gamma/2)$.
Then $c_n^{-1}\asymp n^{2\alpha-\gamma}$, and the error bound in Theorem \ref{Main-theorem::L2-upper-bound}-\eqref{main-theorem::L2-bound::small-lambda} becomes
$$
    n^{-1/2}\cdot \|\hat\beta_\lambda-\beta^*\|_2
    \leq
    \frac{q_n^{-1/2}}{\rho_n}\cdot \sqrt{\frac{\log n}n}.
$$
This error bound coincides with the $\ell_\infty$ bound given in \eqref{eqn::interpret-theorem-1-error-rate}.
Similarly, part \eqref{main-theorem::L2-bound::large-lambda} of Theorem \ref{Main-theorem::L2-upper-bound} can be simplified into
\begin{align}
    n^{-1/2}\cdot \|\hat\beta_\lambda-\beta^*\|_2
    \leq
    n^{4\alpha} \cdot n^{-1/2}\big\|\projperp\beta^*\big\|_2
    +
    n^{\alpha-1}\log^{1/2}n.
    \label{eqn::interpret-theorem-2-large-lambda}
\end{align}
To understand the RHS of \eqref{eqn::interpret-theorem-2-large-lambda}, recall that node degree heterogeneity increases with $\alpha$.
This aligns with our discussion of Theorem \ref{Main-theorem::Linf-bound}: a large $\lambda$ encourages the elements of $\hat\beta_\lambda$ to be similar, yielding better performance when the true $\beta_i^*$'s are themselves similar.

\subsection{Lower bounds}
\label{subsec::theory::lower-bounds}

Readers may naturally wonder how tight the error bounds  presented in Section \ref{subsec::theory::upper-bounds} are.
 Existing work in this direction remains scarce.
In this section, we establish the first local lower bounds in $\ell_p$  norms for $p=0,1,2$.  
Here "local" means that we consider a bounded $\ell_\infty$-neighborhood around the true $\beta^*$.

\begin{theorem}[Local lower bound]
    \label{Main-theorem::local-lower-bounds}
       
     Consider the  
      simplified model \eqref{model::original-beta-model}, \eqref{setting::simplified-chen-kato-leng}, \eqref{assumption-chen-kato-leng} and \eqref{beta-sparsity-assumption}.
    Let ${\cal B}_{\|\cdot\|,r}(\beta')$ be the closed ball with radius $r$, centered at $\beta'$, with a generic norm $\|\cdot\|$. 
    Then we have
    \begin{align}
        \inf_{\hat\beta}
        \sup_{\beta^*\in {\cal B}_{\|\cdot\|_\infty, M}(\beta_0)}
        n^{-1}\cdot \|\hat\beta - \beta^*\|_1
        \gtrsim &~
        (b_n^{-1}n)^{-1/2},
        \label{eqn::new-lower-bound::L1}
    \end{align}
    where $M>0$ is an arbitrary constant.
    
\end{theorem}

Notice that in Theorem \ref{Main-theorem::local-lower-bounds}, the RHS of \eqref{eqn::new-lower-bound::L1} could have been written as $(b_{n,0}^{-1}n)^{-1/2}$, where
$
    b_{n,0}
    :=
    \sup_{\beta\in {\cal B}_{\|\cdot\|_\infty, M}(\beta_0)}
    \max_{1\leq i
    <j\leq n} (1+e^{\beta_i+\beta_j})^2/e^{\beta_i+\beta_j}.
$
However, it is easy to verify that $b_{n,0}\asymp b_n$ always holds for any $\beta^*\in {\cal B}_{\|\cdot\|_\infty, M}(\beta_0)$.
 Hence, we denote this quantity simply by "$b_n$" in \eqref{eqn::new-lower-bound::L1} for cleaner interpretation.

As the inequality $n^{-1}\|x\|_1 \leq n^{-1/2}\|x\|_2 \leq \|x\|_\infty$ holds, \eqref{eqn::new-lower-bound::L1} also implies  lower bounds in $\ell_2$ and $\ell_\infty$ norms:
\begin{align}
    \sup_{\beta^*\in {\cal B}_{\|\cdot\|_\infty, M}(\beta_0)} \|\hat\beta - \beta^*\|_\infty
    \geq
    \sup_{\beta^*\in {\cal B}_{\|\cdot\|_\infty, M}(\beta_0)} n^{-1/2}\cdot \|\hat\beta - \beta^*\|_2
    \gtrsim
    (b_n^{-1}n)^{-1/2}.
    \label{eqn::lp-norm-inequality}
\end{align}
 Under the simplified model \eqref{model::original-beta-model}, \eqref{setting::simplified-chen-kato-leng}, \eqref{assumption-chen-kato-leng} and \eqref{beta-sparsity-assumption}, the upper bound in Proposition \ref{proposition::theorem-1-simplified-model} and the lower bound in \eqref{Main-theorem::local-lower-bounds} nearly match; thus, our method achieves nearly optimal $\ell_p$ error rates (for $p=1,2,\infty$).

Theorem \ref{Main-theorem::local-lower-bounds} is pioneering in the  $\beta$-model literature.
The closest existing results are \citet{wahlstrom2017beta} and \citet{lee2020minimax}.
Section III.A of \citet{wahlstrom2017beta} presents a marginal Cramer-Rao  lower bound $(\sum_{j\neq i}\ep[A_{i,j}])^{-1}$  for estimating $\beta_i$, where $i$ is \emph{fixed} and they have \emph{repeated} network observations generated by the same true $\beta^*$.
To facilitate the comparison with \citet{lee2020minimax}, we present a non-local lower bound  in a similar spirit to theirs.

\begin{theorem}[Non-local lower bounds]
    \label{Main-theorem::lower-bounds}
    Assume the $\beta$-model as in \eqref{model::original-beta-model} and \eqref{eqn::assumption-0}.
    Define $b_{{\cal I}^*} = \sup_{\beta\in {\cal I}^*}\max_{1\leq i
    <j\leq n} (1+e^{\beta_i+\beta_j})^2/e^{\beta_i+\beta_j}$.
    We have
    \begin{align}
        \inf_{\hat{\beta}}
        \sup_{\beta^*\in {\cal I}^*} 
        n^{-1/2}\cdot \ep\big[\big\| \hat{\beta} - \beta^* \big\|_2\big]
        \gtrsim &~
        (b_{{\cal I}^*}^{-1}n)^{-1/2}.
        \label{eqn::theorem::lower-bound-L2}
    \end{align}
    Moreover, \eqref{eqn::theorem::lower-bound-L2} remains valid if $n^{-1/2}\|\cdot\|_2$ is replaced by $n^{-1}\|\cdot\|_1$ or $\|\cdot\|_\infty$.
\end{theorem}
Theorem 7 of \citet{lee2020minimax} established a $C_{{\cal I}^*} n^{-1}$ lower bound on $n^{-1}\cdot \ep[\|\hat\beta - \beta^*\|_2^2]$, where $C_{{\cal I}^*} := \inf_{\beta\in {\cal I}^*}\min_{1\leq i <j\leq n} (1+e^{\beta_i+\beta_j})^2/e^{\beta_i+\beta_j}$. 
Consequently, the bound provided by that theorem is  strictly weaker than our result in \eqref{eqn::theorem::lower-bound-L2}.
Moreover, their proof  is tailored to the $\ell_2$ norm.
 Adapting their proof to the $\ell_0$ and $\ell_1$ norms is not straightforward.
\citet{lee2020minimax} remarked that currently no matching upper bound is known for GLM's.
This is not surprising, since our \eqref{eqn::new-lower-bound::L1} shows that the lower bound depends on the true $\beta^*$ parameter value.
Whether the non-local lower bound in \eqref{eqn::theorem::lower-bound-L2} is attainable remains an open problem.

\subsection{High-dimensional asymptotic normality}
\label{subsec::high-dim-asymp-norm}

We now turn our attention to the central limit theorem (CLT) results. 
We study two representative cases: $\lambda=O(1)$ and $\lambda\to\infty$, which yield distinct asymptotic behaviors of $\hat\beta_\lambda - \beta^*$.
\begin{theorem}[High-dimensional CLT]
    \label{Main-theorem::asymptotic-normality}
    Let ${\cal J}\subset \{1,\ldots,n\}$ be an index set, chosen independently of  the observed adjacency matrix $A$, that satisfies 
    $
        |{\cal J}|
        \ll
        c_n(b_n^{-1}n)^{1/2}.
    $
    Under the conditions of Theorem \ref{Main-theorem::Linf-bound}, we have the following results.
    \begin{enumerate}[(i).]
        \item Set $\lambda=O(1)$.  
        Suppose
        \begin{align}
            \frac{b_n^3}{q_n^2 n^{1/2}}\cdot \log n
            &\to 
            0.
            \label{new-condition::asymptotic-normality-1}
        \end{align}
        Let $D(\beta^*)$ denote the diagonal matrix of $V(\beta^*)$,  where $V(\cdot)$ was defined in Section \ref{section::introduction::subsec::notation}.
        Then
        \begin{align}
            \big( \hat\beta_\lambda-\beta^* \big)_{\cal J}
            \stackrel{d}\to &~
            N\Big[
                0, \{D(\beta^*)\}_{{\cal J},{\cal J}}^{-1}
            \Big].
            \label{eqn::theorem-asymp-norm-part-i}
        \end{align}
        
        \item Set $\lambda\gg n$.  
        We have
        \begin{align}
            \hat\beta_\lambda
            -
            \frac
                {\mathbbm{1}^T  \breve{V}(\beta^*)\beta^*}
                {\mathbbm{1}^T  \breve{V}(\beta^*)\mathbbm{1}}
                \cdot \mathbbm{1}
            \stackrel{d}\to &~
            N\Bigg(
                0,
                \frac
                    {n(n-1)\mathbbm{1}^T V(\beta^*)\mathbbm{1}}
                    {2\big\{\mathbbm{1}^T  \breve{V}(\beta^*)\mathbbm{1}\big\}^2}
            \Bigg)
            \cdot \mathbbm{1},
            \label{eqn::theorem-conclusion::asymptotic-normality-large-lambda}
        \end{align}
        where $\{ \breve{V}(\beta^*)\}_{i,j}$ is defined as follows:
        with $\bar{A}:=\frac{\mathbbm{1}^T A\mathbbm{1}}{n(n-1)}$ and $\mathrm{logit}(u):=\log \frac{u}{1-u}$, set
        \begin{align}
            \{ \breve{V}(\beta^*)\}_{i,j}
            := &~
            \frac{
                \ep[\bar{A}\,] - \ep[A_{i,j}]
            }
            {
                \mathrm{logit} \big(\ep[\bar{A}\,]\big) - \mathrm{logit} \big( \ep[A_{i,j}] \big)
            }
            ,
            \quad \textrm{for all }1\leq \{i\neq j\}\leq n,
            \label{def::V-breve-beta}
        \end{align}
        and $\{ \breve{V}(\beta^*)\}_{i,i} := \sum_{j:j\neq i}\{ \breve{V}(\beta^*)\}_{i,j}$.
    \end{enumerate}
    
\end{theorem}

 We begin by interpreting part (i).
The additional assumption \eqref{new-condition::asymptotic-normality-1} in part (i) of Theorem \ref{Main-theorem::asymptotic-normality} can be  rewritten as 
$
    (b_n^{5/2}/q_n^2)\cdot \log n/(b_n^{-1}n)^{1/2} \to 0.
$
This is stronger than the assumption \eqref{assumption::main-theorem-Linf-1} in Theorem \ref{Main-theorem::Linf-bound} with $\lambda=O(1)$, which simplifies into
$
    (b_n/q_n)^{3/2}\cdot \log n/(b_n^{-1}n)^{1/2} \to 0.
$
This is consistent with our observation that existing $\beta$-model works that we are aware of \citep{yan2013central, chen2019analysis, stein2020sparse} all require considerably stronger assumptions for asymptotic normality than estimation consistency -- this is not surprising as normality is a particularly stronger conclusion than consistency in high-dimensional problems.
Recall  that under the model of \citet{chen2019analysis}, we have $b_n\asymp q_n$.
Then our assumption \eqref{new-condition::asymptotic-normality-1} implies $\rho_n\gg n^{-1/2}$.
This matches the minimum network density required by \citet{chen2019analysis} for achieving estimation consistency, but they required additional conditions such as knowing $(\alpha, \gamma)$ and the $\beta$-sparsity assumption \eqref{beta-sparsity-assumption}, which we do not need.
Our condition also significantly improves  upon the $n\gg n^{-1/6}$ assumption in \citet{stein2020sparse}.

Part (ii) illustrates a very different scenario, where a large $\lambda$ dramatically reduces parameter dimensionality by strongly  shrinking $\hat\beta_{\lambda;i}$'s towards  a common value.
In the extreme case where $\lambda=+\infty$, the $\beta$-model degenerates to an Erdos-Renyi model.
In this case, we can write $\beta^* = \beta_0^*\mathbbm{1}$, and by L'Hopital's rule, $\{\breve{V}(\beta^*)\}_{i,j} = e^{2\beta_0^*}/(1+e^{2\beta_0^*})^2$ for all $i\neq j$.
Then we find that \eqref{eqn::theorem-conclusion::asymptotic-normality-large-lambda} reduces to
\begin{align}
    \hat\beta_\lambda - \beta^*\cdot \mathbbm{1}
    \stackrel{d}\to&~
    N\Big( 0, \big\{2n(n-1)e^{2\beta_0^*}/(1+e^{2\beta_0^*})^2\big\}^{-1} \Big)
    \cdot\mathbbm{1},
    \label{interpretation-eqn::asymptotic-normality-large-lambda}
\end{align}
which recovers the standard one-dimensional MLE asymptotic normality result for estimating $\beta_0^*$ in Bernoulli$\big(e^{2\beta_0^*}/(1+e^{2\beta_0^*})\big)$ distribution with $\binom n2$ i.i.d. observations.

Theorem \ref{Main-theorem::asymptotic-normality} reveals the distinct asymptotic behaviors of $\hat\beta_\lambda$ for small and large  values of $\lambda$.
Part (i) shows that when $\lambda=O(1)$, the elements of $\hat\beta_\lambda$ are asymptotically independent.
This aligns with  the fixed-dimension results of \citet{yan2013central}.
Part (ii) presents new findings with no analogous results in the literature: for a large $\lambda$, the elements of $\hat\beta_\lambda$ tend to be perfectly correlated.
This aligns with the expectation that a large $\lambda$ would heavily penalize differences between estimated $\beta_i$ values, leading $\hat\beta_\lambda$ to approach an asymptotic center parallel to $\mathbbm{1}$.
However, the conclusion of part (ii) is nontrivial, as it  further shows that the random variations of $\hat\beta_\lambda$ elements  are asymptotically perfectly correlated.
This behavior is confirmed by our numerical experiments.
Theorem \ref{Main-theorem::asymptotic-normality} also presents the first \emph{high-dimensional} central limit theorem for the $\beta$-model, allowing $|{\cal J}|$ to grow with $n$; whereas existing works only present fixed-dimensional asymptotic normality \citep{yan2013central, yan2016asymptotics, chen2019analysis, stein2020sparse}.

\section{An AIC-type criterion for data-driven $\lambda$ selection}
\label{section::tuning-parameter}

Selecting the tuning parameter $\lambda$ in our method  poses a challenge.  
Unlike \citet{chen2019analysis} and \citet{stein2020sparse}, where $\lambda$ can be tuned by selecting a proper size of the active set ${\cal S}$, our method does not assume ``$\beta$-sparsity''.
Consequently, the BIC method for tuning $\lambda$ in \citet{chen2019analysis, stein2020sparse} is inapplicable in our setting.

To address this challenge, we propose a novel AIC-type criterion for the $\beta$-model.
Recall that the $\beta$-model can be viewed as a logistic regression with design matrix $X = (X_{(1,2)},\cdots,X_{(i,j)},\cdots,X_{(n-1,n)})^T\in\{0,1\}^{\binom{n}2\times n}$, where $X_{(i,j)}\in\{0,1\}^{n\times 1}$ contains two $1$'s at the $i$th and $j$th positions and is $0$ elsewhere.
The conventional AIC criterion is given by
\begin{align}
    \textrm{AIC}
    =&~
    \textrm{(number of model parameters)}
    +
    \textrm{(negative log-likelihood)}.
    \label{ordinary-AIC}
\end{align}
In a linear regression with ridge penalty, according to Section 1.8.1 of \citet{van2015lecture}, the first term on the RHS of \eqref{ordinary-AIC} equals the trace of the hat matrix $H(\lambda)$.
We draw an analogy by using the $H(\lambda)$ for the $\ell_2$-penalized logistic regression as the first term on the RHS of \eqref{ordinary-AIC}, replacing the number of model parameters as the measure of model complexity.
By Equation (12.3) in \citet{mccullagh2019generalized} and Equations (2.4.4), (2.4.7) and (2.4.13) in \citet{lu1994standardized}, we have
\begin{align}
    H(\lambda)
    = &~
    \tr\big[
        {\cal W}^{1/2} X \Big(X^T {\cal W}X + \lambda I\Big)^{-1} X^T {\cal W}^{1/2}
    \big],
    \label{AIC::raw}
\end{align}
where ${\cal W}:=\textrm{Diag}\big( {\cal W}_{(1,2),(1,2)}, \ldots, {\cal W}_{(1,n),(1,n)}, {\cal W}_{(2,3),(2,3)}, \ldots, {\cal W}_{(n-1,n),(n-1,n)} \big)$, in which, ${\cal W}_{(i,j), (i,j)} = \{V(\beta^*)\}_{i,j}$.
The formula \eqref{AIC::raw} involves multiplications of very large matrices.
Next, we simplify it to facilitate practical use.
One can verify that $X^T{\cal W}X = V(\beta^*)$.
We have
\begin{align}
    H(\lambda)
    = &~
    \tr\big\{
        (X^T{\cal W}X + \lambda I)^{-1}X^T{\cal W}X
    \big\}
    =
    \tr\big\{
        \big(
            I+\lambda\{V(\beta^*)\}^{-1}
        \big)^{-1}
    \big\}
    \notag\\
    \leq &~
    n\lambda_{\max}\big\{
        \big(
            I+\lambda\{V(\beta^*)\}^{-1}
        \big)^{-1}
    \big\}
    =
    \frac{n}{\lambda_{\min}\big(I+\lambda\{V(\beta^*)\}^{-1}\big)}
    =
    \frac{n}{1+\lambda\cdot \{\lambda_{\max}(V(\beta^*))\}^{-1}}.
    \label{AIC::simplification-step}
\end{align}
To speed up computation, we introduce some practical approximations to further simplify \eqref{AIC::simplification-step}.  
Observing that the upper bounds for $\|V(\beta^*)\|_\infty$ and $\|V(\beta^*)\|_{\rm op}$ in \citet{hillar2012inverses} are similar, we can approximate $\lambda_{\max}(V(\beta^*))$ by $\|V(\beta^*)\|_\infty=q_n^{-1}(n-1)$.  
Subsequently, we estimate $q_n^{-1}(n-1)$ by $d_{\max} = \max_{1\leq i\leq n}d_i$.
This leads to our proposed AIC criterion:
\begin{align}
    {\rm AIC}(\lambda)
    =
    \frac{nd_{\max}}{d_{\max}+\lambda}
    +
    {\cal L}_\lambda(\hat\beta_\lambda)
    .
    \label{our-AIC-criterion}
\end{align}
Our AIC-type criterion demonstrates promising performance and usefulness under various settings, see the numerical results in Sections \ref{section::data::data-1} and \ref{section::data::data-2}.

\section{Simulations}
\label{section::numerical-studies}

\subsection{Simulation 1: convergence, consistency and asymptotic normality}
\label{section::simulation::subsec::performance and theorem-validation}

Our first simulation validates our method and theoretical predictions.
Set
$$
    \beta_1^*=\cdots=\beta_{\lfloor n/5\rfloor}^*=\tilde\alpha\log n,
    \quad
    \beta_{\lfloor n/5\rfloor+1}^*=\cdots=\beta_n^*=\tilde\gamma\log n.
$$
We set up  six different settings to investigate  various network sparsity regimes  and choices of $\lambda$, as detailed in Table \ref{tab::simu-1::set-up}.
Among these, settings (i) and (ii) generate relatively dense networks, for which we use a small $\lambda$.
Settings (iii)--(v) simulate sparser networks,  for which we test small and large $\lambda$.
Setting (vi) will be used for validating part (ii) of Theorem \ref{Main-theorem::asymptotic-normality} (asymptotic normality with large $\lambda$).
\begin{table}[h!]
    \centering
    \begin{tabular}{c|ccccc|c}\hline
        Setting & (i) & (ii) & (iii) & (iv) & (v) & (vi)\\\hline
        True $\tilde\alpha$ &       $-1/3$ & $-1/2$ & $-2/3$ & $-2/3$ & $-2/3$ & $-1/3$\\
        True $\tilde\gamma$ &       $1/5$  & $1/5$  & $1/3$  & $1/3$  & $1/3$ & $-1/3+0.05$\\
        Working $\lambda$ &   $0.1$  & $0.1$  & $0.1$  & $10$   & $200$ & $2n$\\\hline
        Network sparsity & \multicolumn{5}{c|}{Dense $\longleftrightarrow$ Sparse} & \multirow{3}{*}{Small $\|\projperp\beta^*\|_\infty$}\\
        Degree heterogeneity & \multicolumn{5}{c|}{Low $\longleftrightarrow$ High} & \\
        Regularization $\lambda$ & \multicolumn{5}{c|}{Small $\longleftrightarrow$ Large} &\\\hline
    \end{tabular}
    \caption{Set up for Simulation 1}
    \label{tab::simu-1::set-up}
\end{table}

We evaluate the following performance  metrics:
(1)
    convergence speed:  quantified by $n^{-1/2}\|\hat\beta^{(t)}-\hat\beta_\lambda\|_2$, where $\hat\beta^{(t)}$ is the estimator $\hat\beta$ at the $t$th iteration;
(2)
    relative error $\|\hat\beta_\lambda - \beta^*\|_2/\|\beta^*\|_2$:  we plot  the log-relative error  against $\log n$;
    and
(3)
    computation time.
We implemented two versions of our algorithm, optimizing \eqref{eqn::reparameterization-1} by gradient descent and Newton's method, respectively.

\begin{figure}[htb]
    \centering
    \makebox[\textwidth][c]{
    \includegraphics[width=0.3\textwidth]{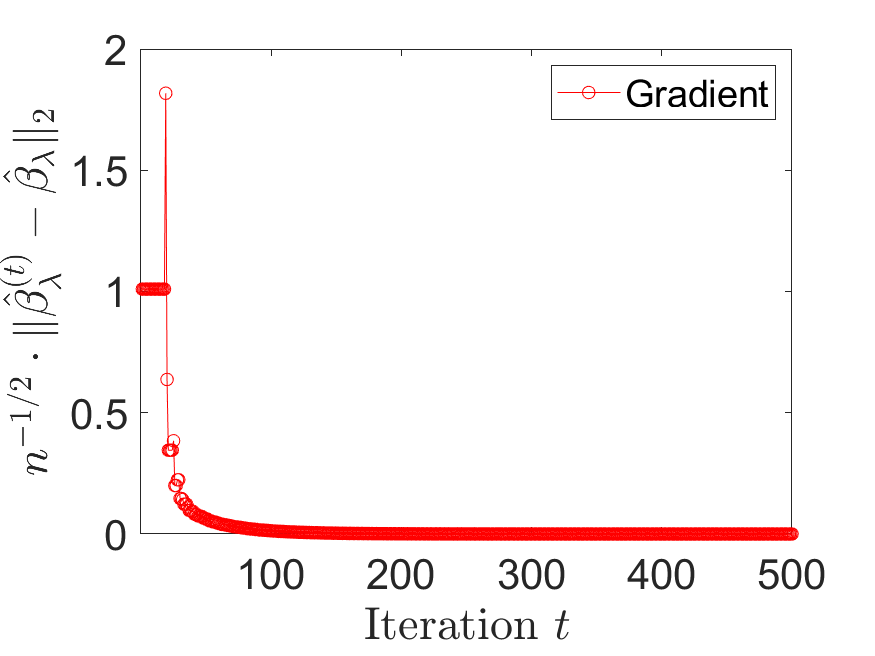}
    \includegraphics[width=0.3\textwidth]{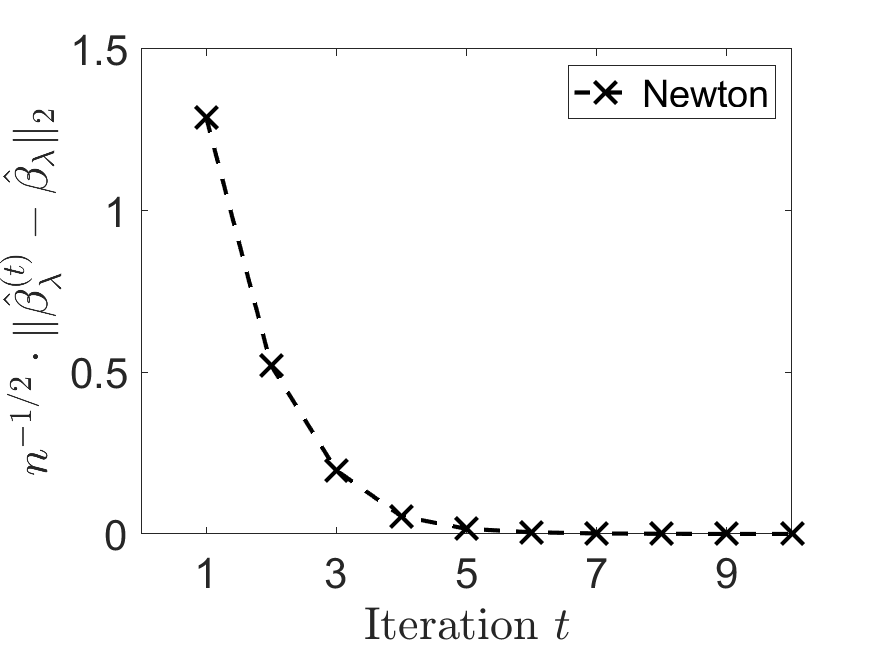}
    \includegraphics[width=0.3\textwidth]{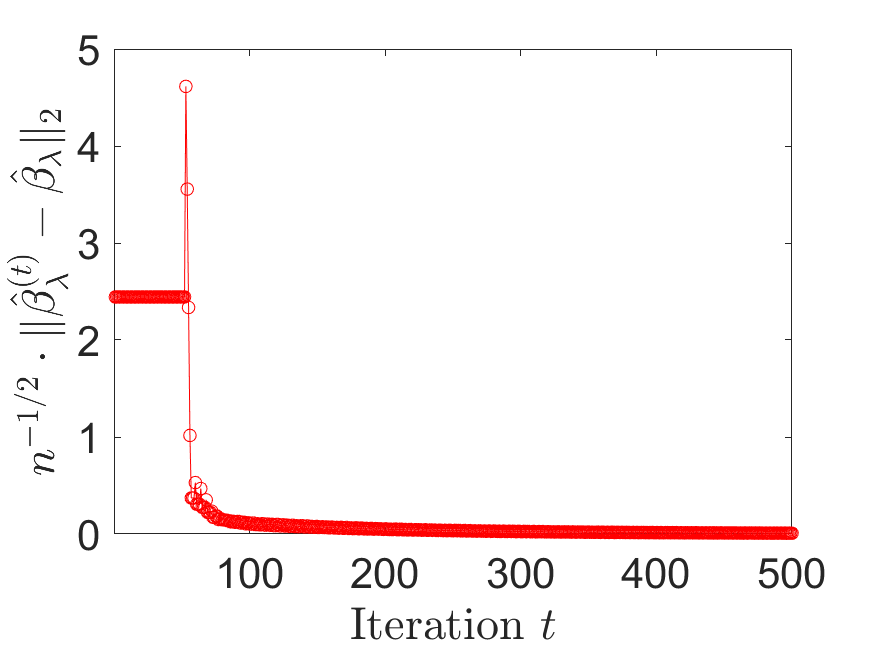}
    \includegraphics[width=0.3\textwidth]{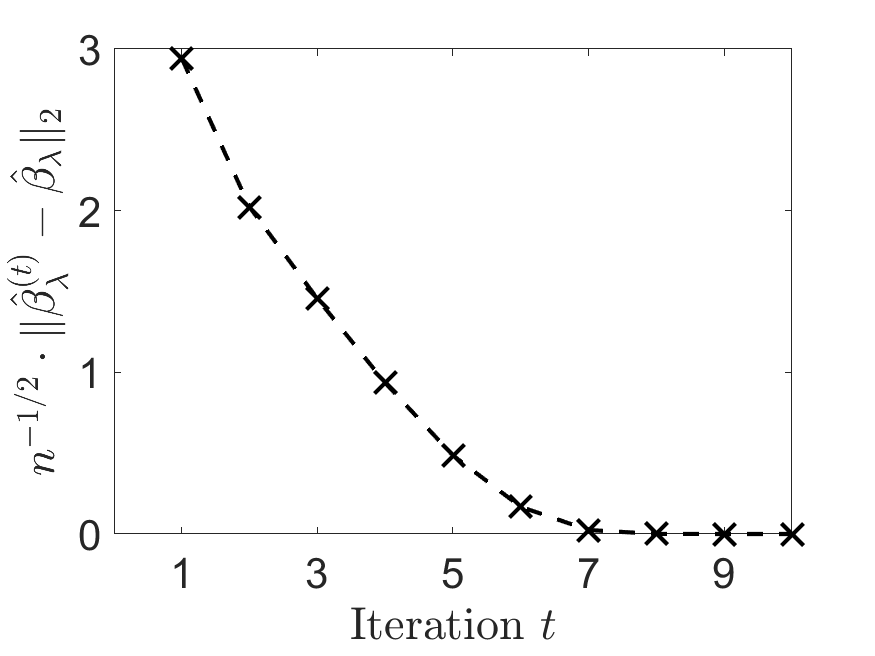}
    }
    \makebox[\textwidth][c]{
    \centering
    \includegraphics[width=0.3\textwidth]{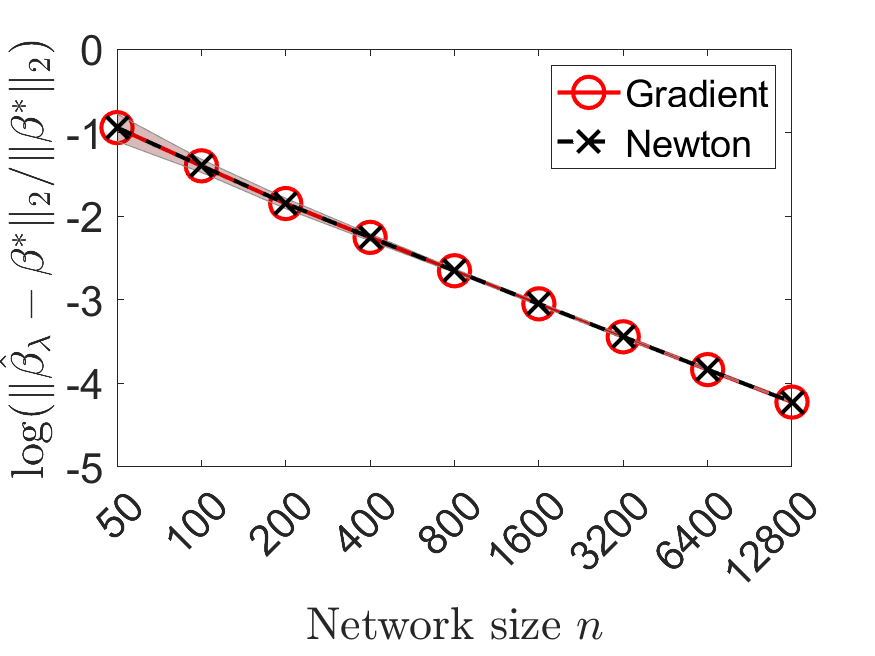}
    \includegraphics[width=0.3\textwidth]{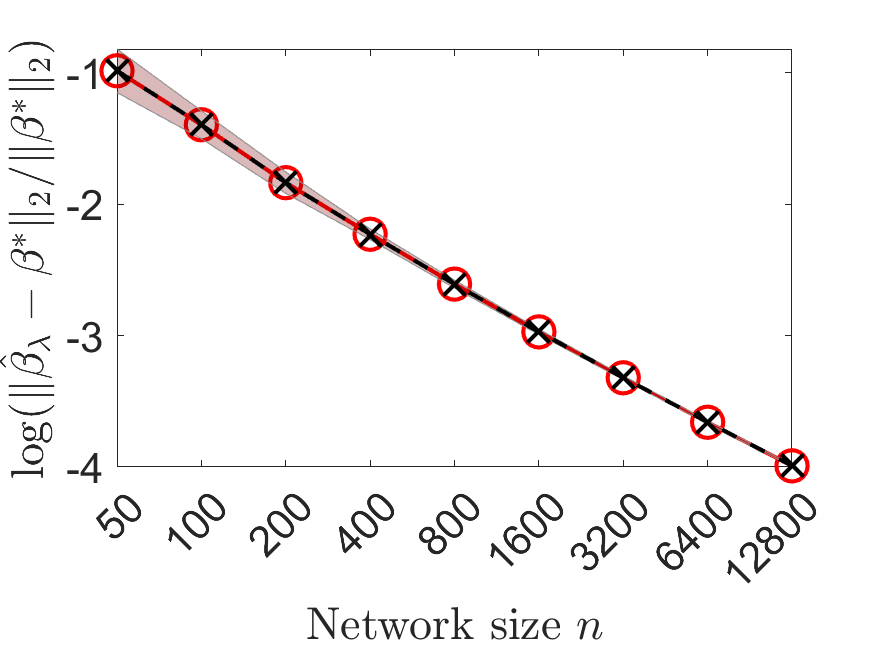}
    \includegraphics[width=0.3\textwidth]{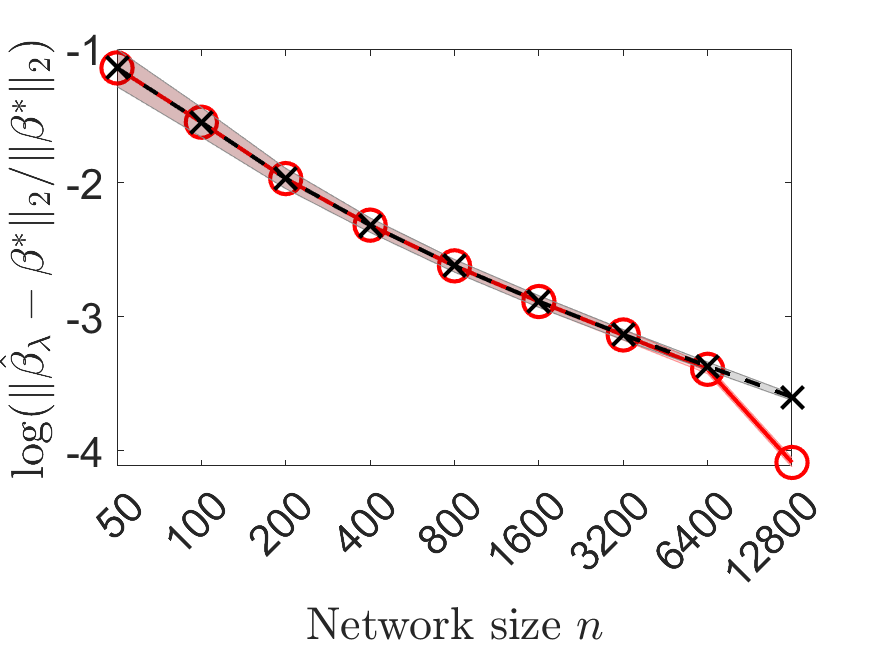}
    \includegraphics[width=0.3\textwidth]{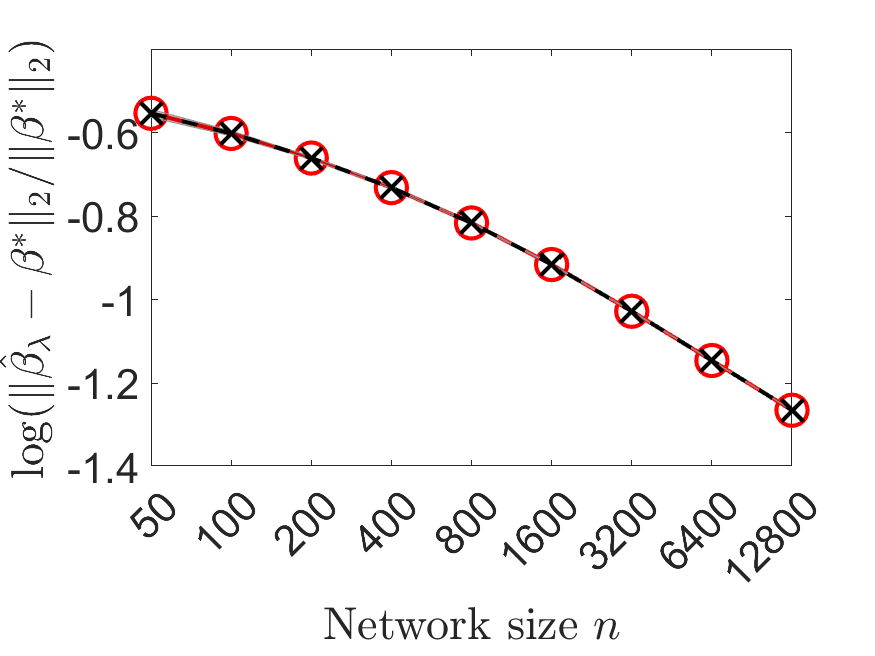}
    }
    \makebox[\textwidth][c]{
    \centering
    \includegraphics[width=0.3\textwidth]{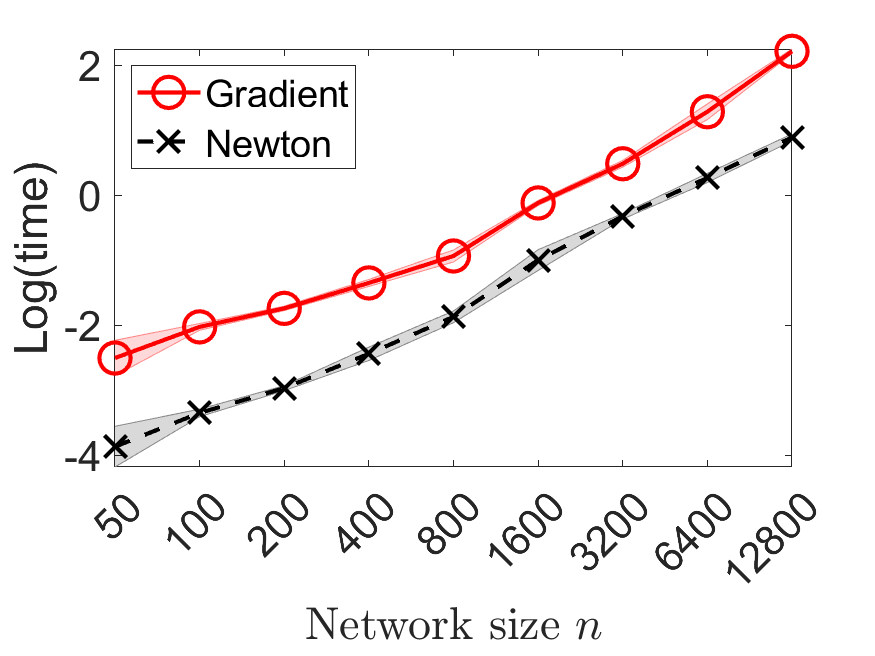}
    \includegraphics[width=0.3\textwidth]{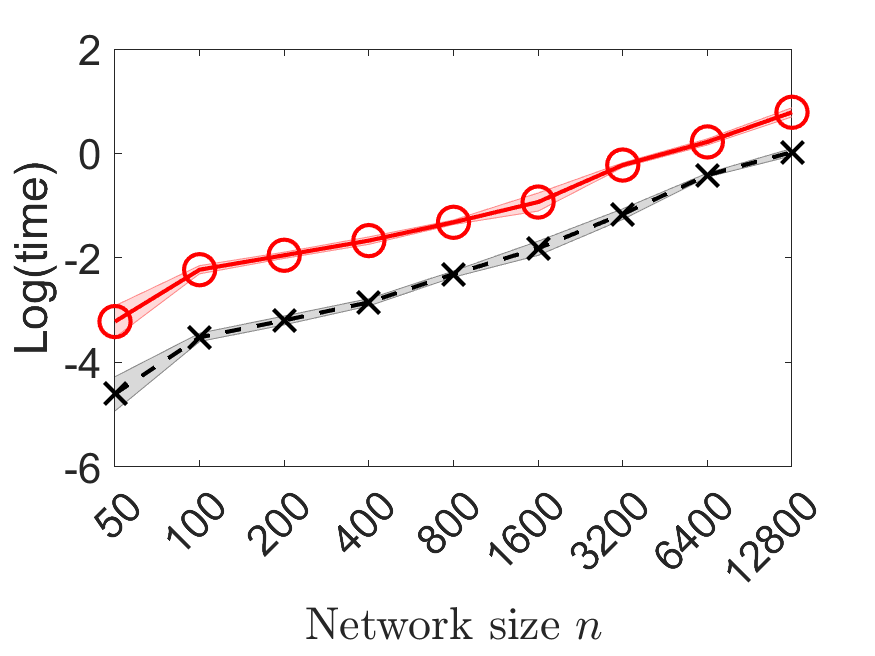}
    \includegraphics[width=0.3\textwidth]{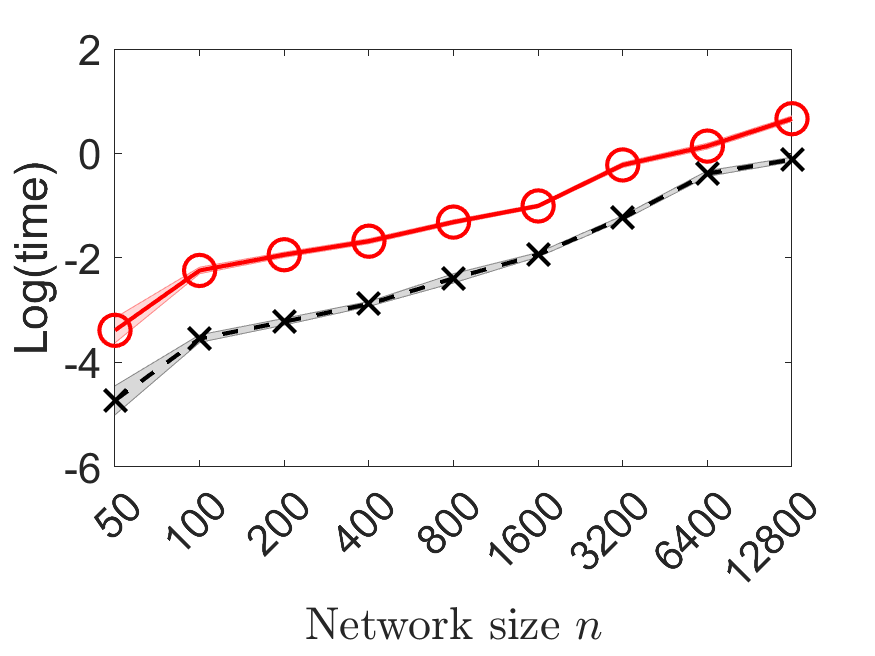}
    \includegraphics[width=0.3\textwidth]{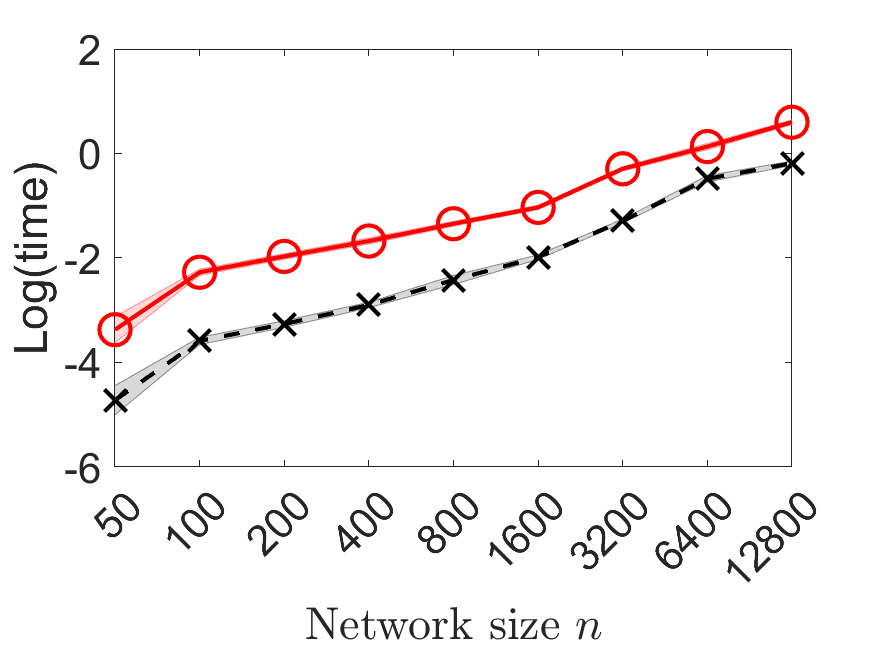}
    }
    \caption{
    Row 1: convergence speed, setting (i), plots 1 \& 2: $n=50$, 3 \& 4: $n=12800$; plots 1 \& 3: gradient, 2 \& 4: Newton;
    Row 2: log average $\ell_2$ errors, from left: settings (i)--(iv);  
    Row 3: computation time, from left: settings (i)--(iv).}
    \label{fig::simu-1-error}
\end{figure}

We first evaluate the point estimation performance of our method.
Figure \ref{fig::simu-1-error} shows the result. 
Row 1 shows that Newton's method needs fewer iterations to converge, but it has a higher per-iteration computation cost than gradient descent  and consequently does not show a significant overall speed advantage.
Row 2 shows an $n^{-1/2}$ decaying rate of the relative error in most settings.
The difficulty of the estimation problem increases from setting (i) to setting (iii), and the error values on the y-axis of the plots in row 2 corroborate this understanding.
Row 3 confirms  that the runtime scales as $O(\rho_n n^2)$  as predicted by our theory.

\begin{figure}[htb]
    \centering
    \makebox[\textwidth][c]{
    \includegraphics[width=0.3\textwidth]{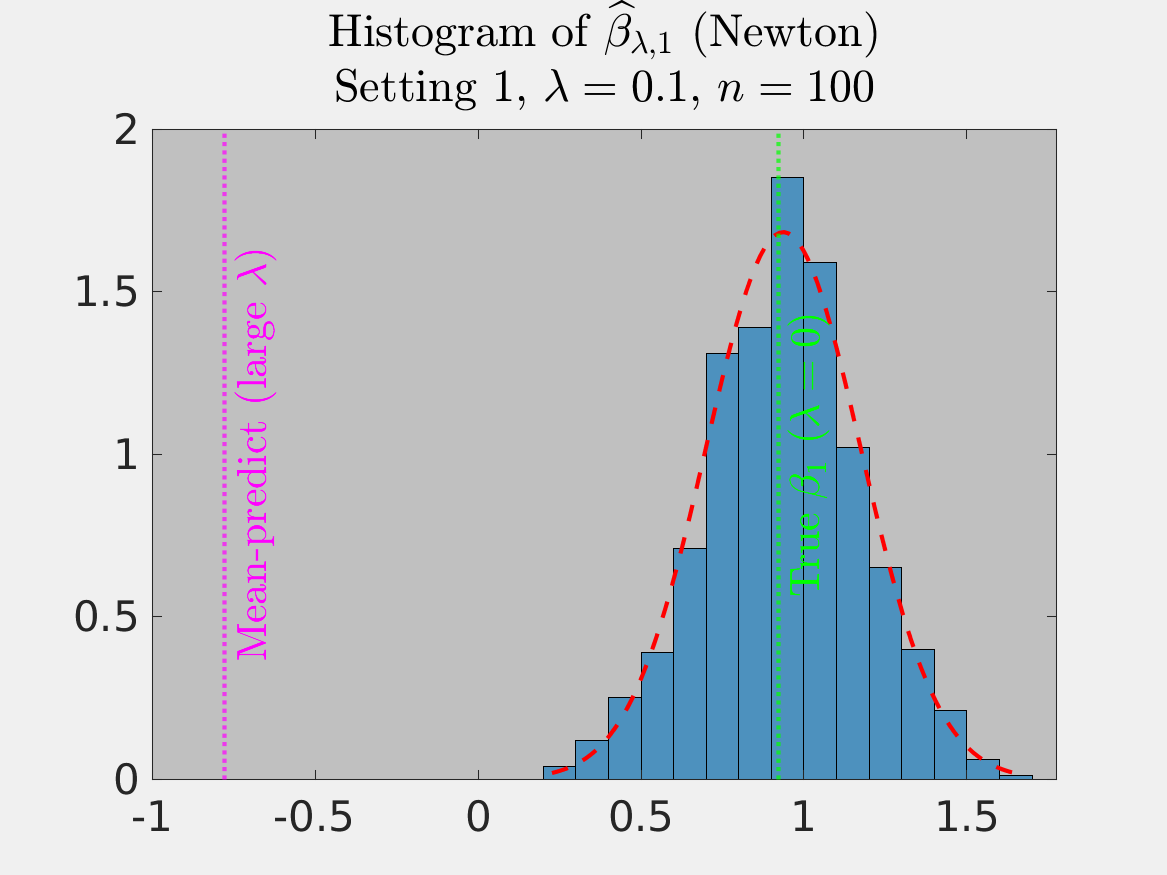}
    \includegraphics[width=0.3\textwidth]{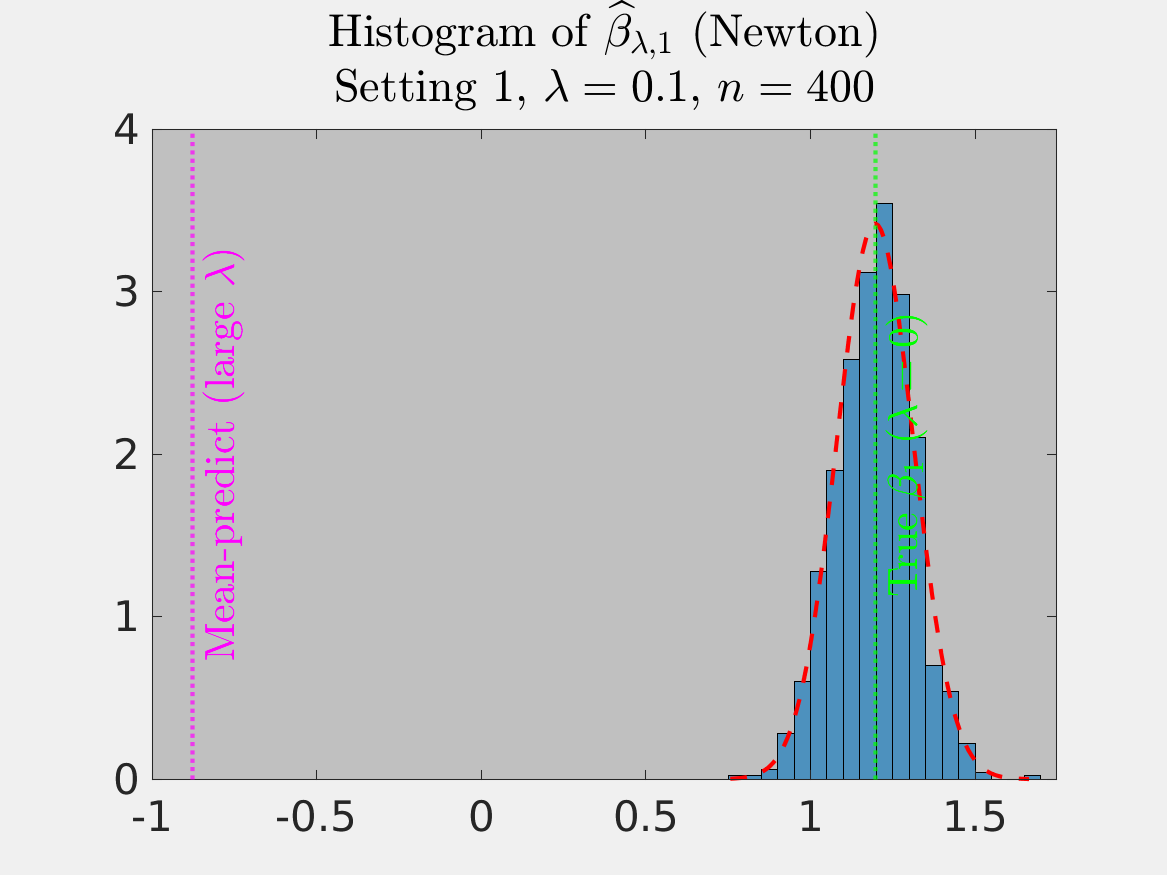}
    \includegraphics[width=0.3\textwidth]{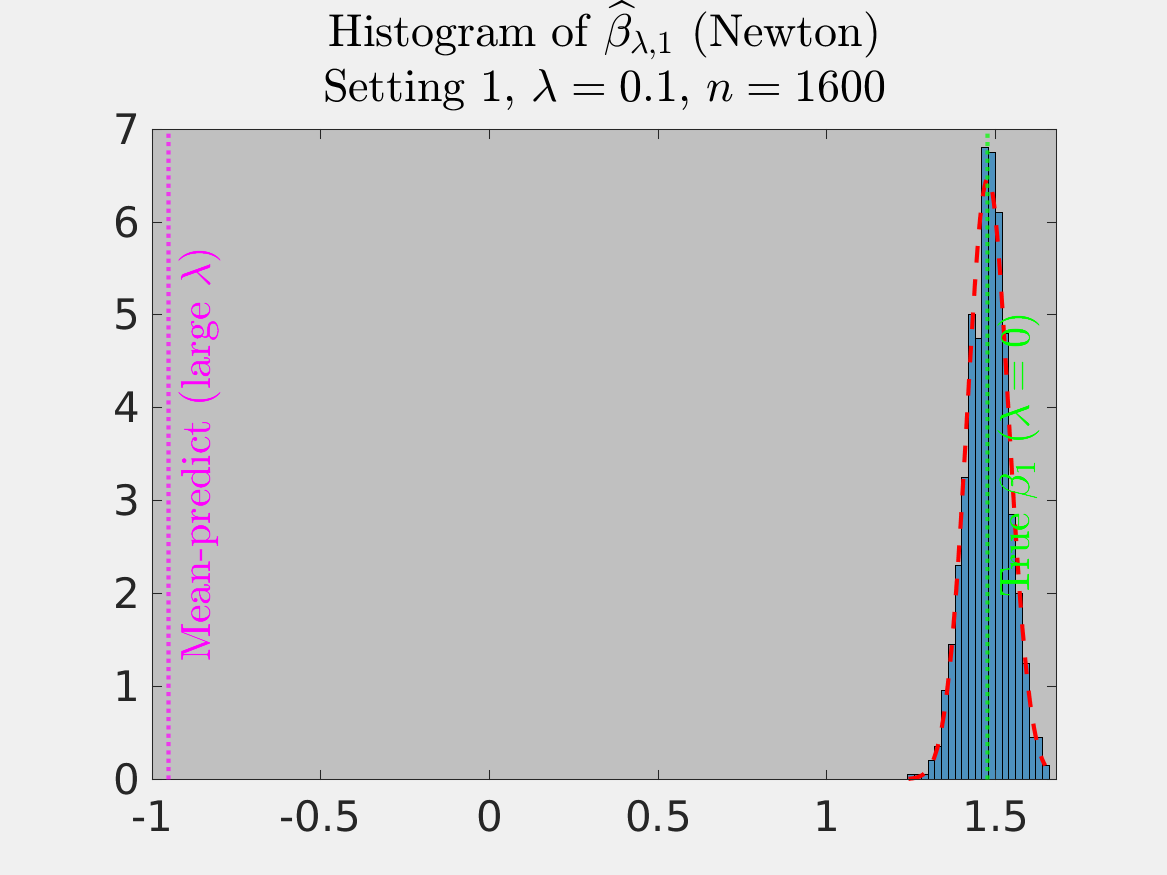}
   \includegraphics[width=0.3\textwidth]{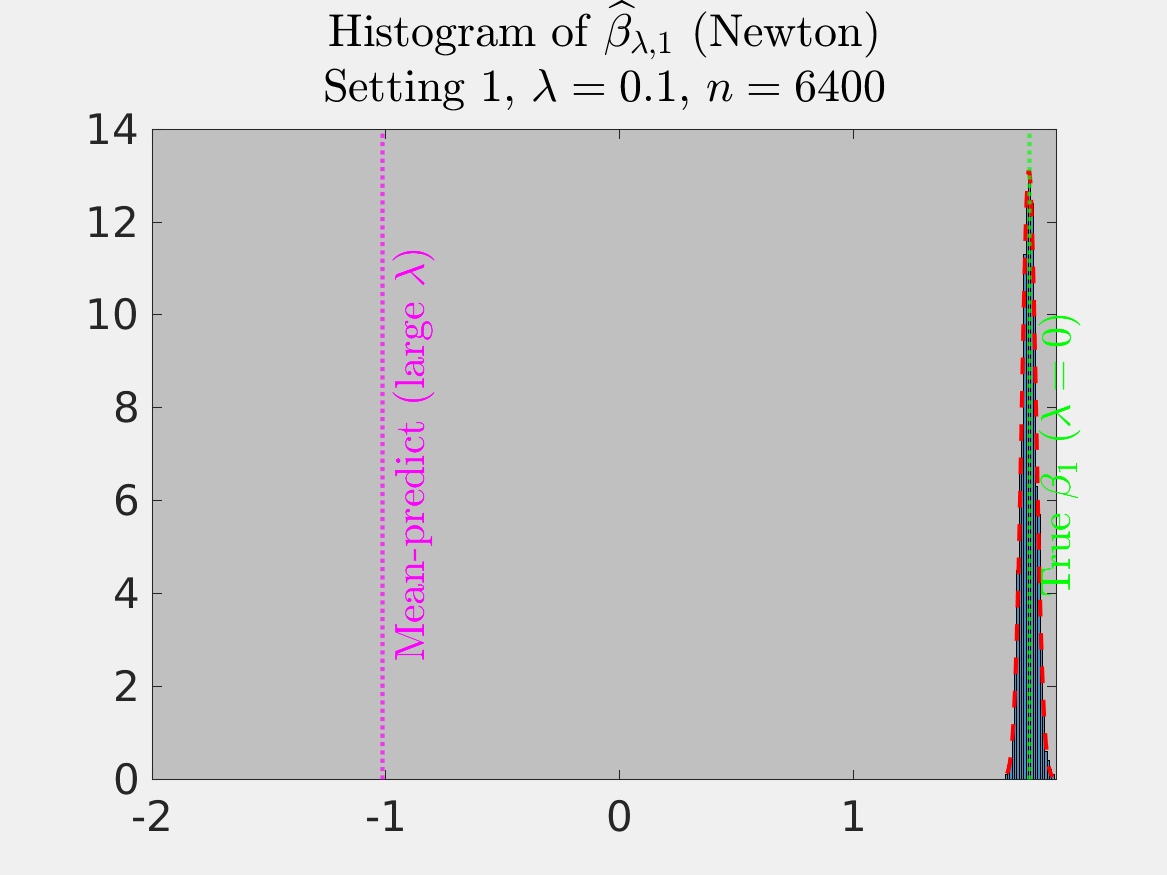}
    }
    \makebox[\textwidth][c]{
    \includegraphics[width=0.3\textwidth]{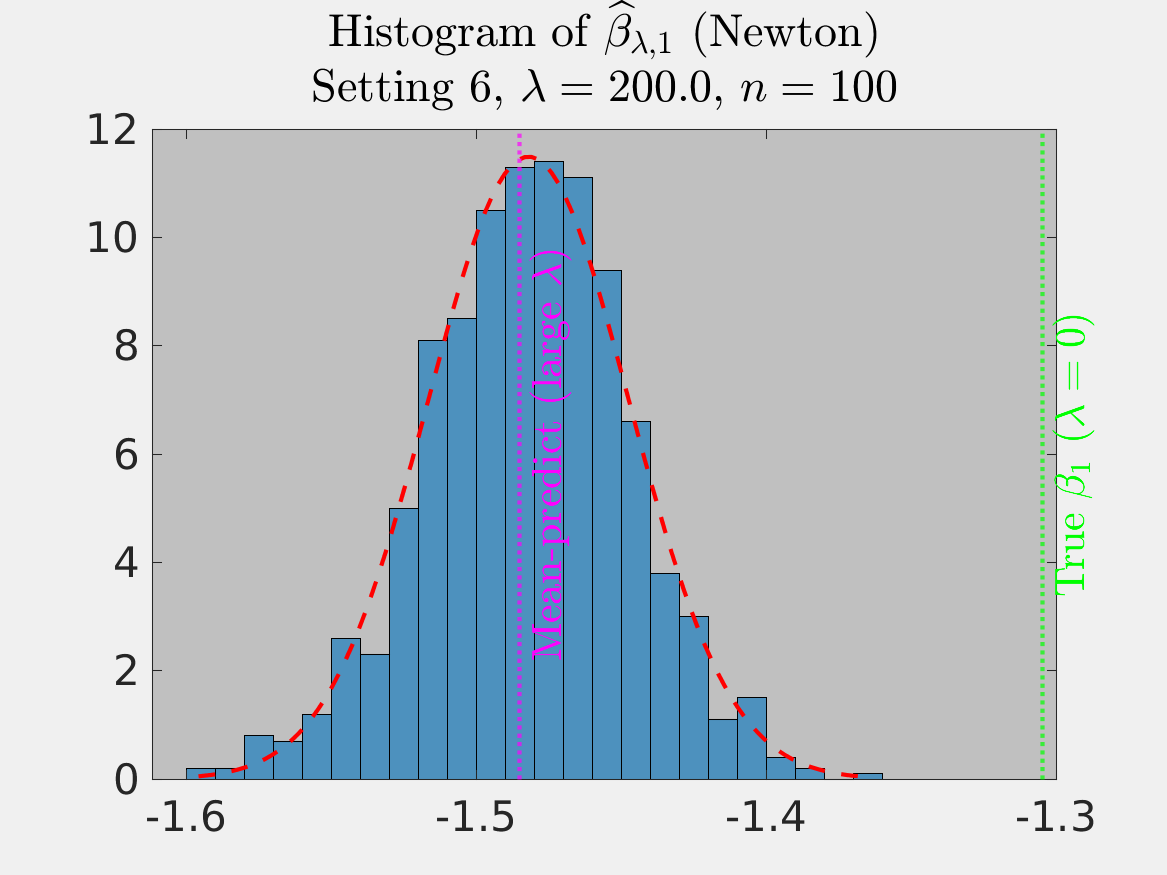}
    \includegraphics[width=0.3\textwidth]{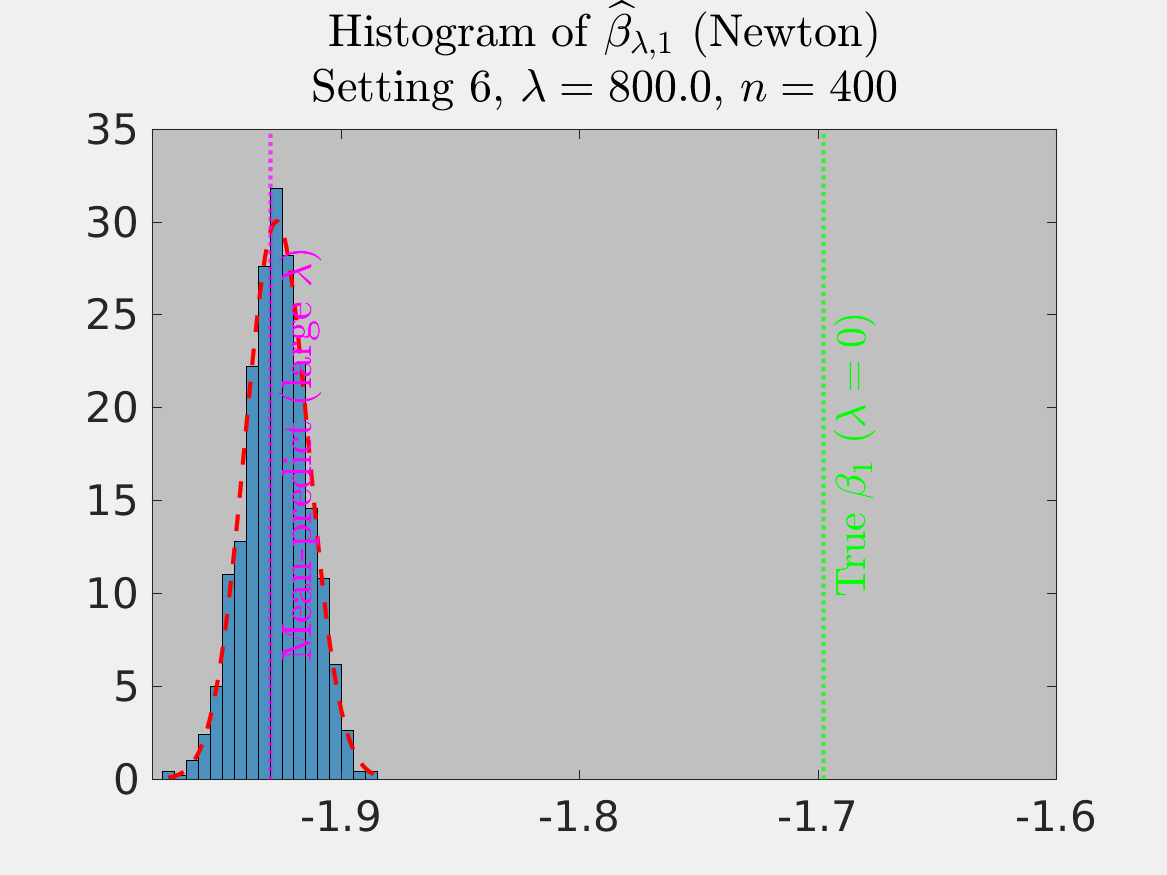}
    \includegraphics[width=0.3\textwidth]{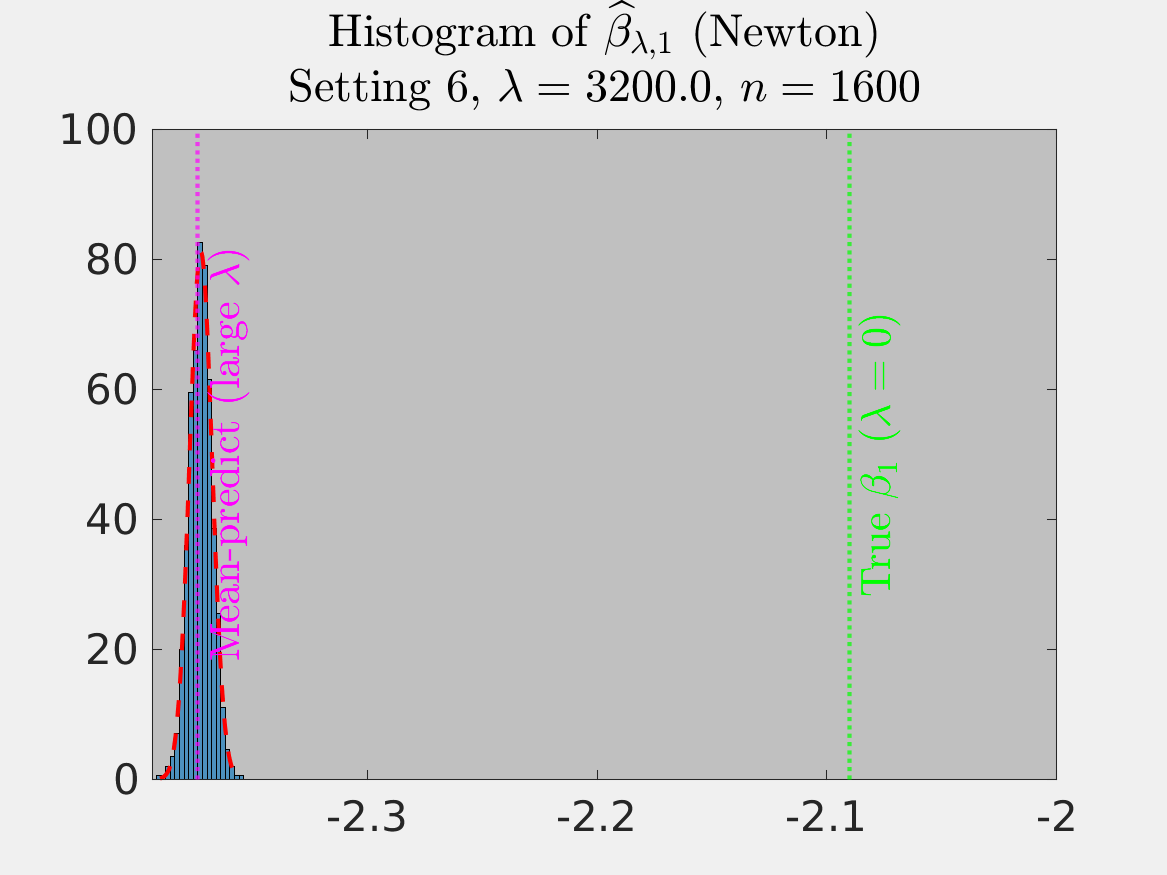}
   \includegraphics[width=0.3\textwidth]{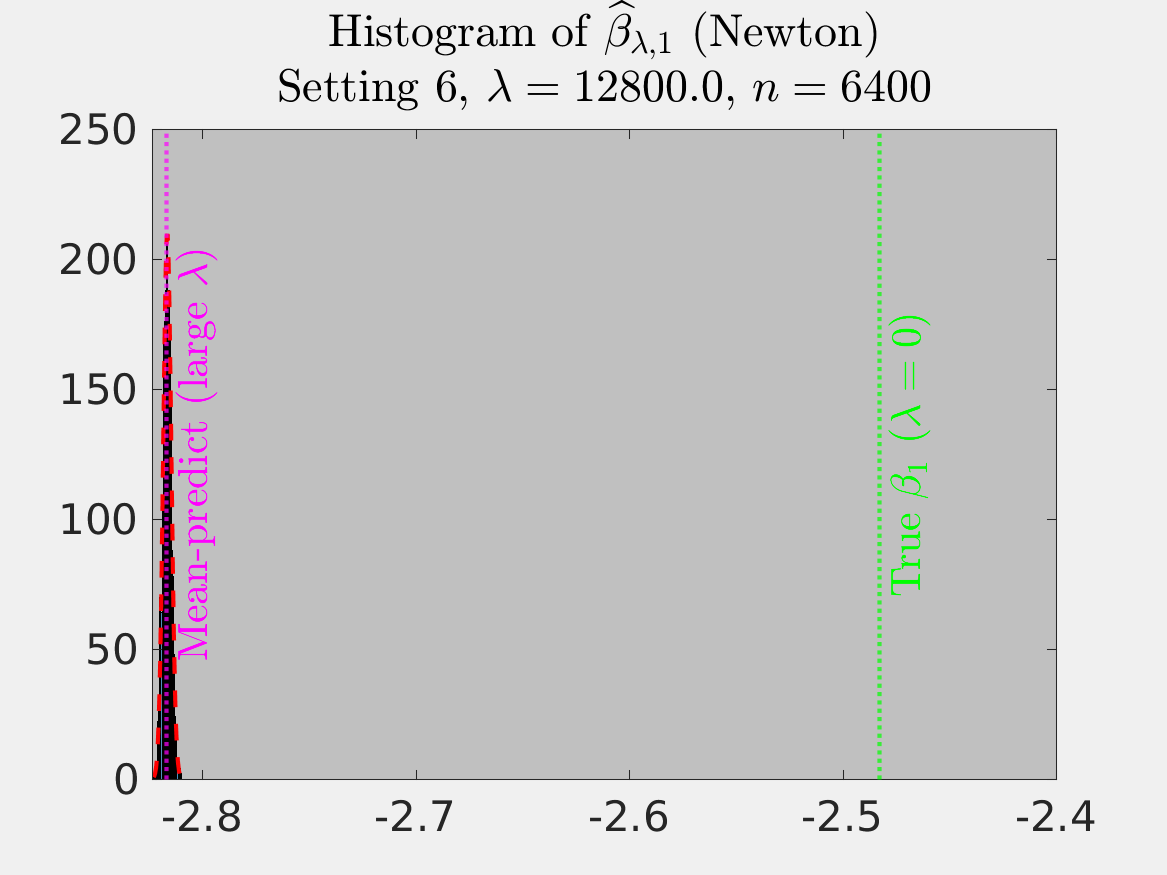}
    }
    \caption{Asymptotic distributions of $\hat\beta_{\lambda;1}$ under various $n$ for tuning parameter  $\lambda$.  Row 1: setting 1 (moderately dense network, small $\lambda$); row 2: setting 6 (sparse network, large $\lambda$).  Green line: mean prediction for small $\lambda$ case; magenta line: mean prediction for large $\lambda$ case.}
    \label{fig::simu-2::normality-pred-mean}
\end{figure}

Next, we validate Theorem \ref{Main-theorem::asymptotic-normality}. 
This is further divided into two sub-tasks: validating the predictions about the center  structure and  the variance structure, respectively.
To validate our theorem's  predicted asymptotic center, we vary $n \in \{100, 400, 1600, 6400\}$ and plot the one-dimensional marginal empirical distributions of $\hat\beta_{\lambda;1}$ under settings (i) and (vi), respectively.  
Figure \ref{fig::simu-2::normality-pred-mean} shows the result.
Row 1 of Figure \ref{fig::simu-2::normality-pred-mean} corresponds to dense networks, where we set a small $\lambda = 0.1$, while in row 2, we set a large $\lambda = 2n$ to address sparse networks.
The result aligns well with the  predicted asymptotic center in Theorem \ref{Main-theorem::asymptotic-normality}.

\begin{figure}[htb]
    \centering
    \makebox[\textwidth][c]{
    \includegraphics[width=0.35\textwidth]{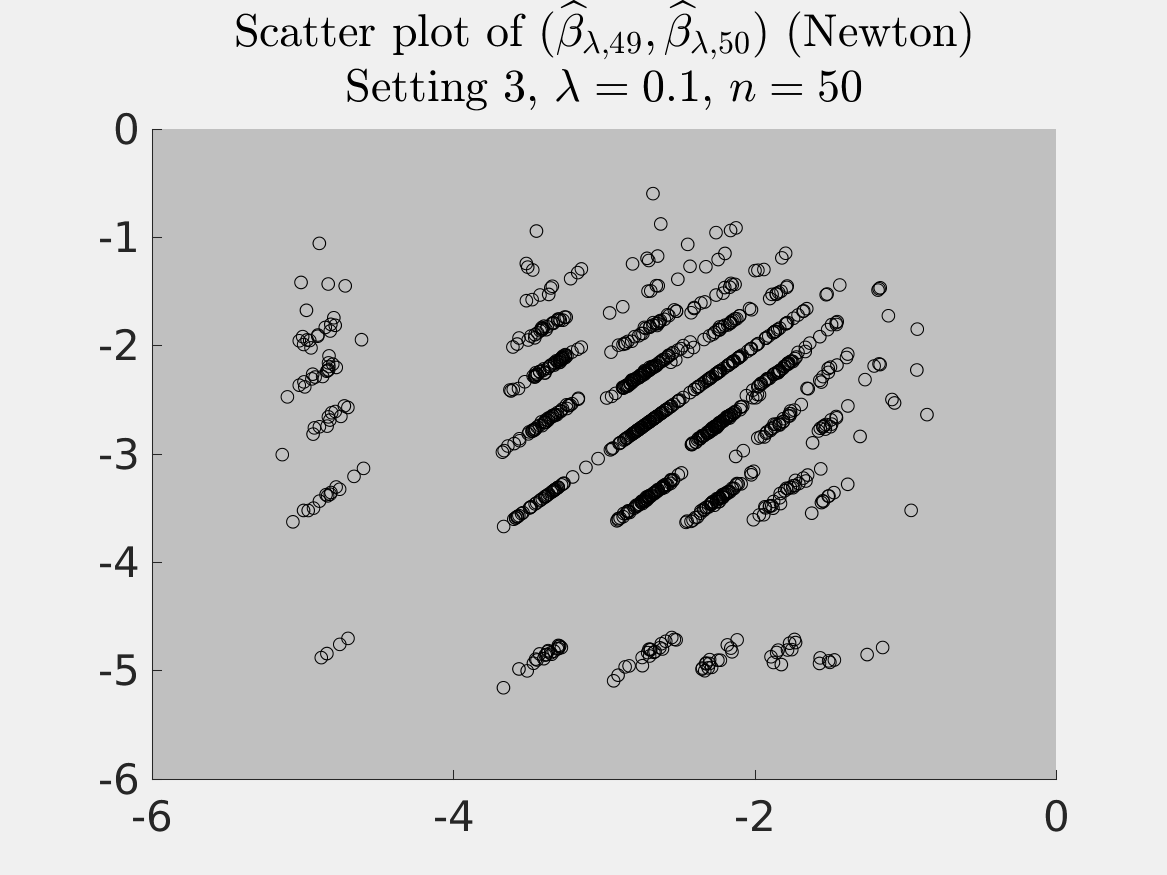}
    \includegraphics[width=0.35\textwidth]{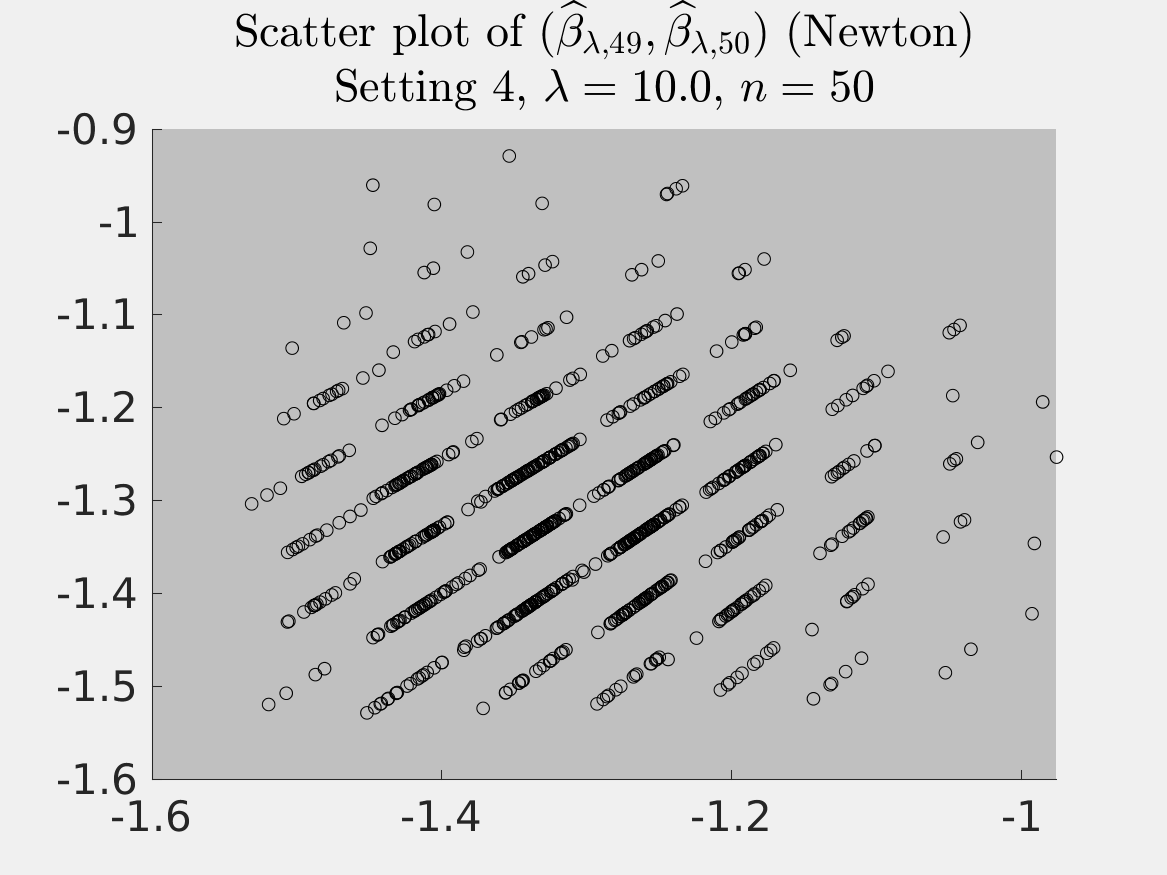}
    \includegraphics[width=0.35\textwidth]{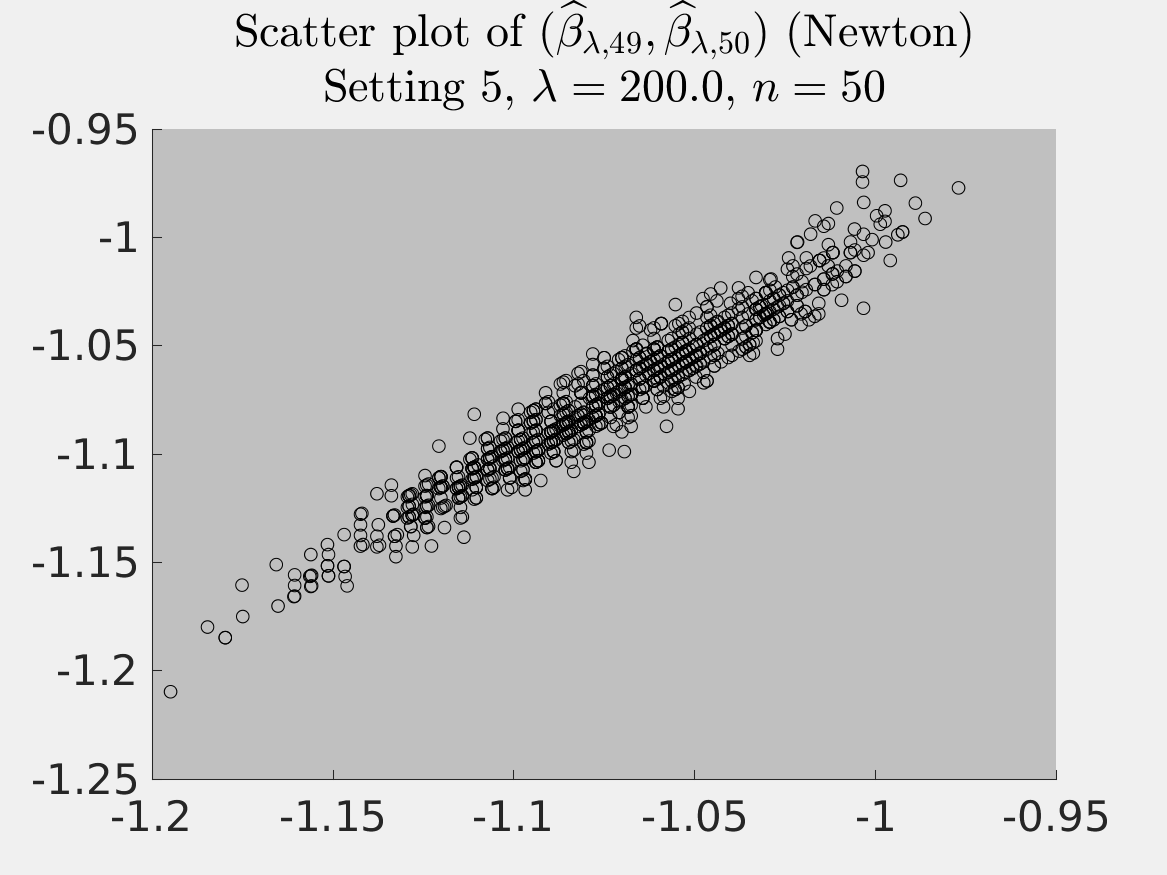}
    }
    \makebox[\textwidth][c]{
    \includegraphics[width=0.35\textwidth]{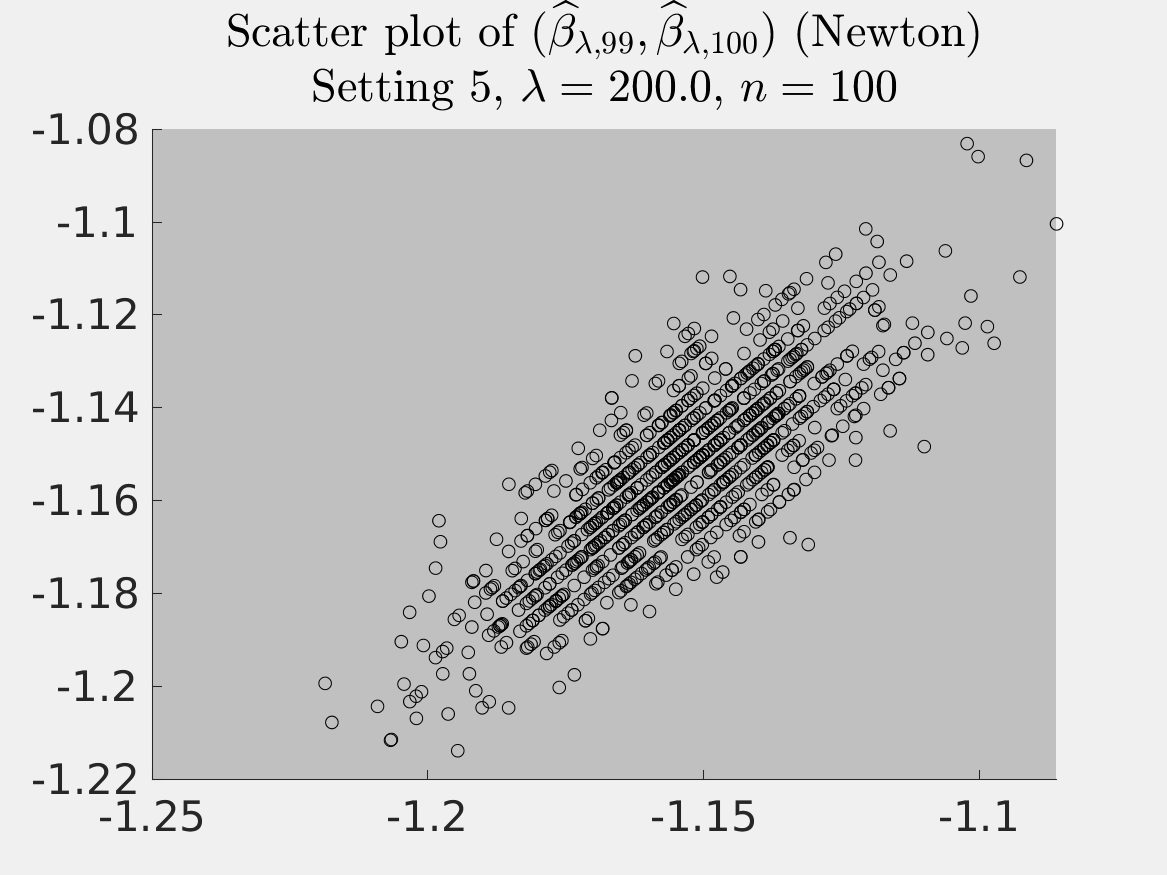}
    \includegraphics[width=0.35\textwidth]{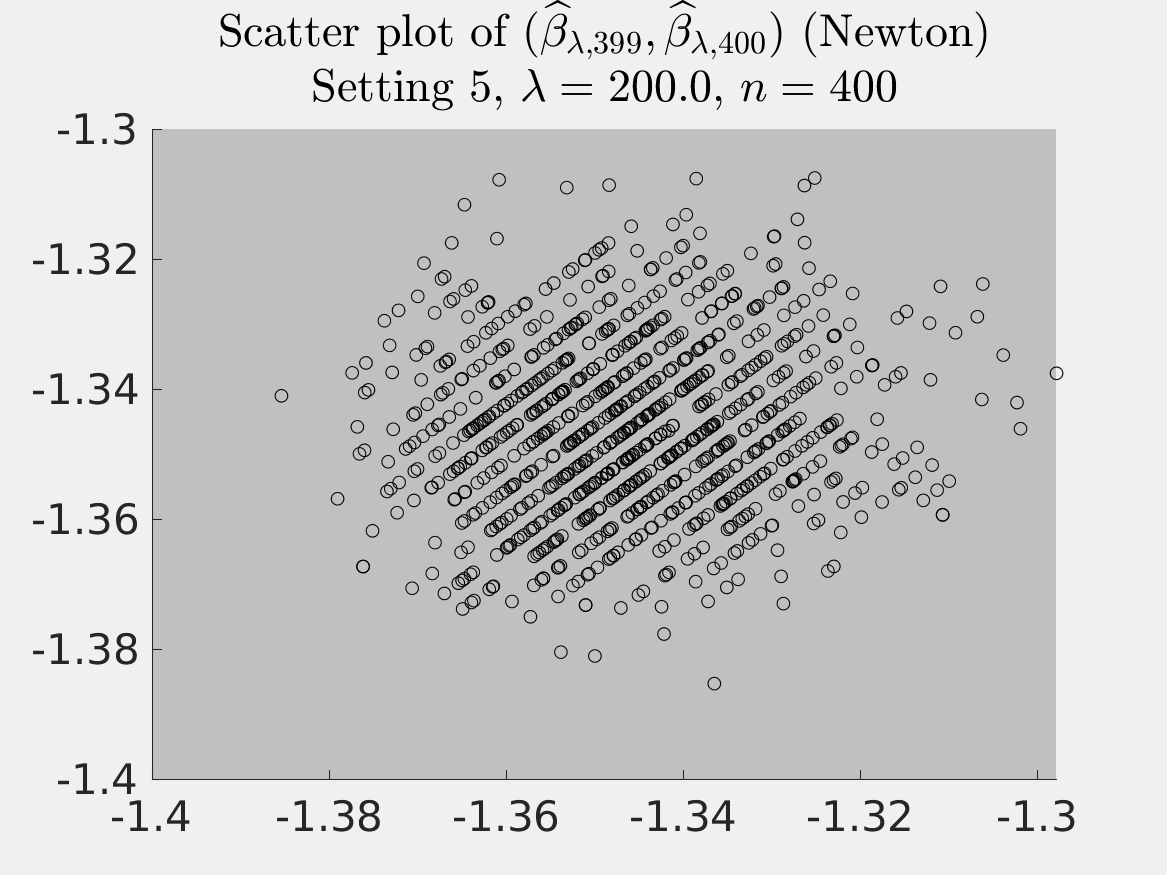}
    \includegraphics[width=0.35\textwidth]{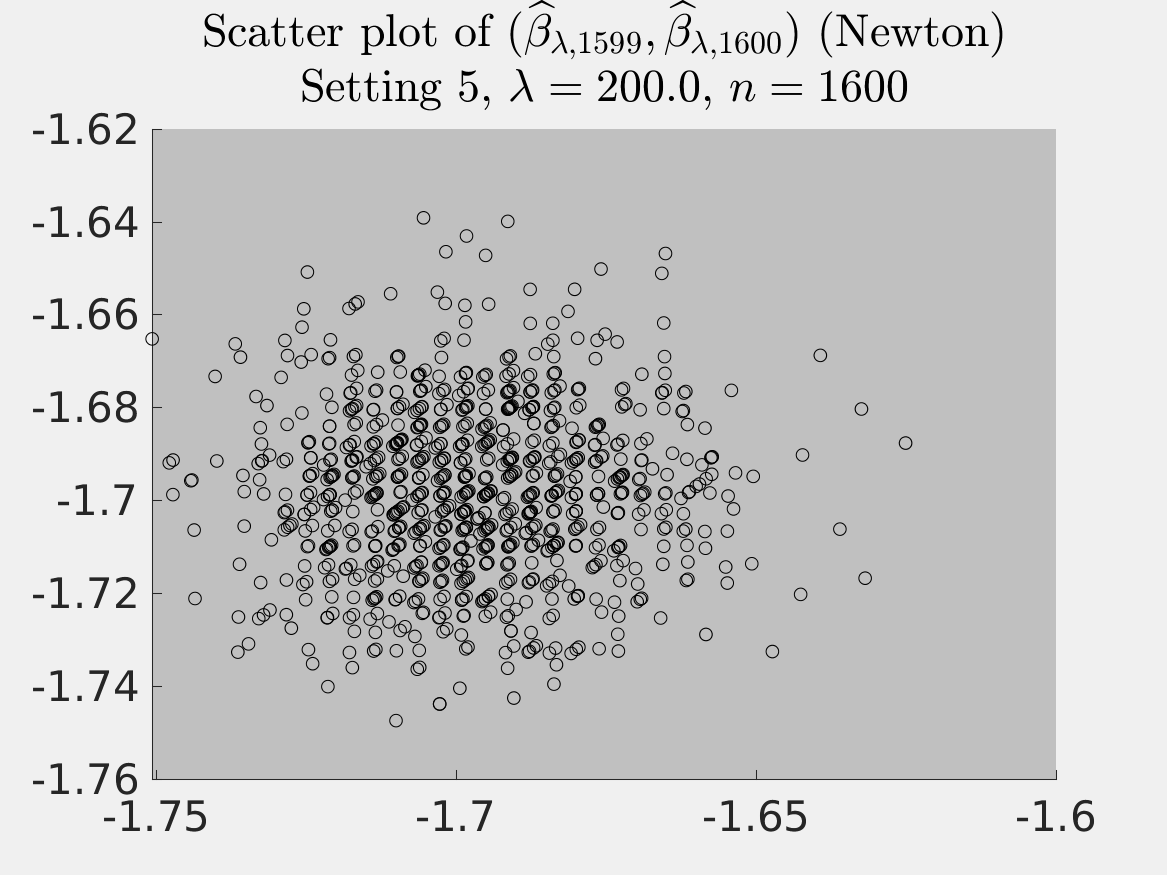}
    }
    \caption{Joint distribution of $(\hat\beta_{\lambda;n-1},\hat\beta_{\lambda;n})$.  Row 1: fix $n=50$ and increase $\lambda\in\{0.1,10,200\}$; row 2: fix $\lambda=200$ and increase $n\in\{100,400,1600\}$; other simulation set-ups are identical across all 6 plots.}
    \label{fig::simu-2::normality-joint-beta1-beta2}
\end{figure}

To validate  the prediction of our theory regarding the variance structure, we evaluate the dependency between $\hat\beta_{\lambda;n-1}$ and $\hat\beta_{\lambda;n}$ under our settings (iii) -- (v).
Figure \ref{fig::simu-2::normality-joint-beta1-beta2} shows the result.
 In row 1, from left to right (as $\lambda$ increases), the correlation between $\hat\beta_{\lambda;n-1}$ and $\hat\beta_{\lambda;n}$  increases, exactly as  predicted by part (ii) of our Theorem \ref{Main-theorem::asymptotic-normality}.
 Next, in row 2 (moving left to right), $n$ increases while $\lambda$ remains fixed.
This  mimics the effect  of decreasing $\lambda$  at fixed $n$.  
We observe  increasing independence between the two $\hat\beta_\lambda$ elements,  again confirming the prediction of Theorem \ref{Main-theorem::asymptotic-normality}.

\subsection{Simulation 2: performance comparison to $\ell_0$ and $\ell_1$ regularization methods}
\label{section::simulation::subsec::comparison-to-other-paper}

We compare our method to the $\ell_0$ \citep{chen2019analysis} and $\ell_1$ \citep{stein2020sparse} regularization methods.
Consider two settings  in the spirit of \citet{chen2019analysis}, emulating the $\beta$-sparsity structures:
\begin{align}
    \textrm{Setting 1: }
    \beta^*_i 
    =&~ 
    (-0.2\log n)/2 + \mathbbm{1}_{[ i\in \{1,\cdots,\lfloor n/10\rfloor \}]}(0.4\log n) + \mathbbm{1}_{[i\in \{\lfloor n/10\rfloor +1,\cdots,\lfloor n/5\rfloor \}]}(-0.4\log n),
    \notag
    \\
    \textrm{Setting 2: }
    \beta^*_i 
    =&~ 
    (-0.6\log n)/2 + \mathbbm{1}_{[i\in \{1,\cdots,\lfloor n/20\rfloor \}]}(0.2\log n) + \mathbbm{1}_{[i\in \{\lfloor n/20\rfloor +1,\cdots,\lfloor n/10\rfloor \}]}(-0.2\log n).
    \notag
\end{align}
Setting 1 generates dense networks, whereas setting 2 produces much  sparser networks.
We vary network size $n \in \{100, 200, \cdots, 3200\}$ and evaluate the performance of the $\hat \beta$'s produced by the three methods in the following five  metrics: 
(1) $n^{-1}\|\hat\beta - \beta^*\|_1$; 
(2) $n^{-1/2}\|\hat\beta - \beta^*\|_2$; 
(3) $\|\hat\beta - \beta^*\|_\infty$; 
(4) the relative error in estimating the active set ${\cal S}$ (${\cal S}=[1:[n/5]]$ in the dense network setting, and ${\cal S}=[1:[n/10]]$ in the sparse network setting),  measured by the Hamming distance between $\hat {\cal S}$ and ${\cal S}$ divided by $n$;
and (5) computation time.
In (4), our method uses Algorithm \ref{algorithm::thresholding} to obtain $\hat{\cal S}$.
In each setting, we repeated the experiment 1000 times for our method and 100 times for the other two methods due to their much higher computational costs.

\begin{figure}[h]
    \centering

    \makebox[\textwidth][c]{
    \includegraphics[width=0.235\textwidth]{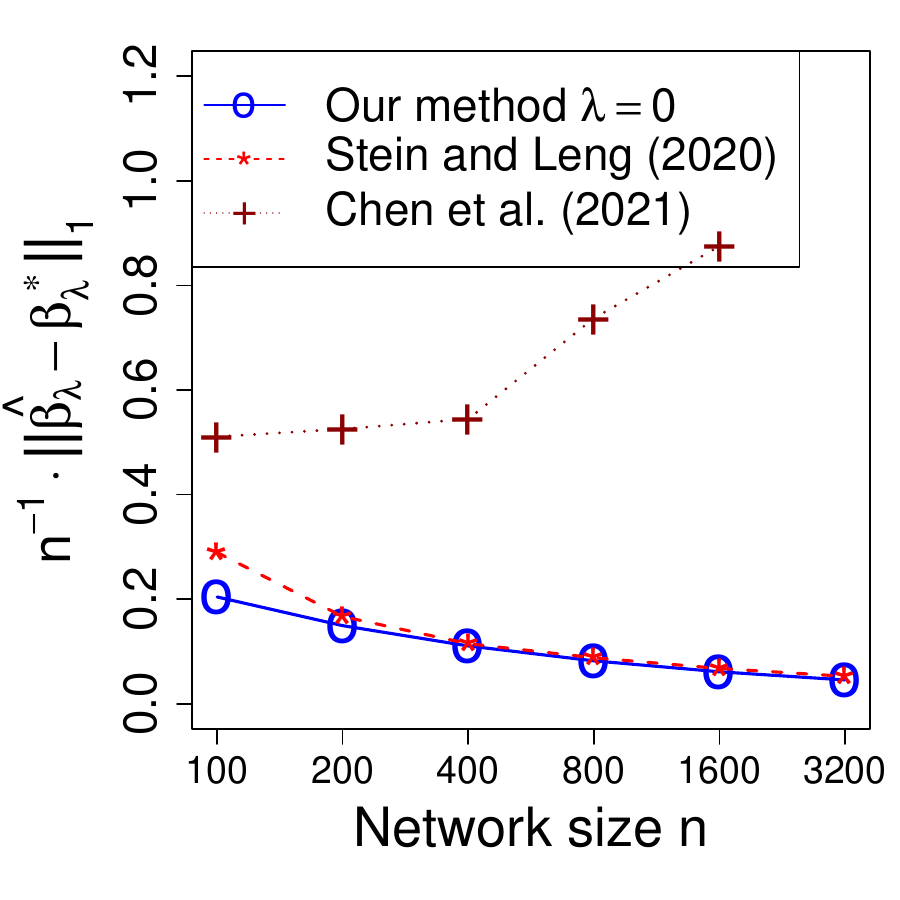}
    \includegraphics[width=0.235\textwidth]{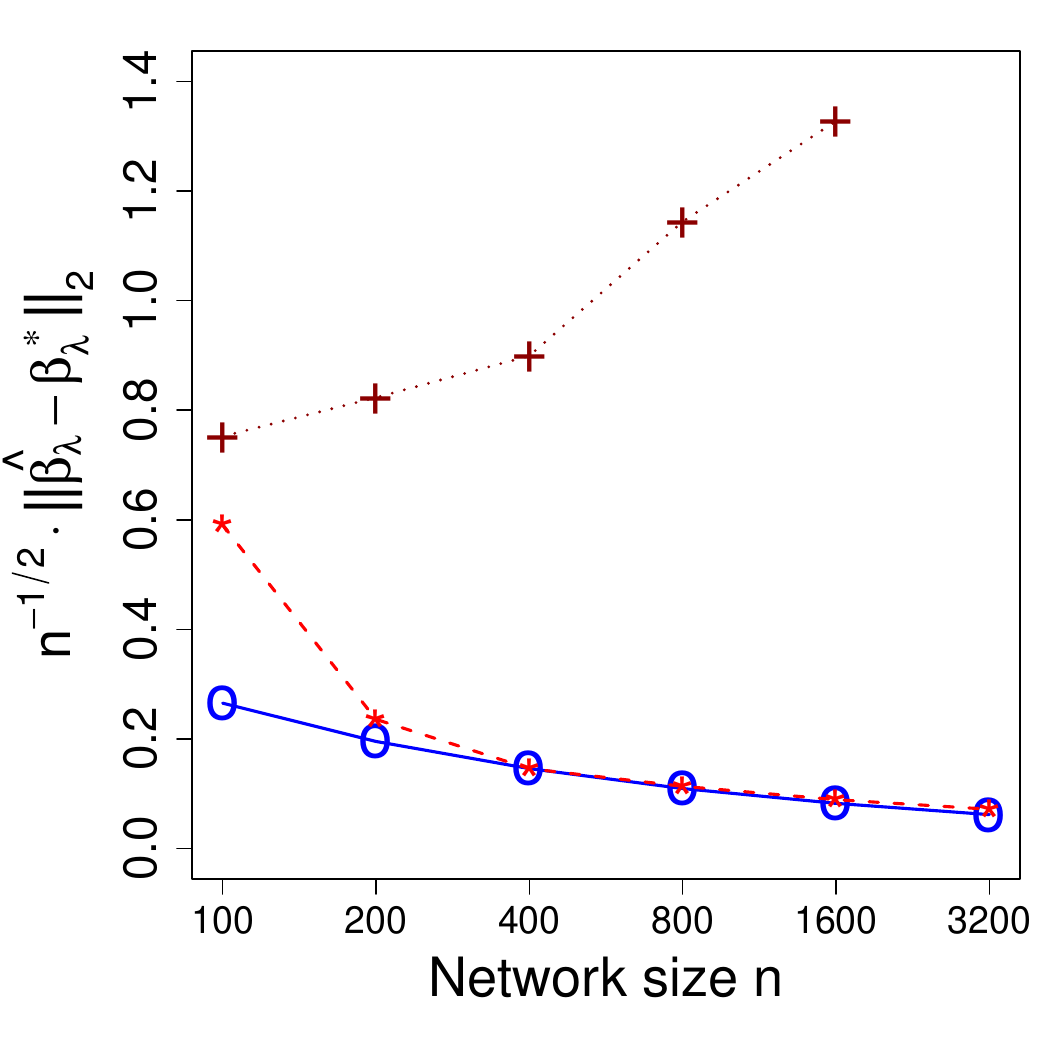}
    \includegraphics[width=0.235\textwidth]{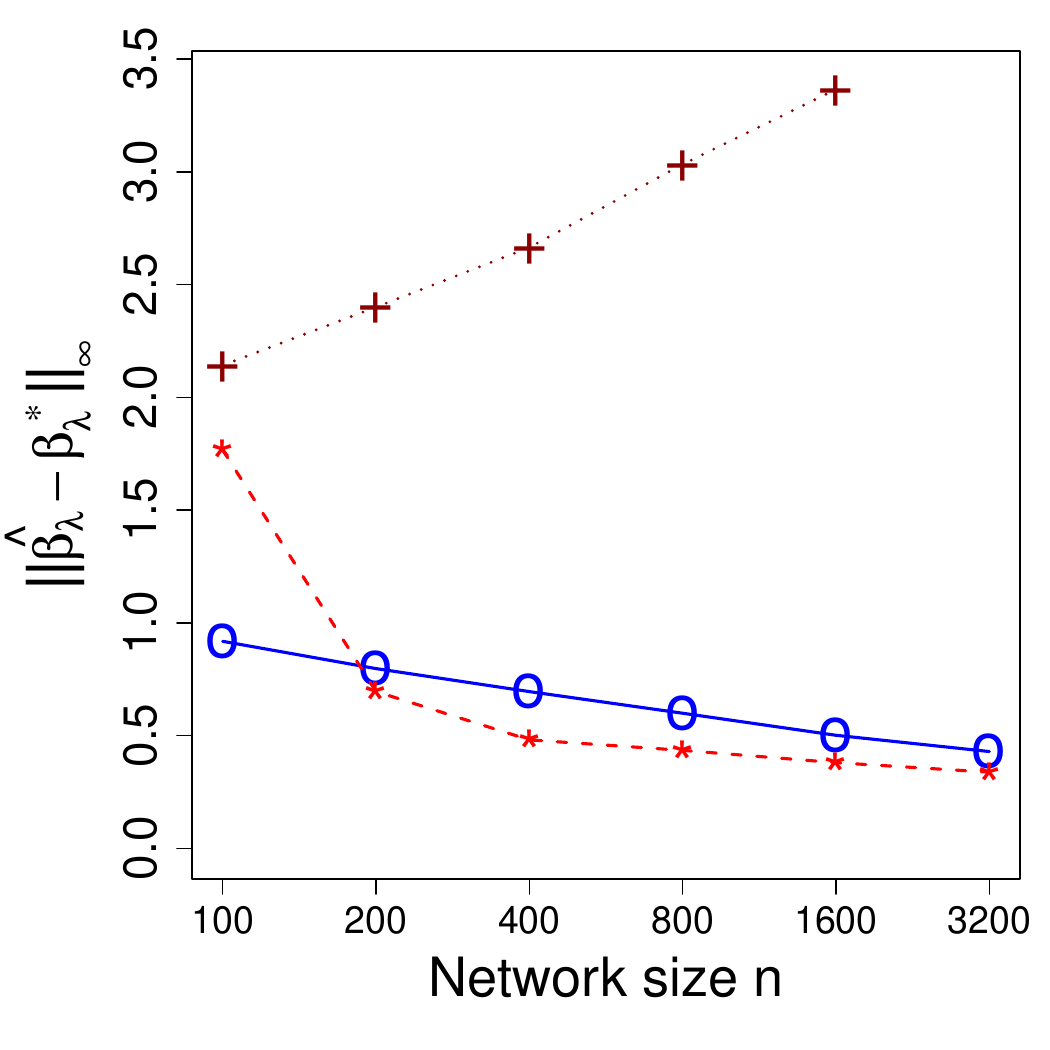}
    \includegraphics[width=0.235\textwidth]{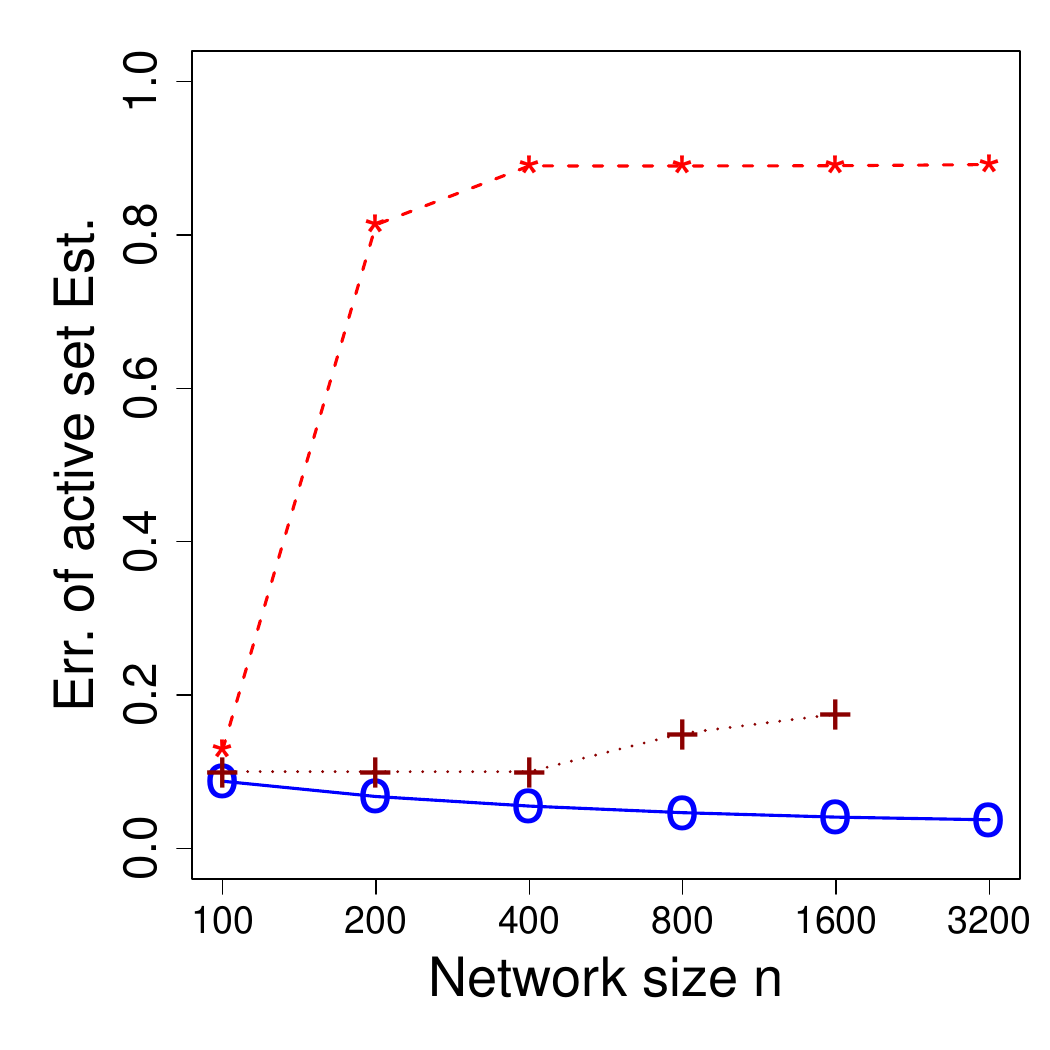}
    \includegraphics[width=0.235\textwidth]{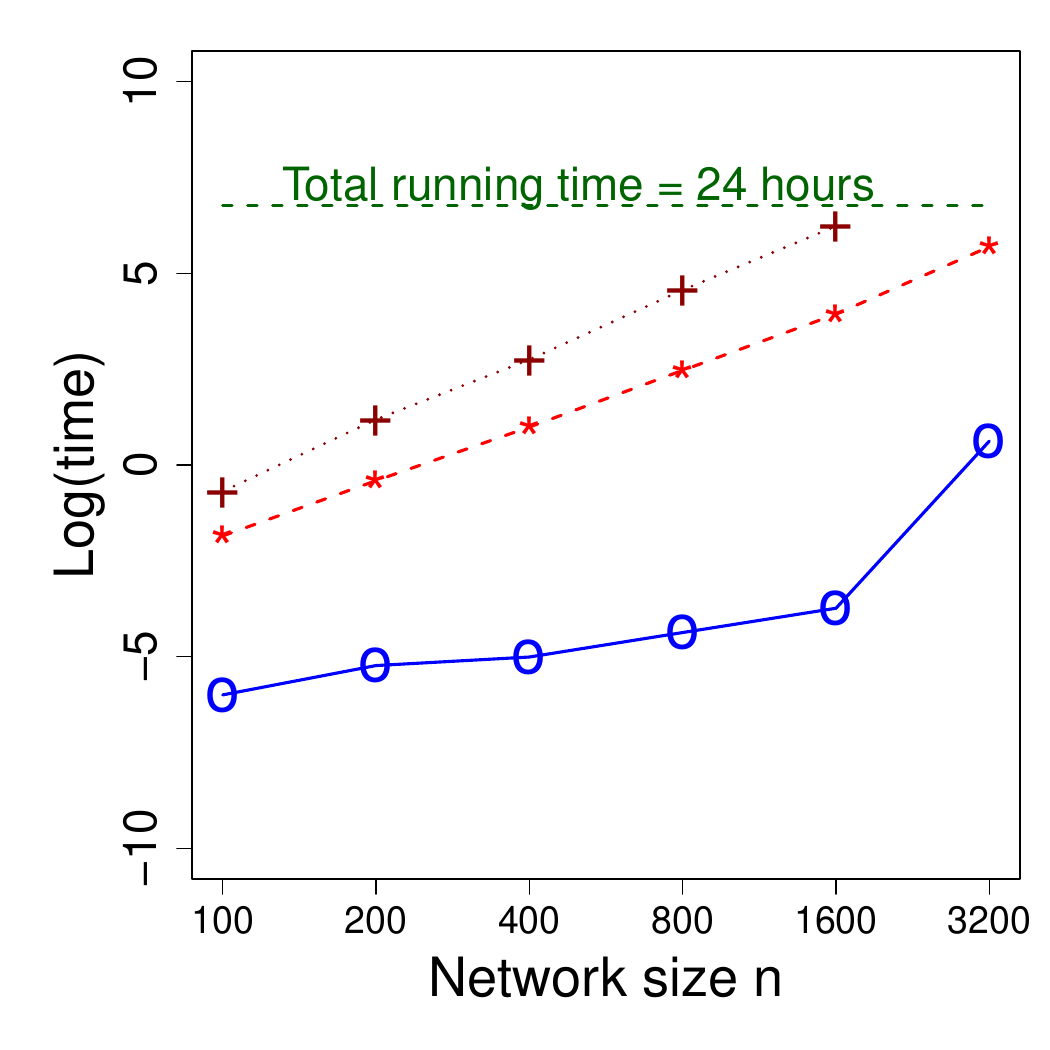}
    }
    \makebox[\textwidth][c]{
    \includegraphics[width=0.235\textwidth]{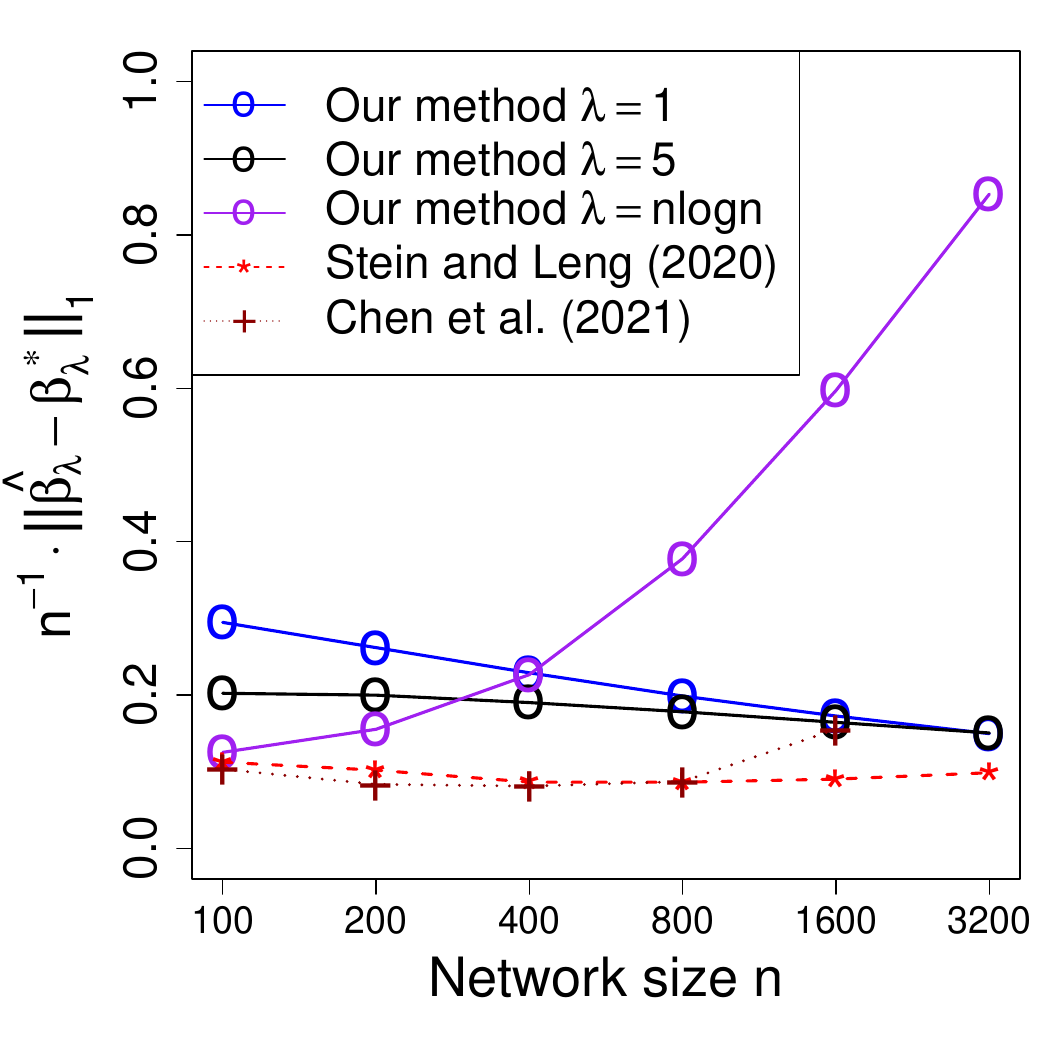}
    \includegraphics[width=0.235\textwidth]{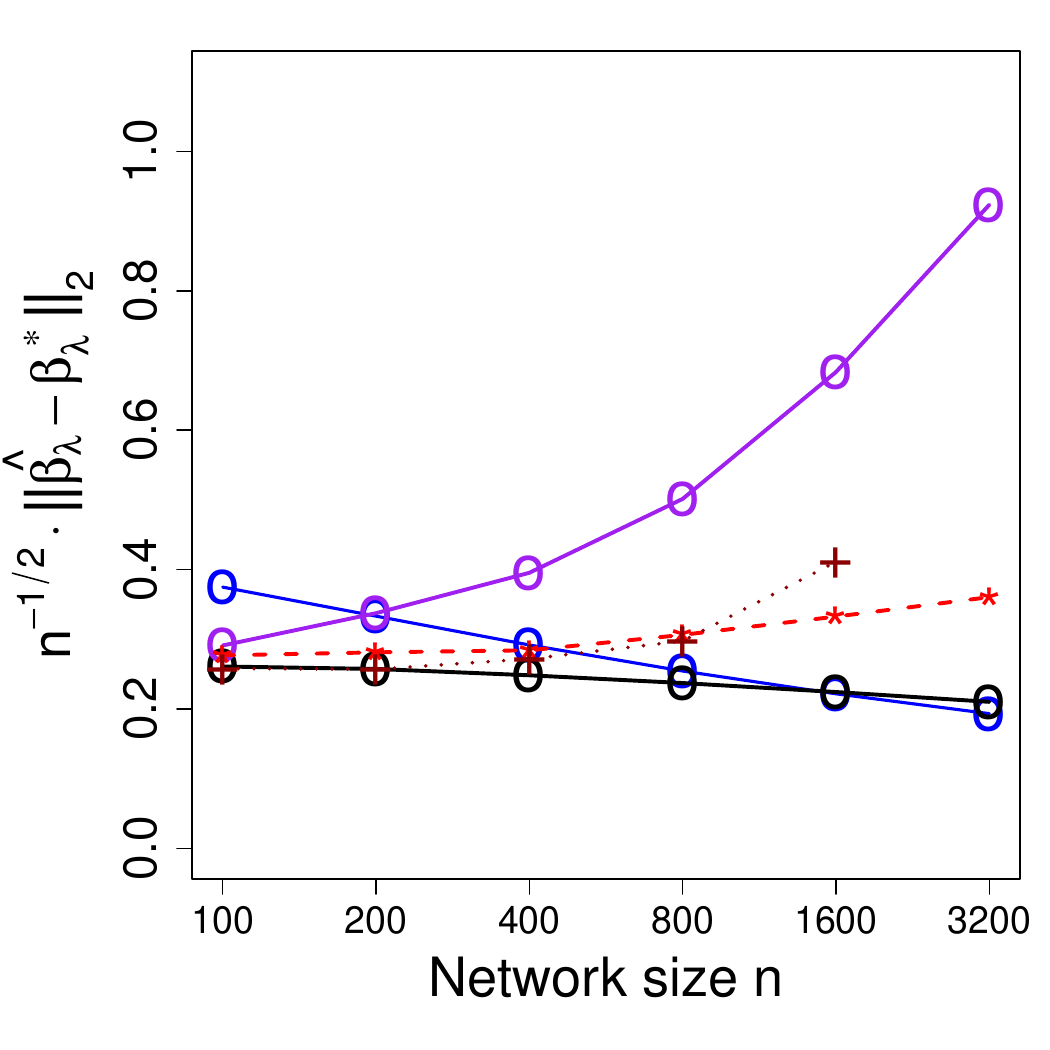}
    \includegraphics[width=0.235\textwidth]{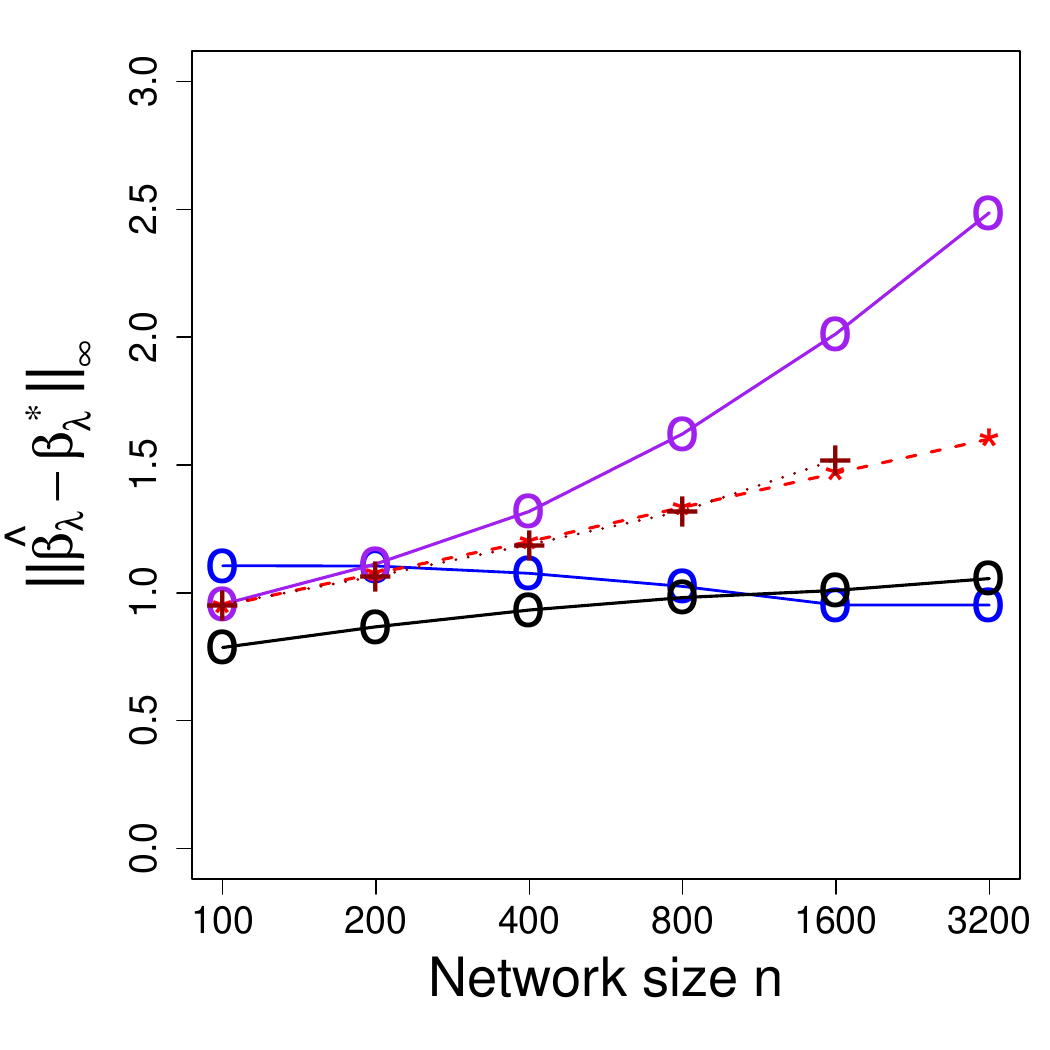}
    \includegraphics[width=0.235\textwidth]{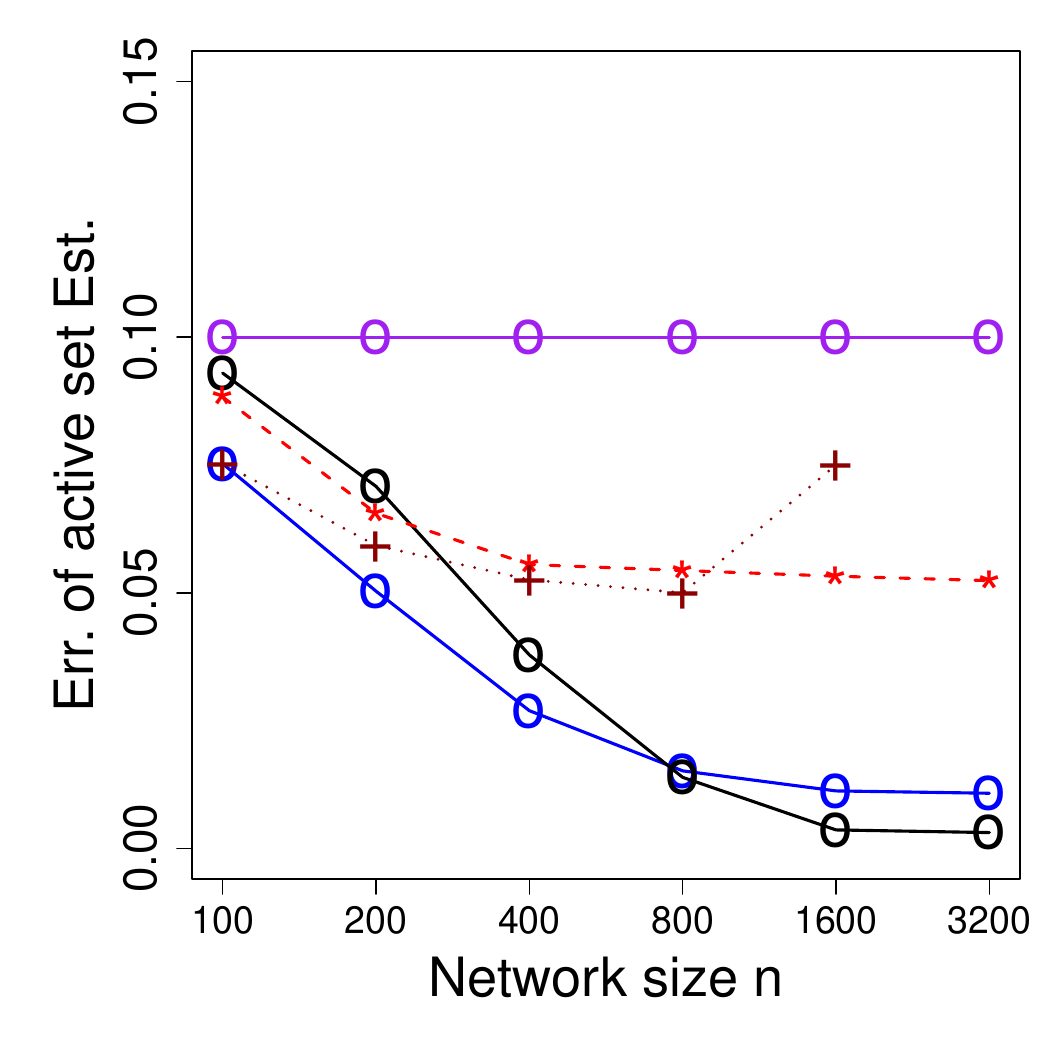}
    \includegraphics[width=0.235\textwidth]{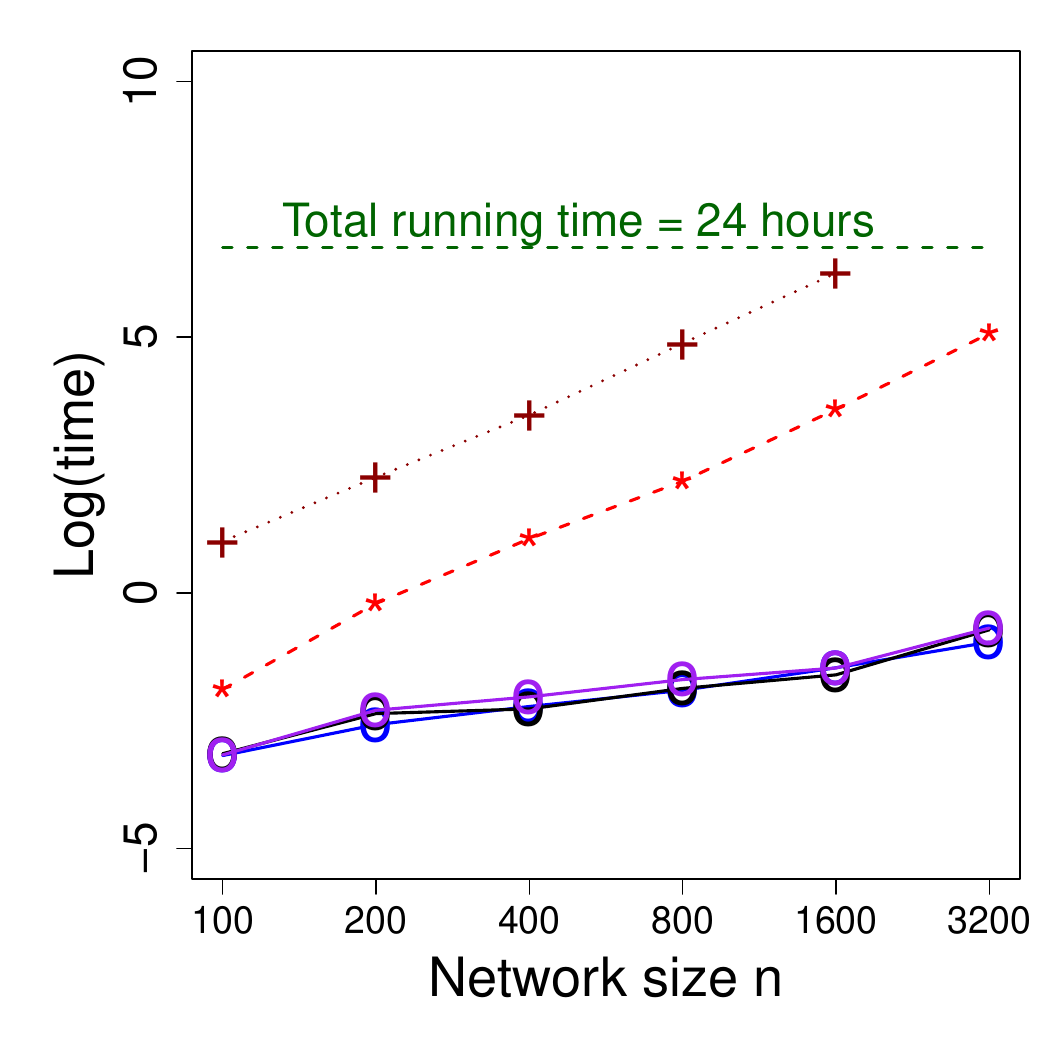}
    }
    \caption{Performance comparison with benchmark methods.  Row 1: setting 1; row 2: setting 2.}
    \label{fig::simu-3::comparison-error-to-others}
\end{figure}

Figure \ref{fig::simu-3::comparison-error-to-others} presents the result. 
In the denser network setting (row 1), where we set $\lambda=0$, our method consistently demonstrates superior or competitive performance in $\ell_1$, $\ell_2$ and $\ell_\infty$ errors, as evident in the first three plots.  
Notably, the fourth plot shows that unlike \citet{chen2019analysis} and \citet{stein2020sparse}, our Algorithm \ref{algorithm::thresholding} can accurately estimate ${\cal S}$ without requiring that $\beta_i^*-\beta_j^*>0$ for all $i\in {\cal S}$, $j\notin {\cal S}$.
The fifth plot demonstrates our method's clear advantage in scalability, while competing methods struggle with networks  having $n>10^3$ and may require an infeasible amount of memory when $n\approx 10^4$.
In our experiments,  the competing methods begin timing out once $n\approx 10^3$.
In row 2, we test different choices of $\lambda$ for our method.  
Even in sparser networks, a small positive $\lambda$ consistently leads to the best performance across all metrics.  
This echoes the practical guidance of our Theorems \ref{Main-theorem::Linf-bound} and \ref{Main-theorem::L2-upper-bound},  which recommend setting $\lambda=O(1)$ in most scenarios, unless the network is extremely sparse and the true $\beta^*$ is nearly parallel to $\mathbbm{1}$.
Setting $\lambda=n\log n$ would lead our method to approximately  fit an Erdos-Renyi model to the data, yielding $\ell_2$ and $\ell_\infty$ errors  growing at the rate of $O(\log n)$, as our theory predicts.

To further assess the robustness of our method in more general settings, we consider two additional settings without $\beta$-sparsity structure:
\begin{align}
    \textrm{Setting 3 (dense network): }
    \beta^*_i 
    =&~ 
    (-0.2\log n)/2 + \{-0.2+0.6 i/n\}\log n,
    \notag
    \\
    \textrm{Setting 4 (sparse network): }
    \beta^*_i 
    =&~ 
    (-0.6\log n)/2 + \{-0.2+0.4 i/n\}\log n.
    \notag
\end{align}
These configurations are similar to settings 1 and 2, but now the true $\beta_i^*$'s are evenly spread across the range.
Notice that the concept of \emph{active set} is not well-defined in this experiment, we therefore do not assess the accuracy of active set recovery.

\begin{figure}[h]
    \centering

    \makebox[\textwidth][c]{
    \includegraphics[width=0.235\textwidth]{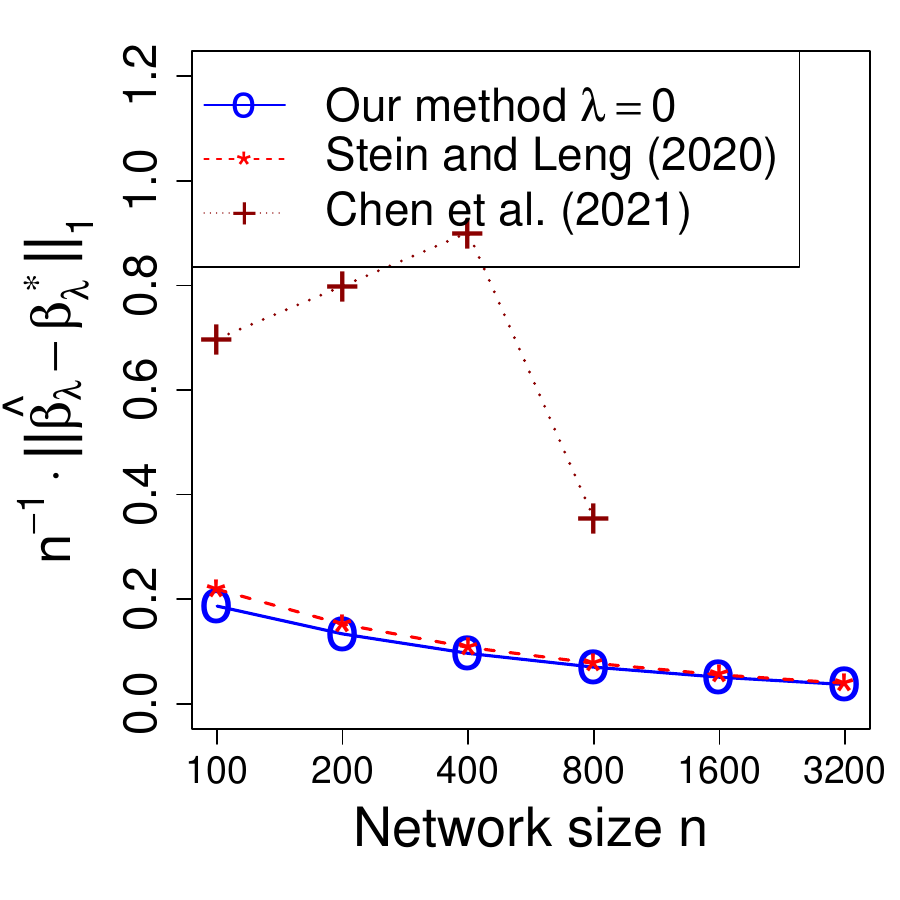}
    \includegraphics[width=0.235\textwidth]{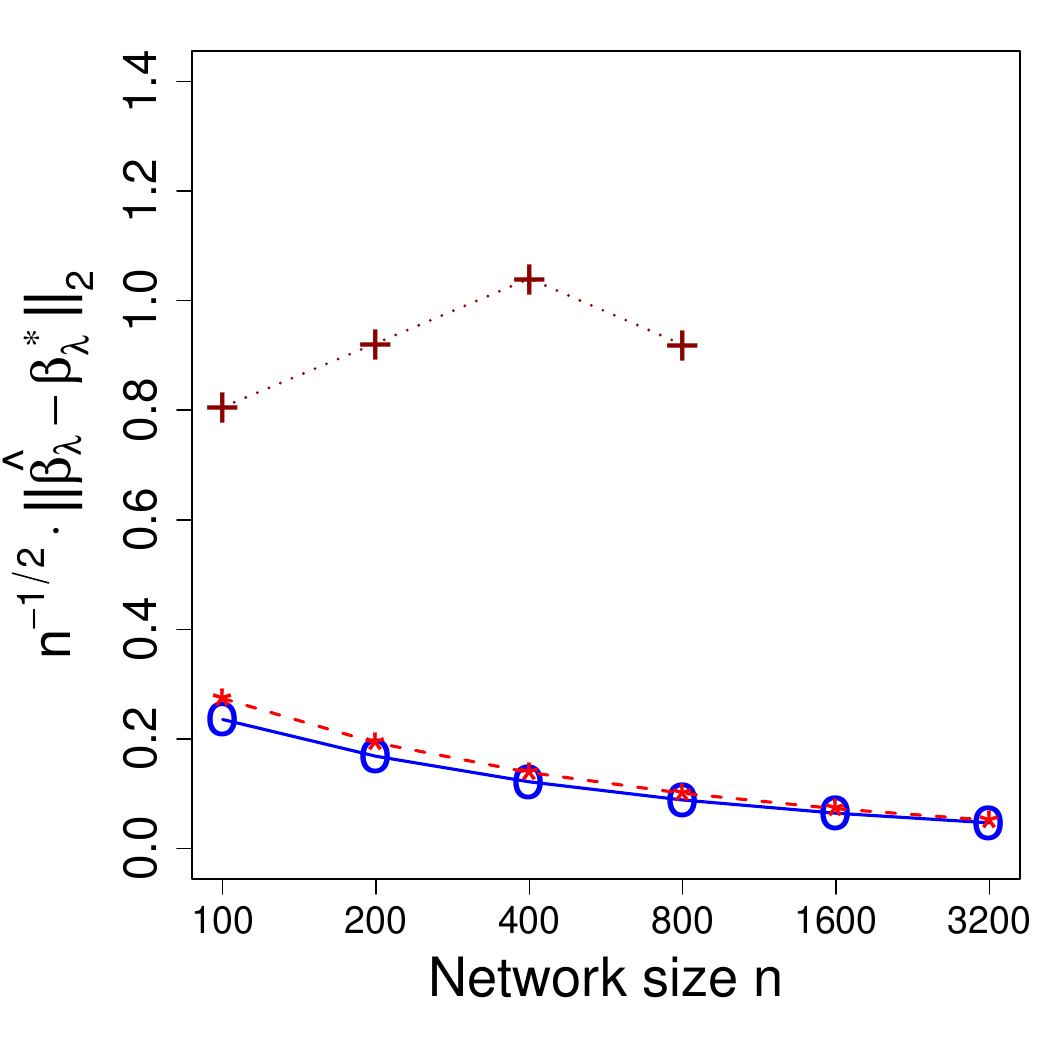}
    \includegraphics[width=0.235\textwidth]{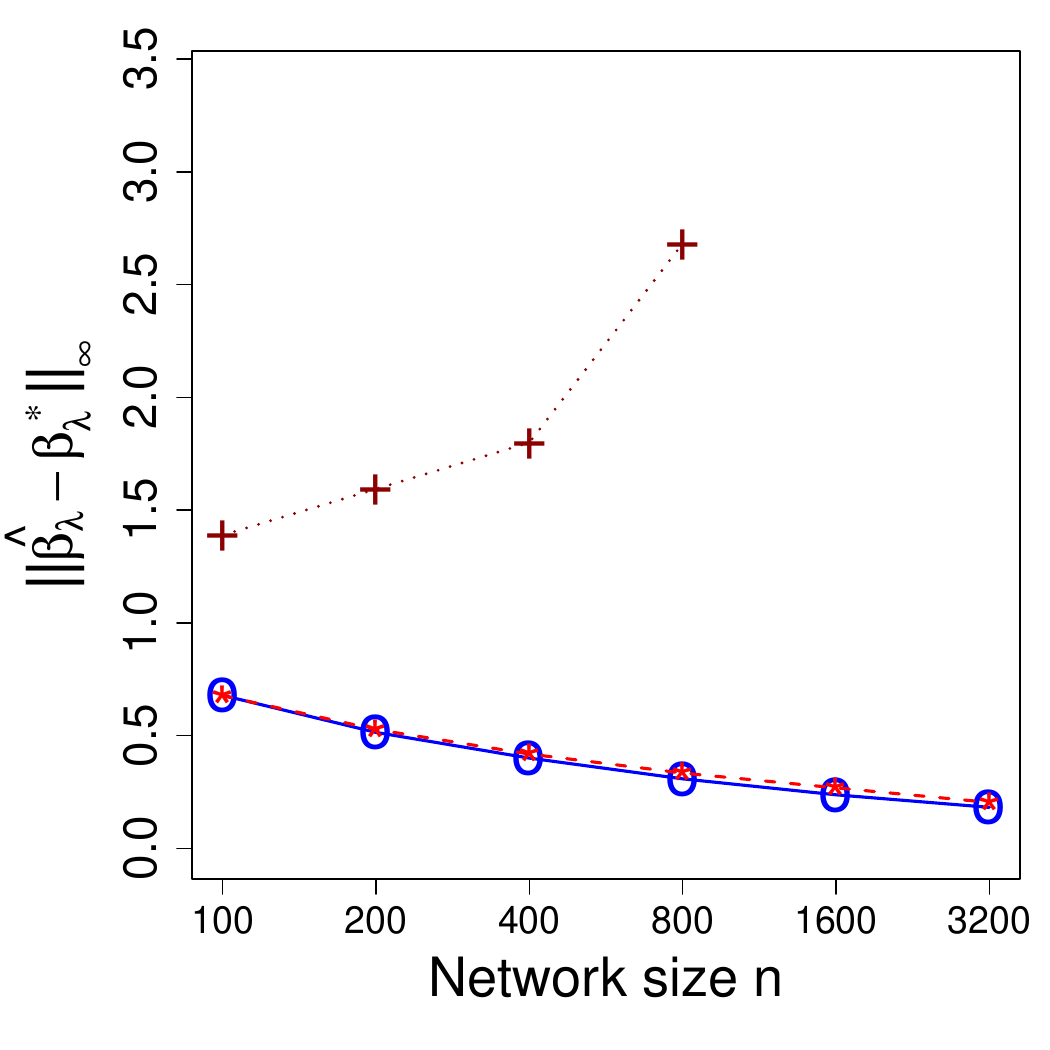}
    \includegraphics[width=0.235\textwidth]{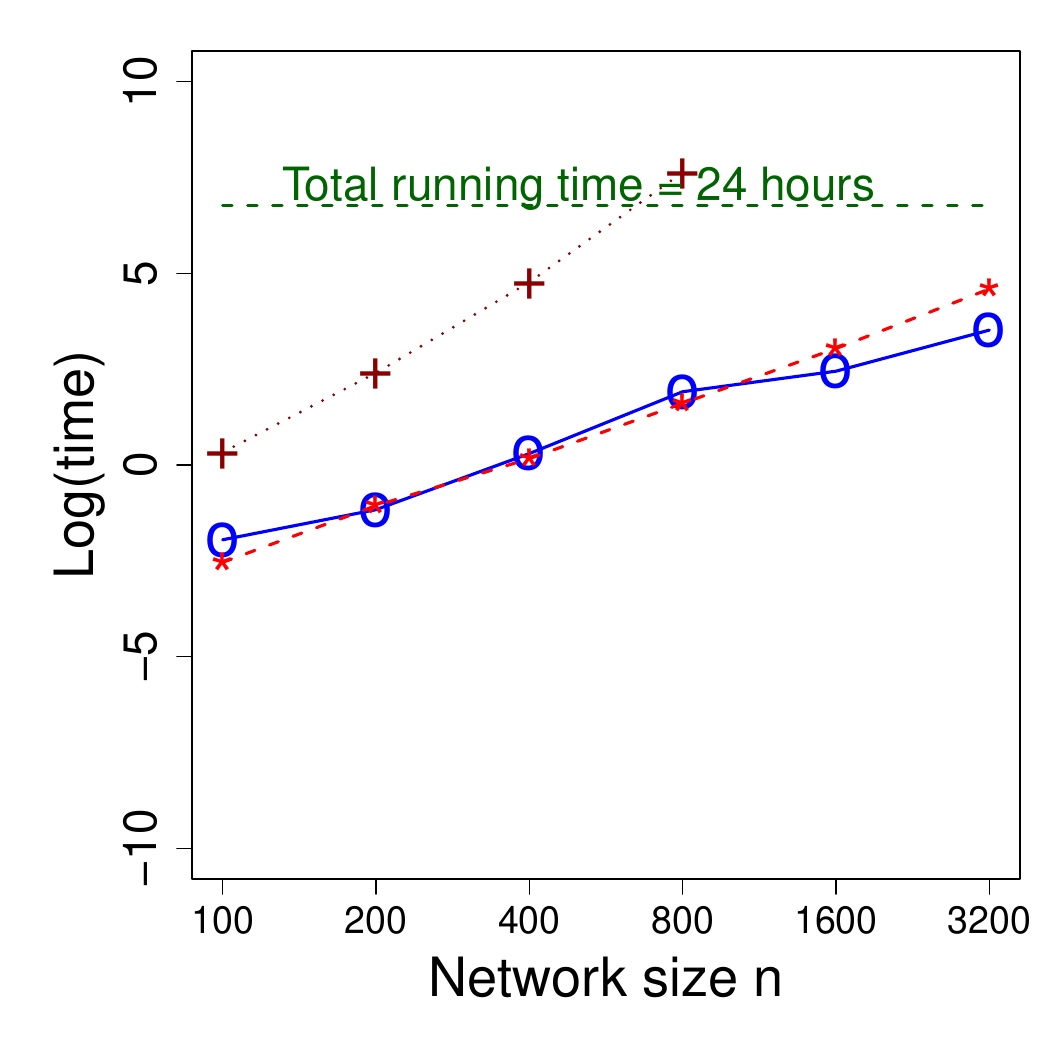}
    }
    \makebox[\textwidth][c]{
    \includegraphics[width=0.235\textwidth]{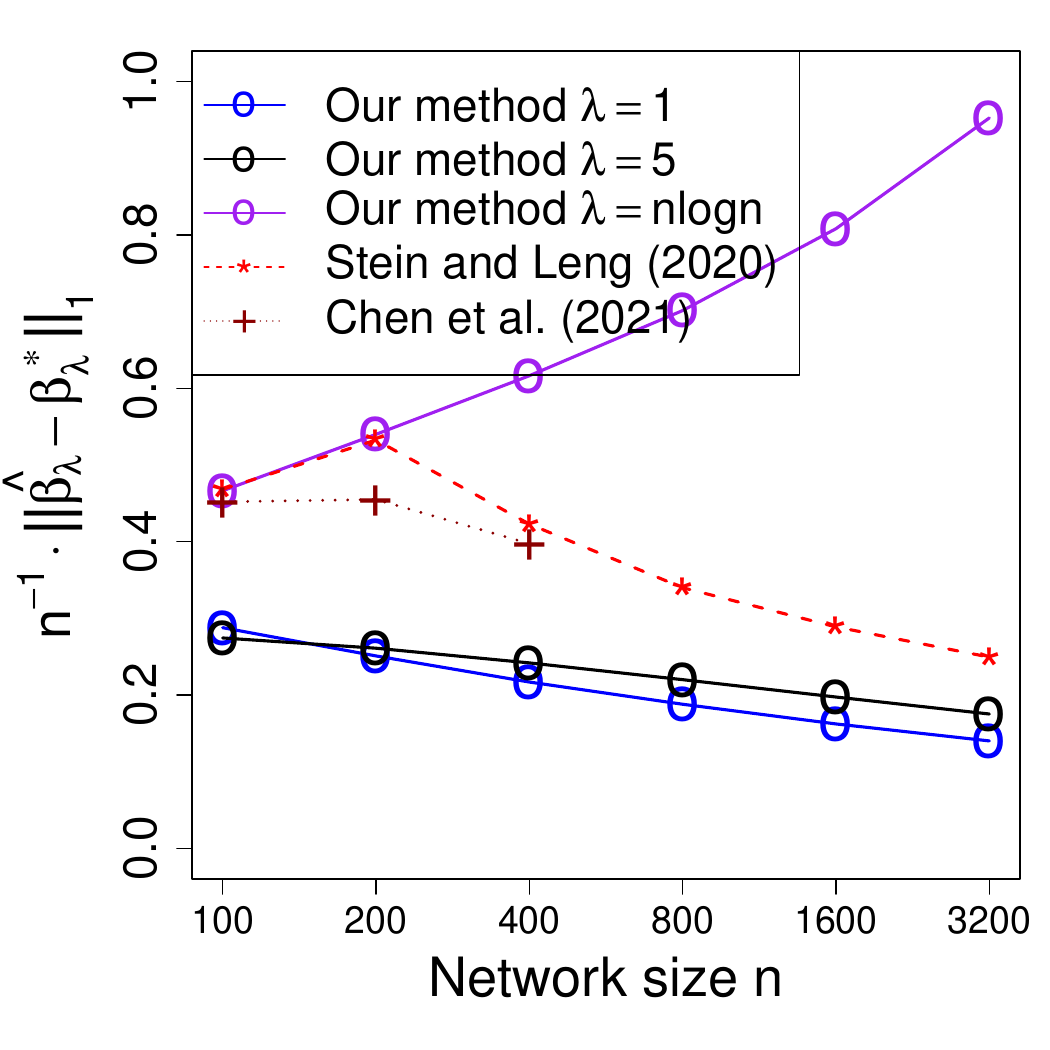}
    \includegraphics[width=0.235\textwidth]{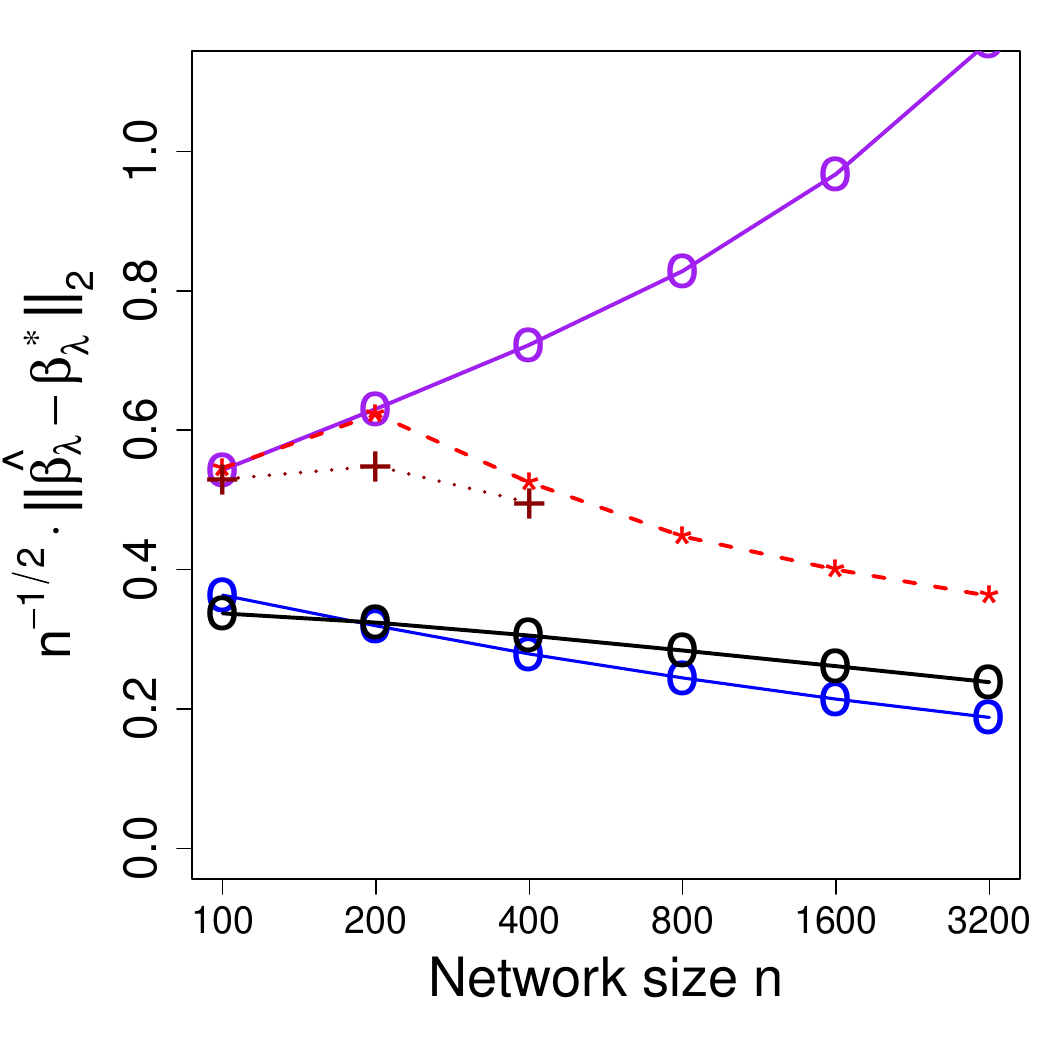}
    \includegraphics[width=0.235\textwidth]{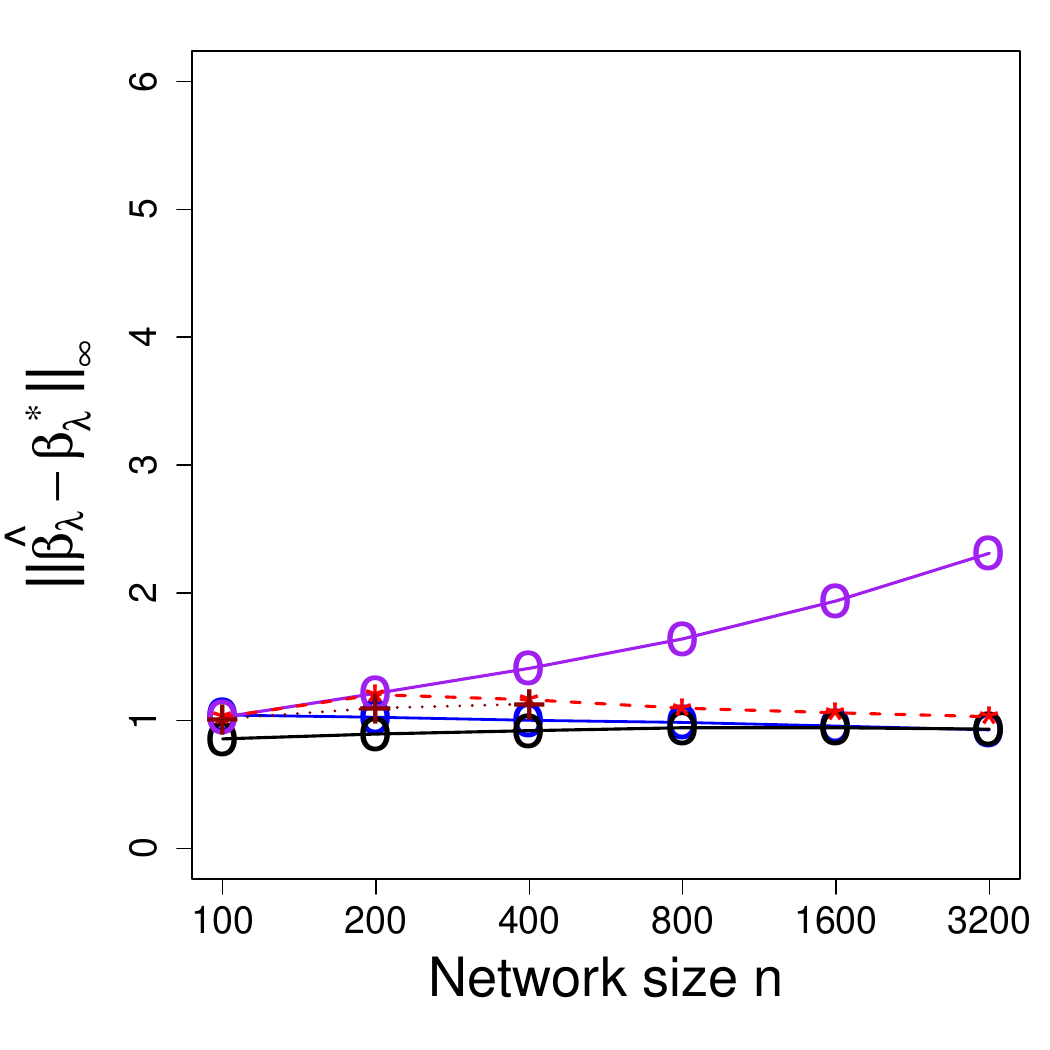}
    \includegraphics[width=0.235\textwidth]{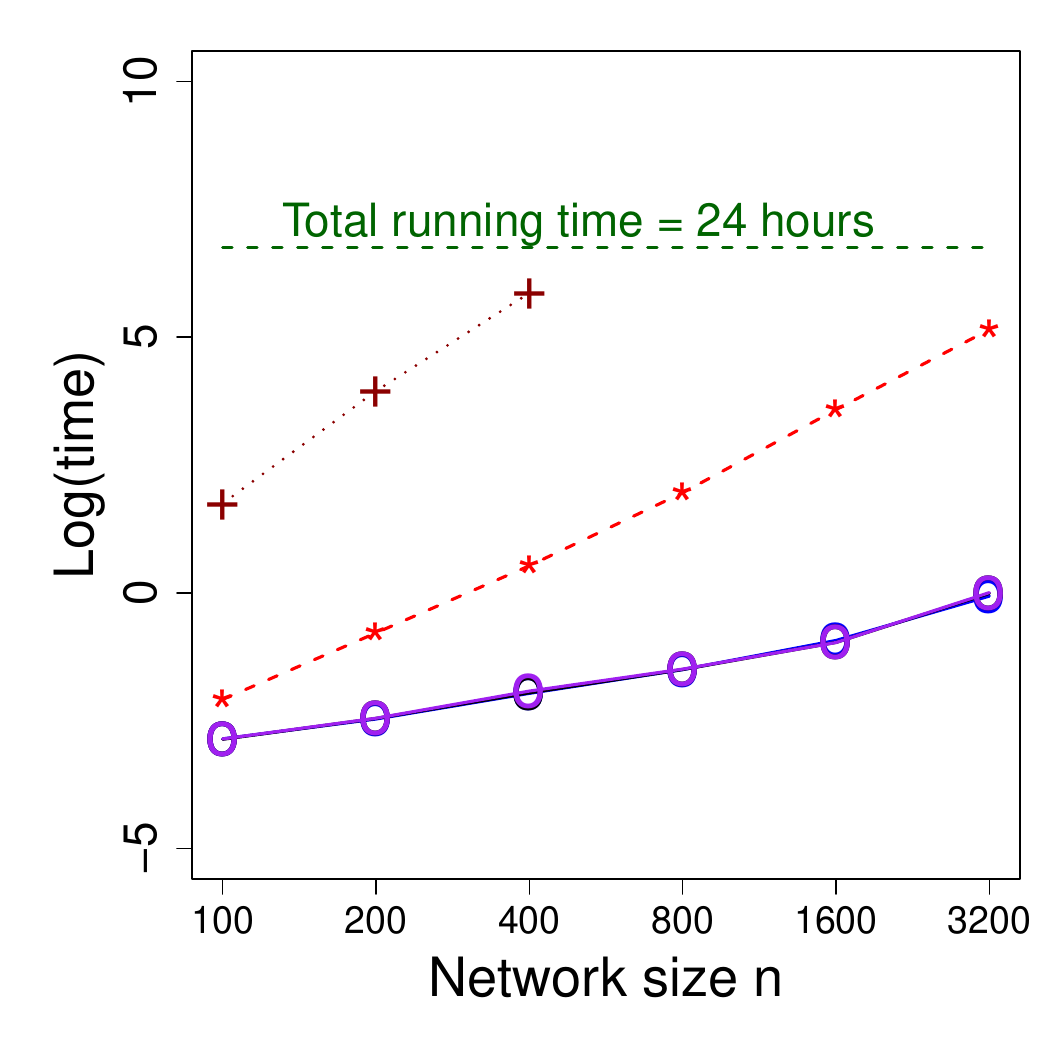}
    }
    \caption{Performance comparison with benchmark methods.   Row 1: setting 3; row 2: setting 4.}
    \label{fig::simu-4::comparison-error-to-others}
\end{figure}

Figure \ref{fig::simu-4::comparison-error-to-others} presents the results. 
Compared to Figure \ref{fig::simu-3::comparison-error-to-others}, here, our method achieves a substantially larger performance advantage over \citet{chen2019analysis,stein2020sparse}. 
This is expected: without the $\beta$-sparsity assumption (Assumption \ref{beta-sparsity-assumption}), those benchmark methods, which rely on first identifying the active set, tend to perform poorly. 
In contrast, our method demonstrates strong versatility and robustness in this more general setting.

\section{Data examples}
\label{section::data-examples}

\subsection{Data example 1: impact of COVID-19 on Swiss student mental health}
\label{section::data::data-1}

The dataset \citet{elmer2020students} contains  the social and psychological measurements from a local group of Swiss students.
The goal is to assess the impact of the COVID-19 lockdown on their mental health. 
The data is divided into three subgroups, dated 2019-04, 2019-09, and 2020-04.
Each subgroup consists of a set of students  (nodes).
On each node, we observe two types of variables:
sociality variables (self-reported connections in the following types: \emph{friendship}, \emph{pleasant interactions}, \emph{emotional support}, \emph{informational support}, and \emph{co-study});
and mental health variables (\emph{depression}, \emph{anxiety}, \emph{stress}, and \emph{loneliness}).
Each sociality variables  represents the degree of the node in the corresponding social network type, while the mental health variables serve as nodal covariates.
Since only node degrees were available, we could not apply benchmark methods  requiring full adjacency matrices, e.g., \citet{yan2019statistical, chen2019analysis, stein2020sparse}.

Next, we describe our data pre-processing steps.
For better comparability, we focus on the 2019-09 and 2020-04 subgroups,  which sample largely overlapping cohorts, while the 2019–04 subgroup surveyed a  completely different cohort.
The 2019-09 and 2020-04 subgroups include 207 and 271 students, respectively,  of whom 202 students  are shared.
To fully utilize the available data, we perform separate marginal analyses on each subgroup, followed by a differential analysis to compare the estimated parameters for the common students.
The networks of both groups are very sparse (Table \ref{tab::data-1::net-stats}).
 In particular, the \emph{emotional support}, \emph{informational support} and \emph{co-study} networks  are the sparsest.
To alleviate over-sparsity, we merged these three networks into a \emph{support} network, by summing  each node's degrees across the three networks, yielding  mean(std.) degrees of $5.51(4.33)$ (2019-09) and $5.46(4.63)$ (2020-04), respectively.

\begin{figure}[h!]
    \centering
    \includegraphics[width=\linewidth]{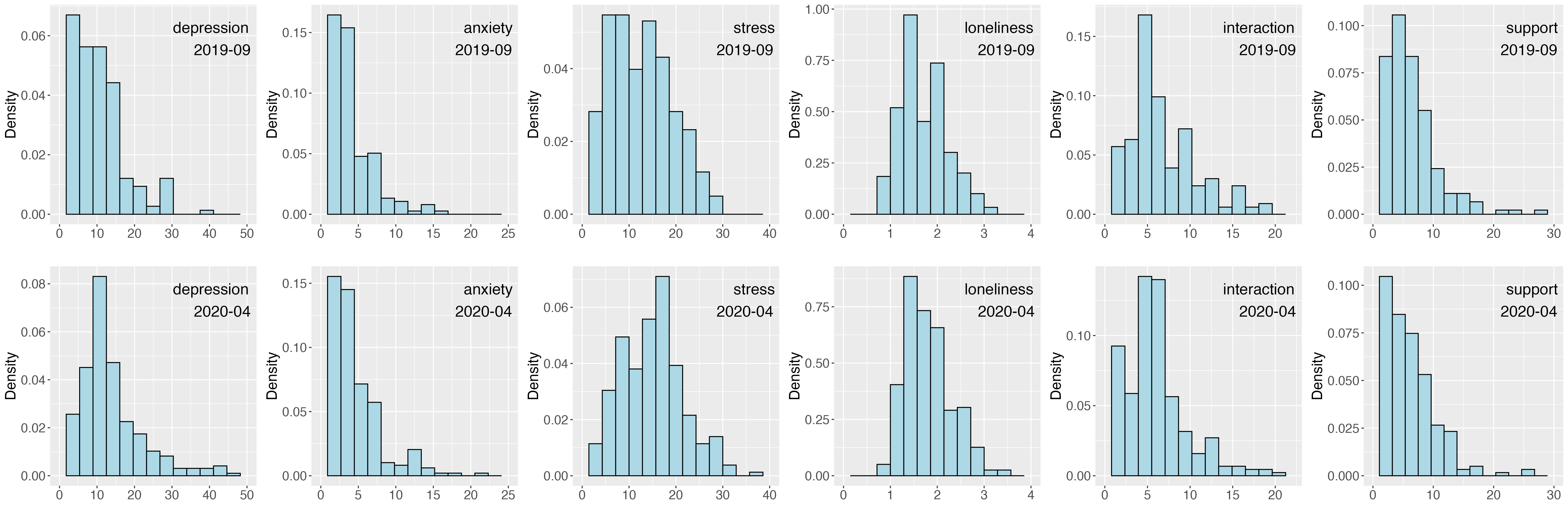}
    \caption{Swiss student data: empirical distributions of mental health variables and degrees.}
    \label{fig::swiss-degree-hist}
\end{figure}

\begin{figure}[h!]
    \centering
    \makebox[\textwidth][c]{
        \includegraphics[width=0.3\textwidth]{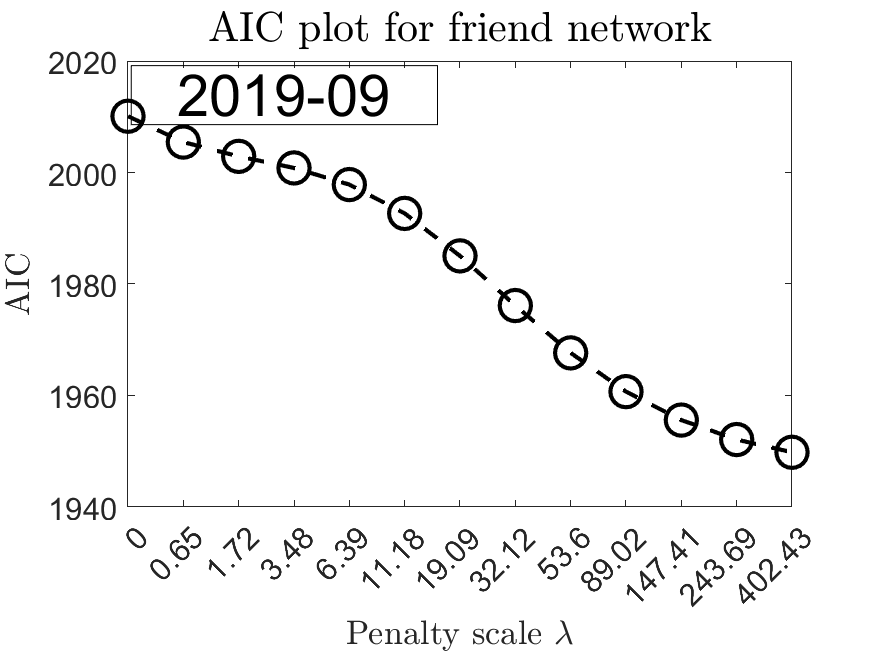}
        \includegraphics[width=0.3\textwidth]{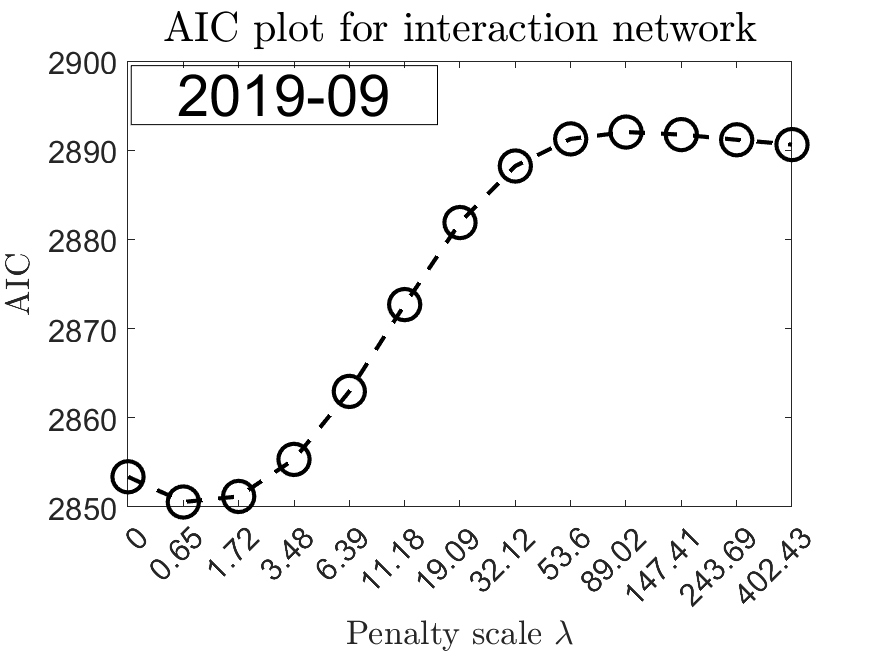}
        \includegraphics[width=0.3\textwidth]{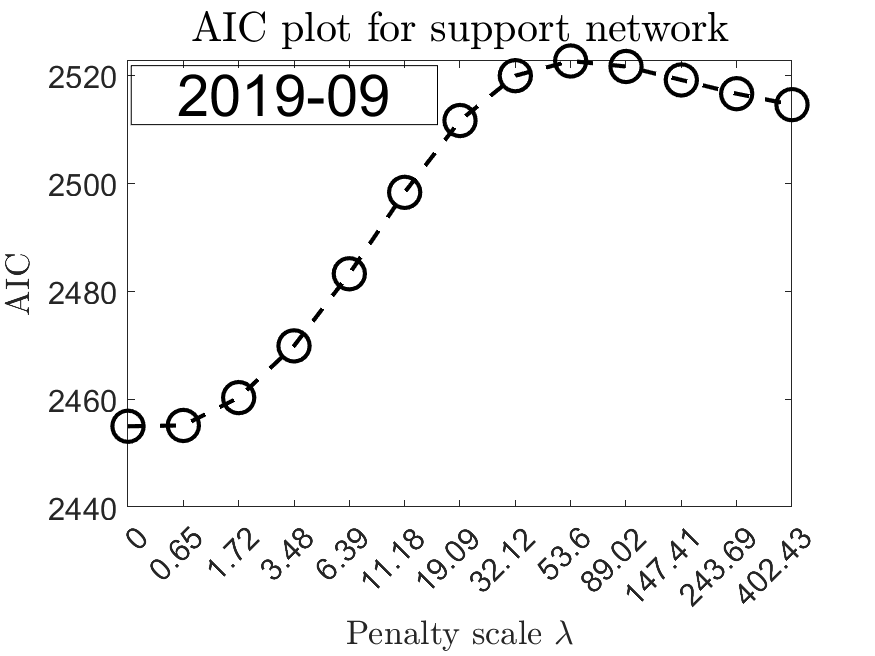}
        \includegraphics[width=0.3\textwidth]{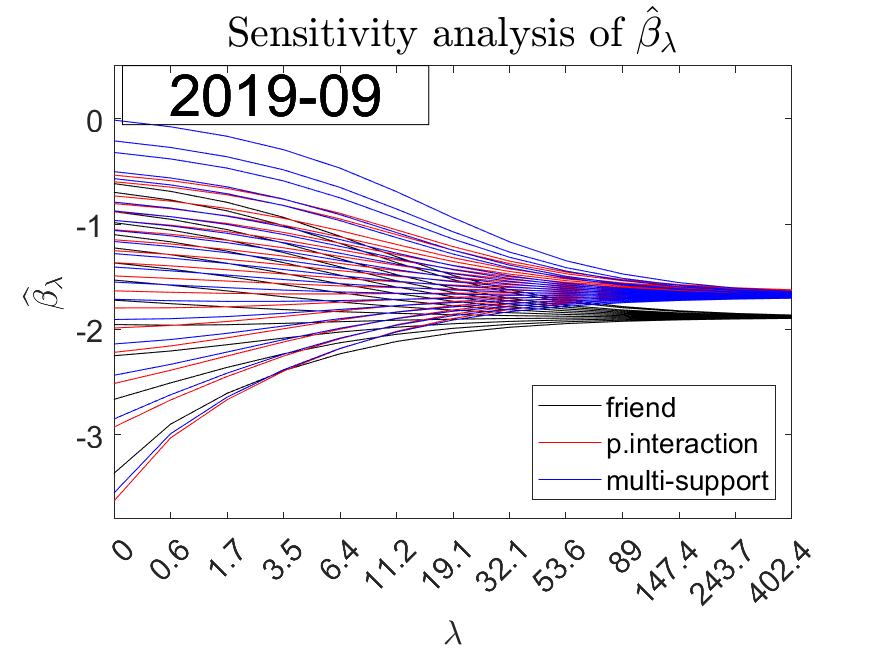}}\\
        \makebox[\textwidth][c]{
        \includegraphics[width=0.3\textwidth]{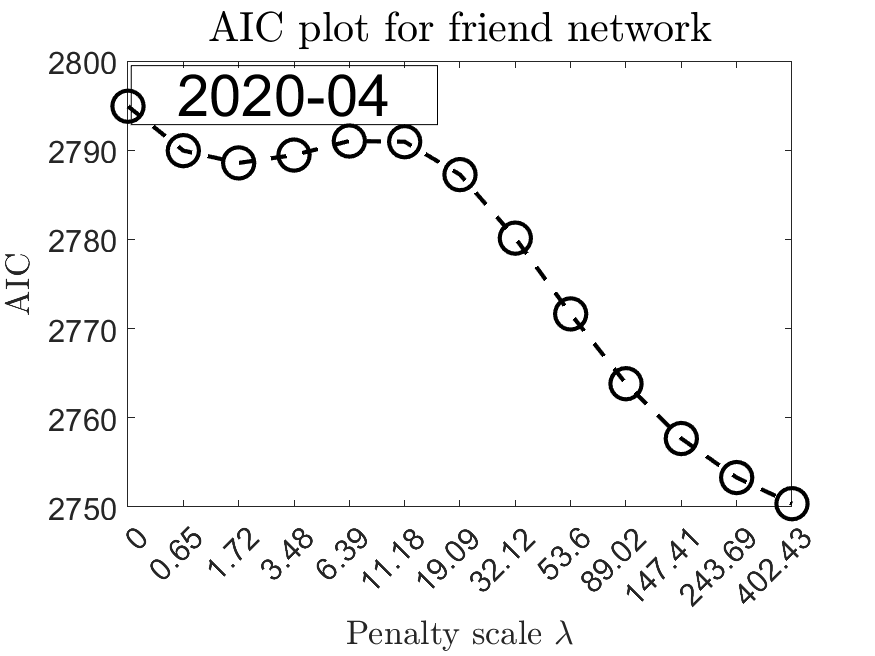}
        \includegraphics[width=0.3\textwidth]{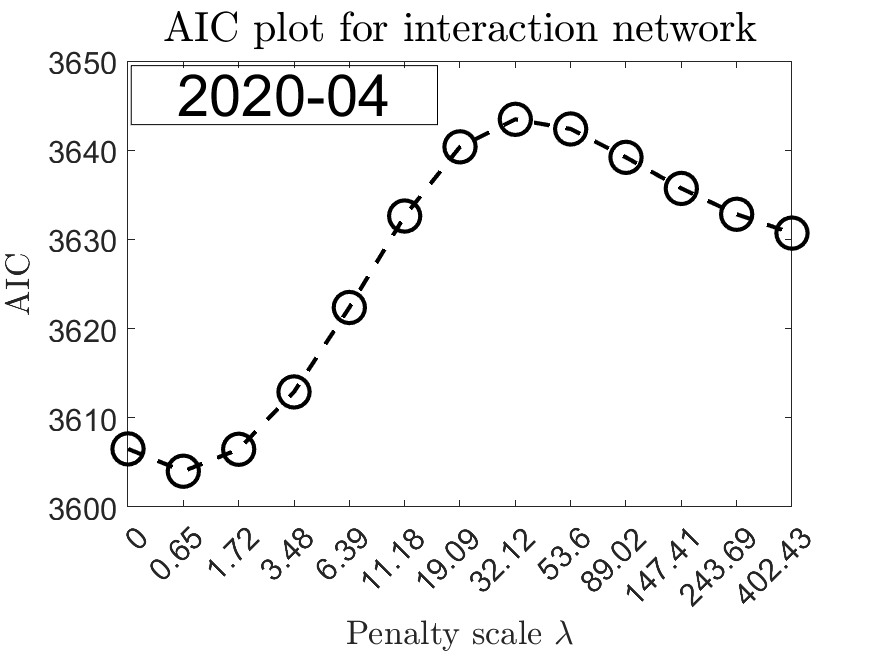}
        \includegraphics[width=0.3\textwidth]{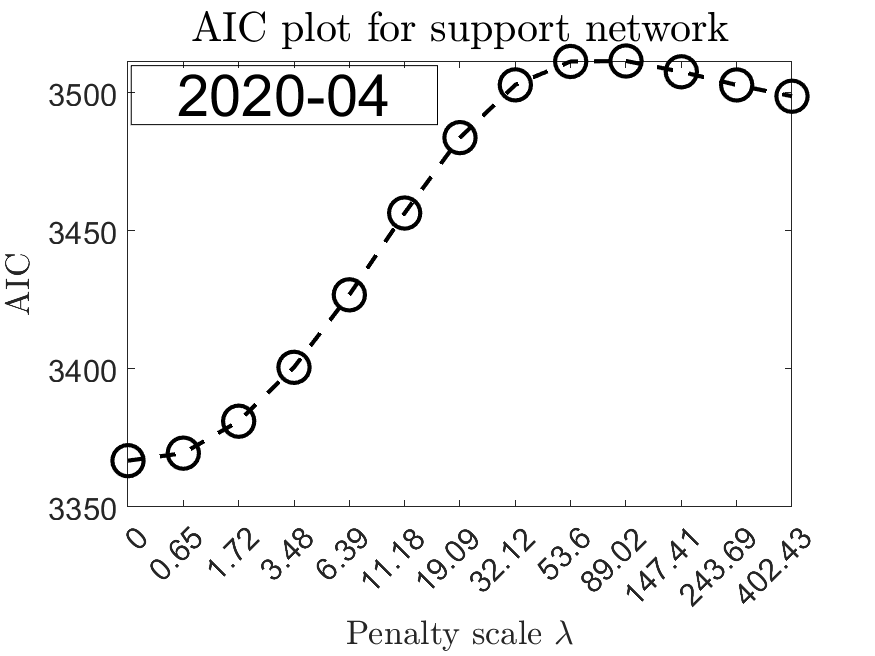}
        \includegraphics[width=0.3\textwidth]{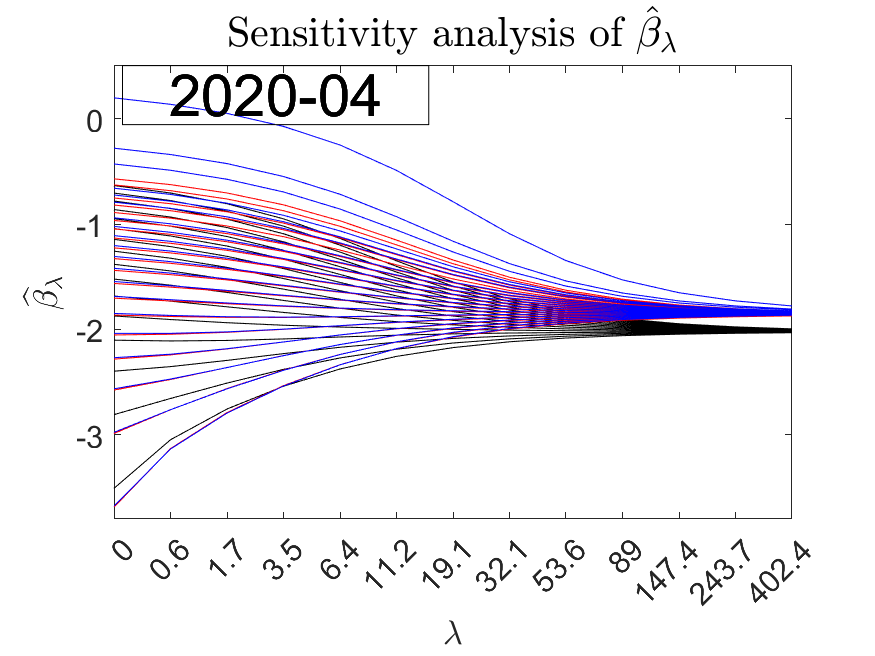}
        }
    \caption{
        Left three panels: AIC plots for tuning $\lambda$; right panel: tracks of $\hat\beta_\lambda$ over different $\lambda$ choices.  The left-most panel suggests that the friendship networks are too sparse for informative $\beta$-model fits.
    }
    \label{fig::data-1::lambda-sensitivity}
\end{figure}

Figure \ref{fig::swiss-degree-hist} shows quite continuous empirical degree distributions in \emph{interaction} and \emph{support} networks, suggesting that the $\beta$-sparsity assumption \citep{chen2019analysis, stein2020sparse} does not hold for this data.
Therefore, our proposed $\ell_2$-regularized MLE  is more appropriate.
We remove isolated nodes in each network and fit $\beta$-models to the social networks, with candidate choices of $\lambda \in \{0, e^{0.5}-1, \ldots, e^6-1\}$.
The first three columns in Figure \ref{fig::data-1::lambda-sensitivity} show the curves of our AIC-type criterion \eqref{our-AIC-criterion} for each network in the two subgroups. 
 We select $\lambda$ by minimizing this criterion.
Our AIC criterion suggests choosing a large $\lambda$ for \emph{friendship} network -- this suggests that this network may be too sparse and does not contain sufficient information for meaningfully fitting a $\beta$-model.
It is also conceptually difficult to combine \emph{friendship} with other types of networks.
Therefore, we exclude the \emph{friendship} network from further analysis.
The AIC plots suggest choosing $\lambda= 0.65$ for the \emph{interaction} network and $\lambda= 0$ for the \emph{support} network, respectively.

\begin{table}[htb]
    \centering
    \begin{tabular}{c|ccccc}\hline
        Variable & friend & p.interaction & e.support & inf.support & co-study  \\\hline
        Mean degree (2019-09) & 3.928  &  6.454  &  1.444  &  2.019  &  2.048 \\
        Std. dev. (2019-09) &  (2.554)  &  (4.225) & (1.503)  &  (1.692)    & (2.004)\\\hline
        Mean degree (2020-04) & 4.007  &  5.657  &  1.590  &  2.173  &  1.694\\
        Std. dev. (2020-04) & (2.878) & (3.810) & (1.749) & (1.835) & (2.071) \\\hline
    \end{tabular}
    \caption{Network degree statistics, each column represents a different sociality network.}
   \label{tab::data-1::net-stats}
\end{table}

Our goal is to study the relationship between mental health and sociality.
We use  the estimates $\hat\beta_\lambda$ from the \emph{interaction} and \emph{support} networks as covariates representing sociality.
To analyze their association with mental health variables, we then perform a sparse canonical correlation analysis (sparse CCA) \citep{witten2009penalized} on both 2019-09 and 2020-04 subgroups,  and on the differences in these covariates among the shared nodes.
We use {\tt CCA} function from the R package {\tt PMA} \citep{witten2009penalized}, which automatically tunes the regularization strength in its $\ell_1$-penalized sparse CCA procedure.
To assess significance, we employ a bootstrap  approach.
We randomly permute the rows of mental health variables (keeping sociality fixed) and compute empirical p-values  for each CCA coefficient.
The idea is similar to the permutation tests for linear models \citep{lei2021assumption, guan2023conformal}.

Table \ref{tab::data::student-mental-health} presents the result for the  first two canonical components.  
Across both time periods, the CCA consistently identified \emph{support} as having a stronger association with mental health variables than \emph{pleasant interaction}.
Before the lockdown, \emph{loneliness} showed the strongest correlation with sociality measures.
 Its significance diminished after the lockdown.
In contrast, \emph{stress}  rose to a leading role in the canonical correlation components after the lockdown.
From Figure \ref{fig::swiss-degree-hist}, we observe a notable change in the distribution of \emph{stress}, whereas the distribution of \emph{loneliness} changed less.
This suggests that the reduced significance in \emph{loneliness} may be due to the competing factor \emph{stress}.
Although \emph{depression} also showed a clear distributional shift, reflected in the differential analysis  of nodes common to both periods, it was never a prominent factor in the marginal analysis for either time period.
This aligns with the understanding that compared to \emph{depression}, the other mental health factors are more ubiquitous.
The differential analysis also reveals two pairs of most significant canonical  correlations: (\emph{loneliness}, \emph{pleasant interaction}) and (\emph{anxiety}, \emph{support}).
This finding, consistent with Figure 2 of \citet{elmer2020students}, suggests that increased \emph{loneliness} and \emph{anxiety} may have led to higher demands for \emph{pleasant interaction} and \emph{support}, respectively.

\begin{table}[htb]
    \adjustbox{max width=1.15\linewidth,width=1.15\linewidth,center=\linewidth}{
    \centering
    \caption{Estimated sparse CCA coefficients with empirical p-values.}
    \vspace{0.5em}
    \begin{tabular}{c|cc|cc|cc|cc|cc|cc}
        \hline
        & \multicolumn{8}{c|}{Marginal}
        & \multicolumn{4}{c}{Common nodes}
        \\\hline
        & \multicolumn{4}{c|}{2019-09}
        & \multicolumn{4}{c|}{2020-04}
        & \multicolumn{4}{c}{Difference}
        \\\hline
        Variable & CC1 & p-val.& CC2 & p-val.& CC1 & p-val. & CC2 & p-val. &  CC1 & p-val. &  CC2 & p-val.\\\hline
depression & $-$0.005 & (0.424) & 0.183 & (0.316) & 0.000 & (0.420) & 0.049  & (0.371) & 0.399 & (0.278) & 0.000 & (0.459)\\
anxiety & 0.453 & (0.288) & 0.139 & (0.377) & 0.529 & (0.289) & 0.181  & (0.361) & 0.000 & (0.460) & 0.941 & (0.179)\\
stress & $-$0.051 & (0.404) & $-$0.065 & (0.396) & 0.848 & (0.217) & 0.960  & (0.161) & $-$0.089 & (0.425) & 0.295 & (0.353)\\
loneliness & $-$0.890 & (0.244) & 0.971 & (0.149) & $-$0.023 & (0.448) & 0.210  & (0.385) & 0.913 & (0.193) & 0.164 & (0.409)\\
\hline
p.interaction & 0.000 & (0.506) & 1.000 & (0.000) & 0.000 & (0.508) & 1.000  & (0.000) & 1.000 & (0.000) & 0.000 & (0.505)\\
support & 1.000 & (0.000) & 0.000 & (0.494) & 1.000 & (0.000) & 0.000  & (0.492) & 0.000 & (0.495) & 1.000 & (0.000)\\\hline
\end{tabular}
    \label{tab::data::student-mental-health}
    }
\end{table}


\subsection{Data example 2: top concept extraction for Amazon COVID-19 knowledge graph}
\label{section::data::data-2}

\begin{figure}[htb]
    \centering
    \includegraphics[width=0.35\linewidth]{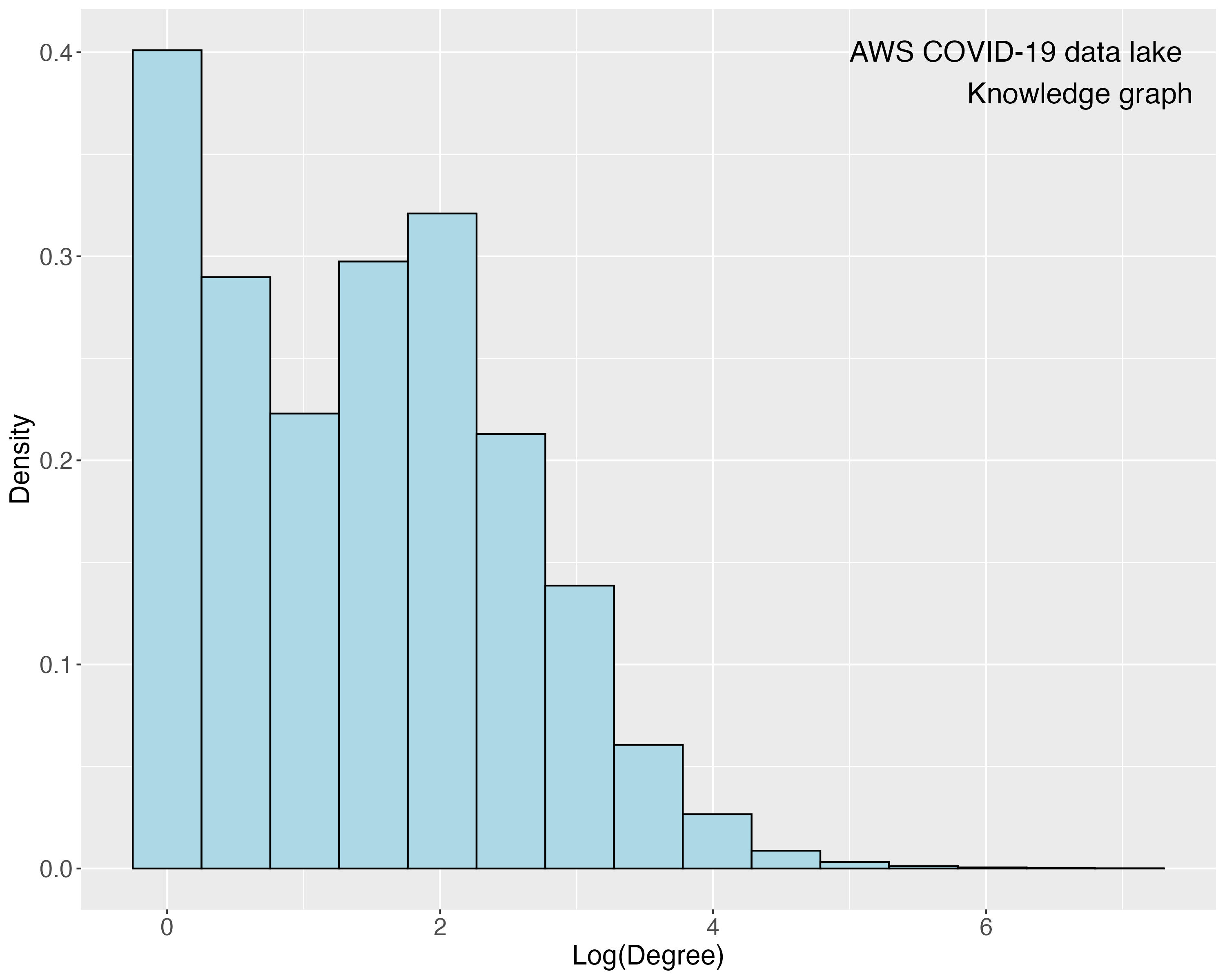}
    \caption{AWS data: empirical degree distribution of the citation network}
    \label{fig::aws-degree-hist}
\end{figure}

Our second example highlights the speed and memory  efficiency of our method.
The dataset was downloaded from the open-access Amazon data lake \citep{AWS_blog_1} at 20:06 UTC on 15th September 2021.
We focused on two files in the raw data:
\begin{itemize}
    \item File 1: \emph{paper citation network}:  following the practice of \citet{li2021undirected} and \citet{liu2019link}, we ignore all edge directions.
    \item File 2: \emph{paper-concept associations}:  each entry is a \emph{relevance score} between paper $i$ and concept $j$, denoted by $s_j^{(i)} \in [0,1]$.
    Each paper may feature several concepts with different relevance.
\end{itemize}
The citation network contains $n=57312$ nodes with only $171277$ edges.
The size and sparsity of this network make the $\beta$-model a natural choice.
The network size is prohibitive for most existing $\beta$-model algorithms \citep{yan2015asymptotic, chen2019analysis, stein2020sparse}.
In contrast, our Algorithm \ref{algorithm::thresholding} can complete  in minutes on a standard personal computer.
Moreover, the empirical degree distribution, illustrated in Figure \ref{fig::aws-degree-hist}, suggests that the $\beta$-sparsity assumption does not hold, and our method  is more suitable than those of \citet{chen2019analysis, stein2020sparse}.

Our goal is to find top concepts in the current literature.  
Intuitively, top concepts are those  that significantly associate with many hub nodes.
Therefore, we propose a \emph{Weighted Accumulative Beta Score (WABS)}, defined as follows.
For each concept $j$, let ${\cal A}_j\subseteq \{1,\ldots,n\}$ be the set of all papers associated with $j$.  
Set
\begin{equation}
    \textrm{WABS}_j = \sum_{i\in {\cal A}_j} e^{\hat\beta_{\lambda;i}}
     \times s_j^{(i)},
    \label{def::WABS}
\end{equation}
where $\hat\beta_\lambda$ was obtained by our method.
We provide some  intuition behind our design of \eqref{def::WABS}.
First, because relevance scores are weighted by the estimated popularity of the associated papers, one influential paper's contribution to WABS could easily outweigh the total contributions from many low-influence papers.
This  aligns well with the spirit of our design.
Second, compared to node degrees, the estimated $\beta$-model parameters are \emph{scale-free}  (i.e., potentially independent of network size).
If more data are collected, we could conveniently combine results across these networks of potentially very different sizes (e.g., by averaging the WABS values of corresponding concepts).

\begin{table}[h!]
  \centering
  \caption{Summary statistics for top- and bottom-50 concepts, ranked by \emph{Weighted Accumulative Beta Score} (WABS)}
    \adjustbox{max width=1.15\linewidth,width=1.15\linewidth,center=\linewidth}{
    \begin{tabular}{llrrr|llrrr}\hline
    \multicolumn{5}{c|}{Top 50}           & \multicolumn{5}{c}{Bottom 50} \\\hline
    \multicolumn{1}{c}{Concept} & \multicolumn{1}{c}{Category} & \multicolumn{1}{c}{\# of papers} & \multicolumn{1}{c}{Avg($d_i$)} & \multicolumn{1}{c|}{WABS} & \multicolumn{1}{c}{Concept} & \multicolumn{1}{c}{Category} & \multicolumn{1}{c}{\# of papers} & \multicolumn{1}{c}{Avg($d_i$)} & \multicolumn{1}{c}{WABS} \\
    \hline
infection & dx name & 18558 & 10.66 & 319.85 & methicilline & dx name & 1 & 3.00 & 0.01\\
respiratory syndrome & dx name & 6951 & 13.49 & 158.09 & icer & dx name & 2 & 2.00 & 0.01\\
death & dx name & 7689 & 10.92 & 126.06 & colostrum & dx name & 3 & 1.67 & 0.01\\
lung & system organ site & 6051 & 12.50 & 121.56 & economic injury & system organ site & 2 & 2.50 & 0.01\\
respiratory tract & system organ site & 5094 & 14.53 & 117.10 & cellmediated immunity & system organ site & 2 & 2.50 & 0.01\\
pneumonia & dx name & 4696 & 14.40 & 115.15 & mesenteric lymphatic & dx name & 2 & 2.00 & 0.01\\
fever & dx name & 5531 & 12.11 & 111.01 & perianal infection & dx name & 2 & 1.50 & 0.01\\
viral infection & dx name & 6669 & 11.02 & 109.37 & psychiatric treatment & dx name & 3 & 1.67 & 0.01\\
culture & test name & 5739 & 11.02 & 92.63 & potassium ion & test name & 1 & 5.00 & 0.01\\
cough & dx name & 3794 & 13.37 & 85.58 & lysine decarboxylase & dx name & 2 & 2.00 & 0.01\\
die & dx name & 4194 & 12.46 & 84.92 & fibrosis progression & dx name & 3 & 1.67 & 0.01\\
vaccine & treatment name & 5328 & 11.22 & 83.53 & gldh & treatment name & 2 & 1.50 & 0.01\\
infect & dx name & 5852 & 11.90 & 82.11 & neurobehavioral disorder & dx name & 1 & 3.00 & 0.01\\
liver & system organ site & 3910 & 11.09 & 69.14 & nr2b & system organ site & 2 & 2.00 & 0.01\\
diarrhea & dx name & 3244 & 12.59 & 67.96 & TRA & dx name & 2 & 2.50 & 0.01\\
hand & system organ site & 5347 & 9.37 & 67.51 & hcv replicon assay & system organ site & 2 & 2.50 & 0.01\\
kidney & system organ site & 3213 & 13.31 & 66.10 & Methylprednisolon & system organ site & 3 & 1.00 & 0.01\\
HIV & dx name & 4678 & 9.16 & 65.55 & platelet index & dx name & 2 & 1.50 & 0.01\\
respiratory infection & dx name & 3081 & 13.46 & 65.07 & immune tissue & dx name & 2 & 2.00 & 0.01\\
respiratory disease & dx name & 2686 & 15.10 & 61.62 & herpetic uveitis & dx name & 2 & 1.50 & 0.01\\
respiratory syncytial virus & dx name & 2900 & 12.91 & 54.82 & cadpr & dx name & 2 & 2.50 & 0.01\\
chest & system organ site & 2159 & 14.99 & 53.93 & rna expression profiling & system organ site & 3 & 1.67 & 0.01\\
rt-pcr & test name & 2613 & 15.71 & 53.55 & demostraron & test name & 2 & 2.00 & 0.01\\
infectious disease & dx name & 4766 & 8.55 & 51.79 & tnfsf4 & dx name & 1 & 3.00 & 0.01\\
throat & system organ site & 2201 & 14.78 & 50.78 & boutonneuse fever & system organ site & 2 & 2.50 & 0.01\\
heart & system organ site & 2955 & 10.64 & 50.61 & mucopolysaccharide & system organ site & 3 & 1.33 & 0.01\\
pcr & test name & 3381 & 11.65 & 50.27 & urethral mucosa & test name & 2 & 1.50 & 0.01\\
lesion & dx name & 2586 & 11.38 & 47.47 & covariance analysis & dx name & 3 & 1.33 & 0.01\\
titer & test name & 2681 & 12.25 & 47.05 & transfusion-associated circulatory overload & test name & 2 & 2.50 & 0.01\\
influenza virus & dx name & 3401 & 10.20 & 45.76 & Chondrex & dx name & 3 & 1.33 & 0.01\\
inflammation & dx name & 3500 & 8.51 & 45.29 & tea & dx name & 3 & 1.67 & 0.01\\
phylogenetic analysis & test name & 1825 & 16.37 & 43.70 & oxidovanadium & test name & 2 & 1.50 & 0.01\\
vomiting & dx name & 1869 & 13.33 & 42.37 & anti-tnf$\alpha$ drug & dx name & 1 & 4.00 & 0.01\\
respiratory tract infection & dx name & 2076 & 13.84 & 41.19 & control assay & dx name & 2 & 2.50 & 0.01\\
penicillin & generic name & 2534 & 9.63 & 40.83 & avanzadas & generic name & 3 & 1.33 & 0.01\\
adenovirus & dx name & 2275 & 12.65 & 40.71 & dystrophic neurite & dx name & 3 & 1.67 & 0.01\\
respiratory distress syndrome & dx name & 1248 & 19.52 & 39.50 & gastric erosion & dx name & 2 & 1.50 & 0.01\\
ribavirin & generic name & 1254 & 18.20 & 39.21 & foetal loss & generic name & 3 & 1.33 & 0.01\\
antibiotic & generic name & 3094 & 9.42 & 38.68 & alloreactive t cell & generic name & 2 & 2.50 & 0.01\\
rsv & dx name & 2345 & 11.79 & 38.49 & cefuroxima-axetilo & dx name & 2 & 2.00 & 0.01\\
serum sample & test name & 1822 & 13.51 & 37.97 & cerebrovascular complication & test name & 1 & 3.00 & 0.01\\
respiratory illness & dx name & 1668 & 15.17 & 37.39 & inadequate tissue oxygenation & dx name & 2 & 2.50 & 0.01\\
respiratory virus & dx name & 2101 & 13.47 & 36.86 & il-1$\beta$ concentration & dx name & 1 & 3.00 & 0.01\\
streptomycin & generic name & 2139 & 9.93 & 35.78 & hypertransfusion & generic name & 1 & 3.00 & 0.01\\
brain & system organ site & 2556 & 8.88 & 35.67 & vascular constriction & system organ site & 2 & 2.50 & 0.01\\
respiratory symptom & dx name & 1561 & 15.74 & 34.37 & flu peptide & dx name & 3 & 1.33 & 0.01\\
outbreak & dx name & 2472 & 12.09 & 33.75 & collagenous colitis & dx name & 2 & 1.50 & 0.01\\
membrane & system organ site & 2187 & 12.52 & 33.67 & facial nucleus & system organ site & 2 & 1.50 & 0.01\\
respiratory failure & dx name & 1143 & 17.85 & 33.39 & tricyclic compound & dx name & 2 & 1.50 & 0.01\\
bacterial infection & dx name & 2123 & 9.81 & 33.37 & ischemic heart failure & dx name & 2 & 2.00 & 0.01\\
    \hline
    \end{tabular}%
  \label{tab::data-2::AWS-WABS}%
  }
\end{table}%

Following the suggestion of the AIC track in Figure \ref{fig::data-2::choosing-lambda}, we select $\lambda=0$ for our method to obtain $\hat\beta_\lambda$ and then compute WABS scores.
Table \ref{tab::data-2::AWS-WABS} reports the top  50 and bottom 50 topics ranked by their WABS ratings.  
The outcome aligns well  with our intuitive understanding.  
For instance, the top-ranked list contains relevant organs such as \emph{respiratory tract}, \emph{lung} and \emph{throat} that are directly related to COVID-19 as a respiratory disease.
 It also includes organs like \emph{liver} and \emph{kidney} that are now widely believed also to be main  targets of the virus \citep{zhang2020liver, fan2020clinical, hirsch2020acute, pei2020renal}.  
Top-ranked concepts related to testing methods and treatments, including \emph{culture} (meaning \emph{viral culturing}), \emph{PCR}, \emph{phylogenetic analysis} (related to backtracking ancestors of the virus and monitoring latest variants) and \emph{vaccine}, also reflect the current mainstream approaches.
The other top entries cover important symptoms and related viruses.
In contrast, most concepts ranked at the bottom seem to lack either specificity, such as \emph{dx:pain}, \emph{procedure:surgery} and \emph{organ:skin}; or relevance, such as \emph{dx:injury}, \emph{treatment:chemotherapy}, \emph{dx:tumor} and \emph{organ:bone}.  
Some clearly irrelevant entries in the lower-ranked list, such as \emph{injury}, \emph{tumor} and \emph{cancer}, have comparable  comparable paper counts to some top-tier  concepts including \emph{respiratory illness}, \emph{phylogenetic analysis} and \emph{asymptomatic}.  
Our criterion WABS, based on our efficient $\beta$-model fitting for large networks, effectively recognizes the consistent presence of  these lower-ranked concepts in high-influence research papers.

\section{Discussion}
\label{section::discussion}

\begin{figure}[htb]
    \centering
    \makebox[\textwidth][c]{
    \includegraphics[width=0.3\textwidth]{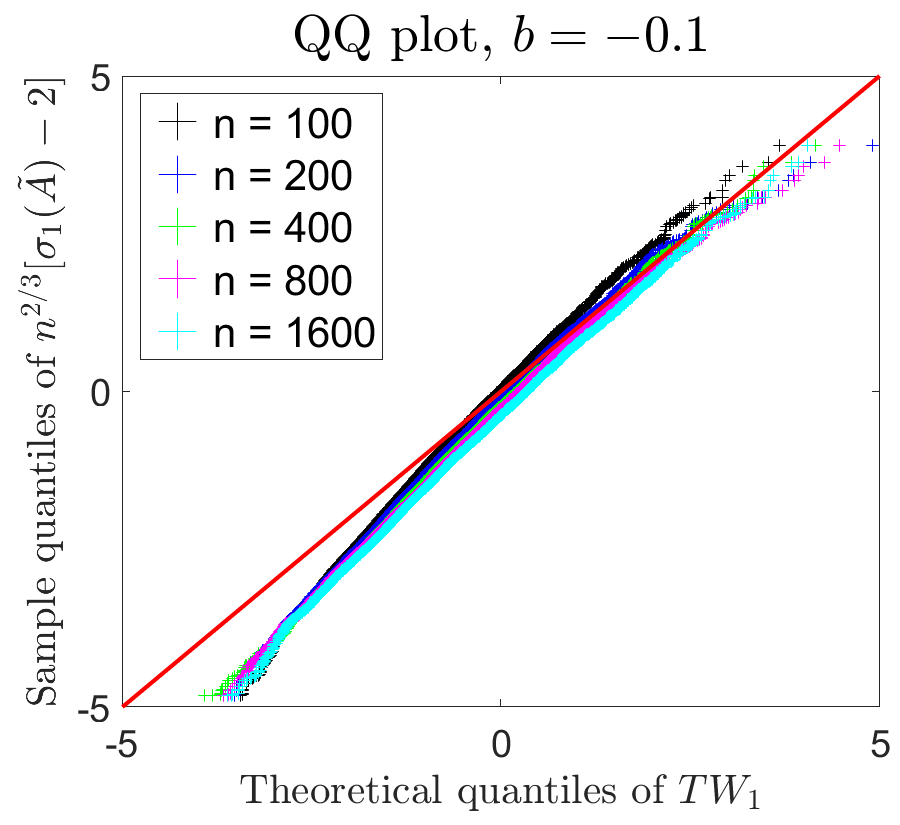}
    \includegraphics[width=0.3\textwidth]{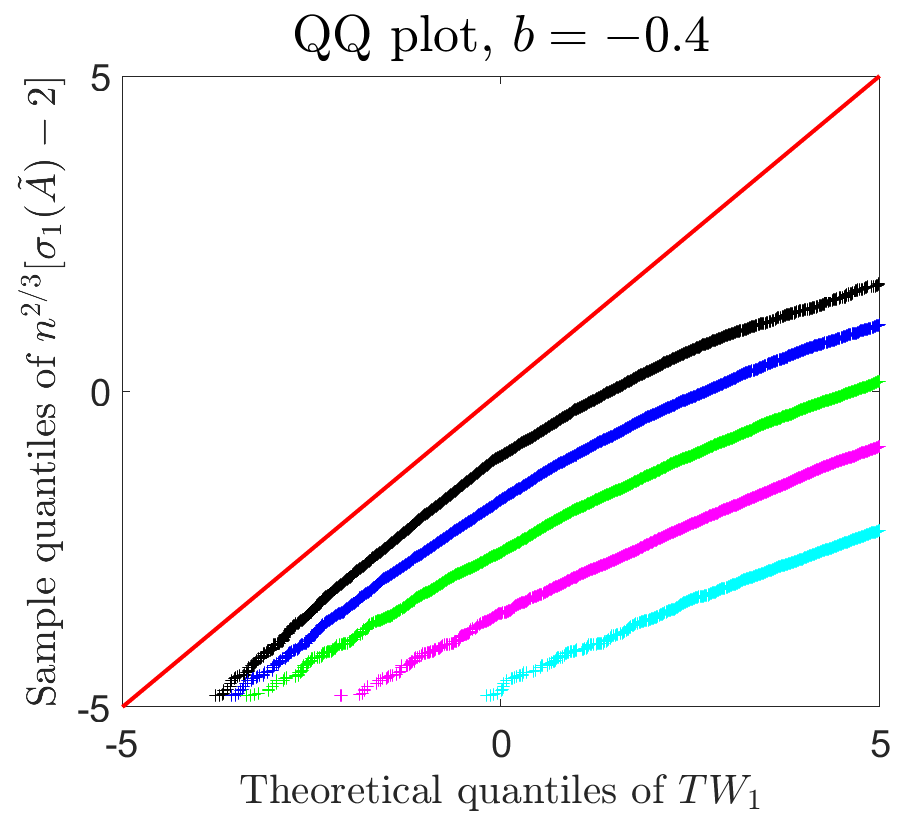}
    }
    \caption{QQ plot of empirical distribution of $T$ using $\hat P$ in \eqref{discussion::tilde-A} vs. ${\rm TW}_1$ distribution.}
    \label{fig::discussion::compare-to-tracy-widom}
\end{figure}

Goodness-of-fit test for $\beta$-models is an interesting but challenging topic  for future research.
For stochastic block models (SBM), this problem was addressed  by \citet{lei2016goodness}.
The largest singular value $\sigma_1(\tilde A)$ of a matrix $\tilde A$, defined by
\begin{align}
    \tilde{A}_{i,j} = \frac{A_{i,j}- P_{i,j}}{\sqrt{(n-1) P_{i,j}(1- P_{i,j})}},
    \quad\textrm{and}\quad \tilde A_{i,i} = 0,
    \label{discussion::tilde-A}
\end{align}
satisfies that $T := n^{2/3}(\sigma_1(\tilde{A})-2)\stackrel{d}\to {\rm TW}_1$, where ${\rm TW}_1$ is a Tracy-Widom distribution with parameter 1 \citep{erdHos2012rigidity, lee2014necessary}.
The validity of the method of \citet{lei2016goodness} critically depends on the fact that $\hat P_{i,j}$'s can be \emph{very accurately} estimated under SBM.
Consequently, the $P_{i,j}$'s in \eqref{discussion::tilde-A} can be replaced by their estimates without altering the limiting distribution of $\sigma_1(\tilde A)$.
Unfortunately, in $\beta$-models, our Theorem \ref{Main-theorem::local-lower-bounds} shows that it is \emph{impossible} to achieve the same level of estimation accuracy as the SBM setting. 
Intuitively, this is because a $\beta$-model has $n$ continuous parameters, compared to the $O(K^2)$ parameters in an SBM of $K$ communities\footnote{An SBM has $O(n)$ community labels, but under mild assumptions, they can be estimated very accurately.}.
\citet{lubold2021spectral} suggests using the plug-in test statistic  anyway, that is, replacing $P_{i,j}$ in \eqref{discussion::tilde-A} by $\hat P_{i,j} = e^{\hat\beta_{\lambda;i}+\hat\beta_{\lambda;j}}/\big(1+e^{\hat\beta_{\lambda;i}+\hat\beta_{\lambda;j}}\big)$.
To assess the validity of this approach, we generated data with $\beta_i^*=b\log n$ for $i\leq \lfloor 0.4n \rfloor$ and $\beta_j^*=0.1\log n$ for $j\geq \lfloor 0.4n \rfloor +1$.  
We varied $b\in\{-0.1, -0.4\}$ and $n \in \{100, 200, \cdots, 1600\}$,
and set $\lambda=0$.
For each $(n, \beta^*)$, we compared the empirical distribution of the test statistic $T$ (with $P$ replaced by $\hat P$) with TW$_1$ via $n_{MC} = 10^4$ Monte-Carlo repetitions. 
Figure \ref{fig::discussion::compare-to-tracy-widom} shows a  significant discrepancy between these two distributions and  suggests against the plug-in method by \citet{lubold2021spectral}.

In this paper, we focus exclusively on analyzing network data without nodal covariates.
 We restrict our scope for three reasons.
First, we believe that  a thorough understanding of a simpler model paves the way for future investigation of more complex ones.  
Research on the $\beta$-model without covariates still faces several open problems that would require considerable future effort to resolve.
Second, the joint modeling approach \citep{yan2019statistical,stein2020sparse,stein2021sparse} may encounter substantive difficulty in large and sparse networks, where the response is highly imbalanced, with  mostly zeros and few ones.  
Some treatments may be necessary to properly address this issue; see analogous discussions in classification \citep{sun2009classification}.
 Third, there is the issue of computation. 
As pointed out by \citet{stein2021sparse}, the monotonicity lemma would not hold for a joint model involving nodal covariates.  
Existing methods  seem insufficient for addressing the serious challenge of scalability.
 Handling large, sparse networks will require novel computational approaches.

Another  avenue for future work is to extend our method to bipartite and directed networks.  
Extension to bipartite networks should be relatively easy, as Lemma \ref{lemma::degree-monotonicity} also applies to the bipartite setting,  with minor modifications.  
Extension to directed networks, however, is more difficult, because there, the lack of symmetry $A_{i,j}\not\equiv A_{j,i}$ breaks Lemma \ref{lemma::degree-monotonicity}.  

Finally,  we focus exclusively on $\beta$-models with \emph{independent} edge generation.  
 A body of work addresses dependent edges \citep{frank1986markov,hunter2012computational,schweinberger2020concentration}.
 Introducing edge dependencies complicates estimation, inference, and even data generation from the model.
Much  effort is required to ensure method validity and scalability.

\section*{Computer code}
The computer code, including the full details for reproducing the simulation and data analysis results, is available at Meijia Shao's GitHub: \url{https://github.com/MjiaShao/L2-beta-model}.

\section*{Acknowledgements}
We thank the Editor, the Associate Editor and two anonymous referees for insightful comments that helped us improve the content and presentation of this paper.
We thank Chenlei Leng for sharing the code for \citet{chen2019analysis,stein2020sparse}.
We also thank him and Stefan Stein for providing insights on some technical details in \citet{stein2020sparse}.
We thank Mingli Chen, David S. Choi, Yoonkyung Lee, Elizaveta Levina and Subhabrata Sen for constructive discussions, and Steven MacEachern and Ji Zhu for warm encouragement.

\clearpage

\begin{center}
	\huge{\ \\\bigskip Supporting Information for\\``$L_2$ Regularized maximum likelihood for $\beta$-model estimation in large and sparse networks''}
\end{center}

\bigskip\bigskip
\section{Appendix:  Additional details on computation}
\label{section::additional-details-computation}
The gradient of $\tilde{\cal L}_\lambda(\beta)$ is
\begin{align}
    G_k(\delta)
    := &
    \dfrac{\partial \tilde{\cal L}_\lambda(\beta)}{\partial \delta_k}
    =
    \sum_{\substack{1\leq \ell\leq m\\\ell\neq k}} n_k n_\ell \dfrac{e^{\delta_k+\delta_\ell}}{1+e^{\delta_k+\delta_\ell}}
    +
    n_k(n_k-1) \dfrac{e^{2\delta_k}}{1+e^{2\delta_k}}
    -
    n_k d_{(k)}
    +
    n_k(\delta_k-\tilde \delta)\lambda,
    \label{eqn::IDEA::unknown-mu::gradient-element}
\end{align}
and its Hessian, denoted by $J$, is
\begin{align}
    J_{k\ell}(\delta)
    = 
    \dfrac{\partial^2 \tilde{\cal L}_\lambda(\delta)}{\partial\delta_k \partial\delta_\ell}
    = &~
    n_k n_\ell \dfrac{e^{\delta_k +\delta_\ell}}{\big(1+e^{\delta_k +\delta_\ell}\big)^2} - \dfrac{n_kn_\ell}{n}\lambda,
    \quad
    \textrm{ for }
    1\leq \{k\neq \ell\} \leq m,
    \\
    J_{kk}(\delta)
    = 
    \dfrac{\partial^2 \tilde{\cal L}_\lambda(\delta)}{\partial\delta_k^2}
    = &~
    \sum_{\substack{1\leq \ell\leq m\\\ell\neq k}}
    n_k n_\ell \dfrac{e^{\delta_k +\delta_\ell}}{\big(1+e^{\delta_k +\delta_\ell}\big)^2}
    +2n_k(n_k-1)\dfrac{e^{2\delta_k}}{\big(1+e^{2\delta_k}\big)^2}
    \notag\\
    &
    + n_k\Big(1-\dfrac{n_k}{n}\Big)\lambda,
    \quad
    \textrm{ for }
    1\leq k\leq m.
    \label{eqn::IDEA::unknown-mu::Jacobian-element}
\end{align}

\section{Appendix:  Analysis of the Kaggle CORD-19 data set}
\label{section::additional-data-kaggle}

This data set \citet{covid_kg} was transcribed from the well-known Kaggle CORD-19 data challenge in 2020,  
available from \url{https://www.kaggle.com/group16/covid19-literature-knowledge-graph}, which contains $n=1304155\approx$~1.3 million nodes.
Compared to the Amazon data set \citet{AWS_blog_1} that we analyzed in Main Paper, \citet{covid_kg} has a larger citation network but fewer nodal covariates and no map between papers and topics (keywords).  Therefore, its analysis is comparatively simpler.
However, in this ISWC transcription of Kaggle data, most of the 1.3 million nodes in the complete network here are not on COVID-19 but general medical literature.  We would use the complete network to estimate $\hat\beta_\lambda$;  afterwards, we use {\tt metadata.csv} in Kaggle CORD-19 open challenge downloaded from \url{https://www.kaggle.com/datasets/allen-institute-for-ai/CORD-19-research-challenge}
to filter and only keep COVID-19 papers, similar to the treatment in \citet{kg_notebook}.

\begin{figure}[htb]
    \centering
    \makebox[\textwidth][c]{
    \includegraphics[width=0.29\textwidth]{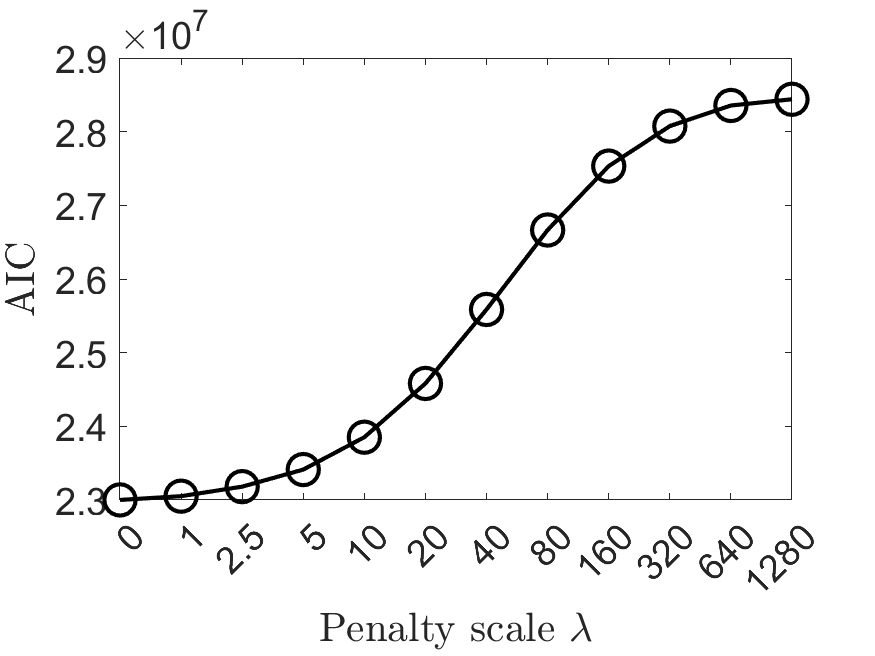}
    \includegraphics[width=0.29\textwidth]{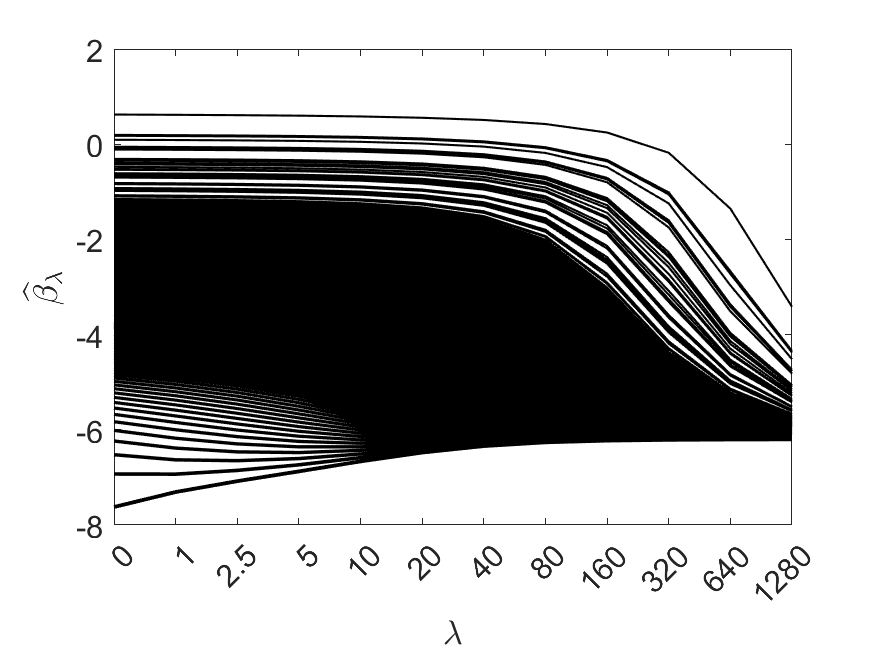}
    \includegraphics[width=0.29\textwidth]{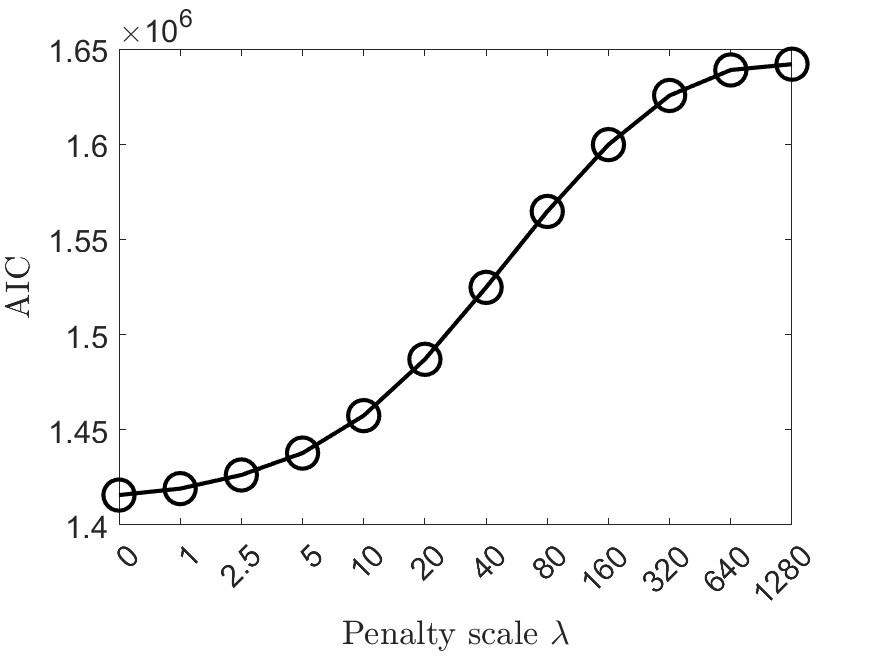}
    \includegraphics[width=0.29\textwidth]{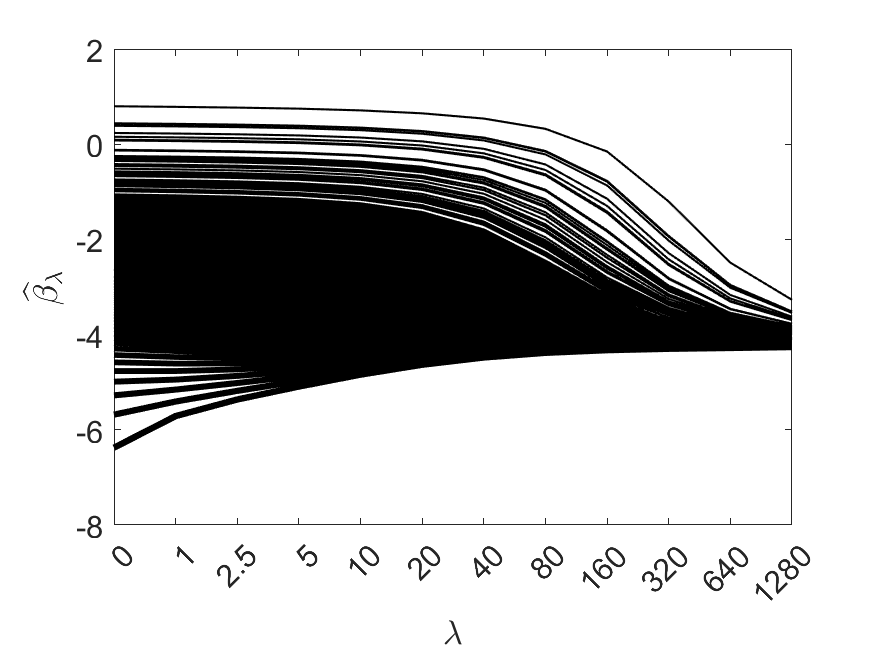}
    
    }
    \caption{Tracks of AIC$(\lambda)$ and $\hat\beta_\lambda$, Left two panels: \citet{covid_kg}; right two panels: \citet{AWS_blog_1}}
    \label{fig::data-2::choosing-lambda}
\end{figure}

Here, we focus on the nodal covariate \emph{country} and compare the empirical distributions of $\hat\beta_\lambda$ entries corresponding to different countries.  
We selected 6 representative countries/regions in the study of pandemic: UK, China, USA, EU (we counted France, Germany, Italy, Spain, Switzerland and Netherlands, which constitute the overwhelming majority of papers from EU), Japan plus South Korea, and India.
All other papers are collected by the ``Other'' category.
To choose a proper tuning parameter, we vary $\lambda\in\{0,1,2.5,\cdots,1280\}$ and plot the track of AIC$(\lambda)$ and $\hat\beta_\lambda$ in Figure \ref{fig::data-2::choosing-lambda}.  
Our AIC-type criterion suggests choosing $\lambda=0$. 
Then we run our accelerated Newton's method on the complete network, and filter $\hat\beta_\lambda$ entries using CORD-19 metadata as aforementioned.

\begin{table}[h!]
    \centering
    \caption{Estimated $\hat\beta_\lambda$ by region}
    \begin{tabular}{c|ccccccc}\hline
    Region & Other & China & US & UK & EU & JpKr & India \\\hline
Entry count & 53929 & 6096 & 15169 & 4747 & 11180 & 2394 & 1450\\
mean($\hat\beta_\lambda$) & $-$5.426 & $-$4.426 & $-$4.477 & $-$4.590 & $-$4.553 & $-$4.507 & $-$4.737 \\
std($\hat\beta_\lambda$) & (1.417) & (1.018) & (1.104) & (1.099) & (1.084) & (1.048) & (1.172) \\\hline
    \end{tabular}
    \label{tab::data-2::beta-est-by-region}
\end{table}

\begin{figure}[h!]
    \centering
    \makebox[\textwidth][c]{
    \includegraphics[width=0.4\textwidth]{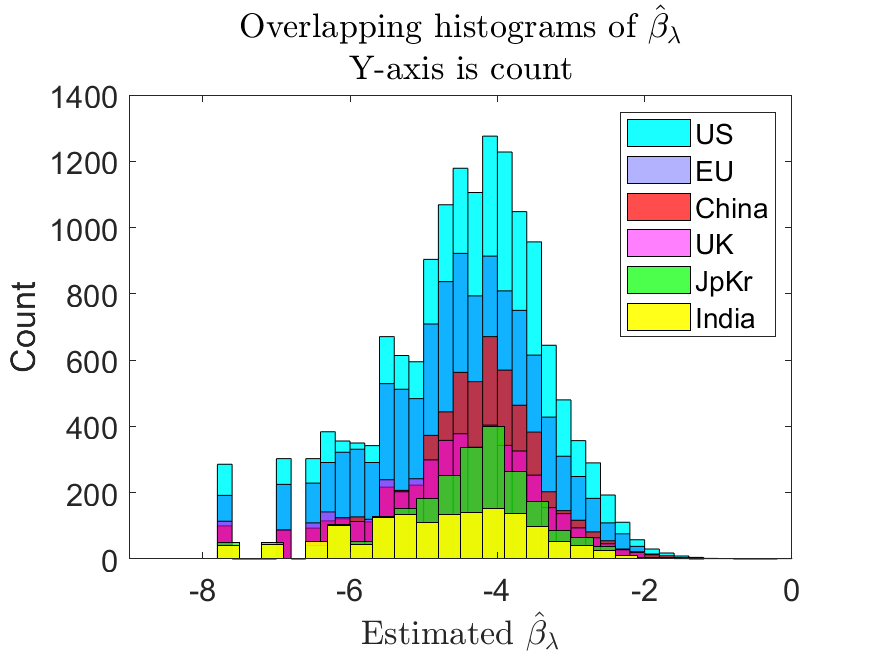}
    \includegraphics[width=0.4\textwidth]{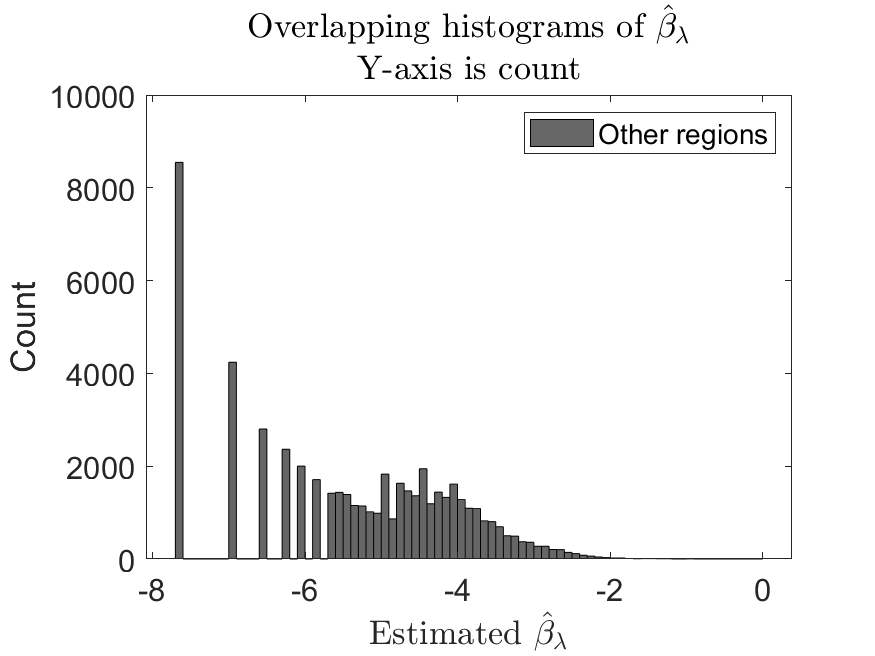}
    }
    \caption{Overlapping histograms of estimated $\hat\beta_\lambda$ by region}
    \label{fig::data-2::histogram-beta-est-by-region}
\end{figure}

Table \ref{tab::data-2::beta-est-by-region} reports the numerical summary of estimation results and Figure \ref{fig::data-2::histogram-beta-est-by-region} shows the region-wise histograms.  
Second, despite different total paper counts, the $\hat\beta_\lambda$ distributions across the 6 regions we studied show similar marginal distributions.  Inspecting the raw data, we understand that this can be partially attributed to the active international collaboration and mutual citation.  Overall, we see from Figure \ref{tab::data-2::beta-est-by-region} the clear evidence of solidarity and impartiality among scientists and researchers studying COVID-19.

\bibliography{all-ref}
\bibliographystyle{plainnat}

\end{document}